\documentclass{ucetd}

\usepackage{amsfonts,amssymb,amsmath,graphicx,color,bm,epsfig}
\usepackage[normalem]{ulem}
\usepackage{multirow}

\definecolor{ultramarine}{rgb}{0.07, 0.04, 0.56}
\definecolor{cadmiumgreen}{rgb}{0.0, 0.42, 0.24}
\definecolor{indigo(dye)}{rgb}{0.0, 0.25, 0.42}
\usepackage[linktocpage=true]{hyperref}
\hypersetup{
colorlinks=true,
citecolor=black,
linkcolor=black,
urlcolor=black,
}

\newcommand{\Mpl}{M_{\rm Pl}}
\renewcommand{\d}{\delta}
\newcommand{\e}{\epsilon_H}

\newcommand{\C}{\mathcal{C}}
\renewcommand{\O}{\mathcal{O}}
\newcommand{\Ha}{\mathcal{H}}
\renewcommand{\L}{\mathcal{L}}
\newcommand{\pa}{\partial}
\newcommand*\diff{\mathop{}\!d}
\newcommand*\diffcubed{\mathop{}\!d^3}
\newcommand*\difffour{\mathop{}\!d^4}
\newcommand*\re[1]{\mathop{}\!\mathrm{Re}\left[#1 \right]}
\newcommand*\im[1]{\mathop{}\!\mathrm{Im}\left[#1 \right]}
\newcommand\numberthis{\addtocounter{equation}{1}\tag{\theequation}}

\newcommand{\End}{\rm end}
\newcommand{\curv}{\zeta}

\def\fncb{$\overline{\mathrm{FNC}}$}
\DeclareMathOperator\Arg{Arg}
\newcommand{\fNL}{f_{\rm NL}}
\newcommand{\Long}{{L}}
\newcommand{\Short}{{S}}
\newcommand{\Peak}{{\rm Peak}}
\newcommand{\kLong}{k_{\Long}}
\newcommand{\kShort}{k_{\Short}}
\newcommand{\kPeak}{k_{\Peak}}
\newcommand{\Squeezedks}{\kLong,\kShort,\kShort}
\newcommand{\USR}{\rm USR}
\newcommand{\SR}{\rm SR}
\newcommand{\Msun}{M_\odot}

\newcommand{\GeV}{{\rm \ GeV}}
\newcommand{\Mpc}{{\rm \ Mpc}}

\newcommand{\hcr}{h_{\rm max}}
\newcommand{\hcl}{h_{\rm cl}}
\newcommand{\hend}{h_{\rm end}}
\newcommand{\hrescue}{h_{\rm rescue}}
\newcommand{\SM}{{\rm SM}}
\newcommand{\BSM}{{\rm BSM}}
\newcommand{\Kelvin}{{\rm K}}

\newcommand{\VT}{V^T}
\newcommand{\widebar}[1]{\mkern 1.5mu\overline{\mkern-1.5mu#1\mkern-1.5mu}\mkern 1.5mu}
\newcommand{\medbar}[1]{\mkern 3mu\overline{\mkern-3mu#1\mkern-3mu}\mkern 3mu}
\newcommand{\N}{\mathcal{N}}

\title{The Black Hole Window on Cosmic Inflation}

\author{Samuel Passaglia}
\department{Astronomy and Astrophysics}
\division{Physical Sciences}
\degree{Doctor of Philosophy}
\date{December 2020}

\dedication{\large
\vspace*{.1\textheight} \begin{center} \begin{minipage}{\textwidth} \centering{To my grandparents\\ Abraham and Diana Friedman, \\ Martin and Alfreda Passaglia.} \end{minipage} \end{center} \vfill 
}

\epigraph{\large
\vspace*{.1\textheight} \begin{center} \begin{minipage}{\textwidth}\begin{quote} When the mariner, sailing over tropic seas, looks for relief from his weary watch, he turns his eyes toward the southern cross, burning luridly above the tempest-vexed ocean. As the midnight approaches, the southern cross begins to bend, the whirling worlds change their places, and with starry finger-points the Almighty marks the passage of time upon the dial of the universe, and though no bell may beat the glad tidings, the lookout knows that the midnight is passing and that relief and rest are close at hand.  

Let the people everywhere take heart of hope, for the cross is bending, the midnight is passing, and joy cometh with the morning. \end{quote} \begin{quote}{\hfill  --- E. V. Debs, Statement to the Court}\end{quote}\vfill  \end{minipage} \end{center} \vfill 
}

\committee{\large
\vfill \begin{center} \begin{minipage}{\textwidth} \centering{{\bf Committee in charge}\\
Professor Wayne Hu, Chair \\
Professor Daniel Hooper \\
Professor Edward W. Kolb \\
Professor Thomas Crawford} \end{minipage} \end{center} \vspace*{.1\textheight}
}

\begin{document}
\maketitle


\makecommittee
\makededication
\makeepigraph

\tableofcontents

\listoffigures
\listoftables

\acknowledgments

This thesis is the product of years of sweat. If at any point the logic blurs and the arguments become confused, charge it only to me and to my faults; but if anything of value is found herein, render the credit to my advisor Wayne Hu. I thank him for his kindness, which I will remember always.

Without my steadfast collaborator Hayato Motohashi, I would have collapsed long ago. Scott Dodelson brought me to Chicago and equipped me for the climb, and I was carried to the summit by my committeemen Dan Hooper, Rocky Kolb, and Tom Crawford. Laticia Rebeles ensured I arrived on time, or nearly so.

The students and postdocs of Wayne's group over the years, that is to say Andrew Long, Austin Joyce, Chen Heinrich, David Zegeye, Giampaolo Benevento, Jose Ezquiaga, Macarena Lagos, Marco Raveri, Meng-Xiang Lin, and especially Pavel Motloch, all received patiently my weekly progress reports and each offered in return their insight and their friendship. 

My officemates of the later period, Dimitrios Tanoglidis and Georgios Zacharegkas, supported me hour-by-hour through every trial. My officemates in the earlier period, Alessandro Manzotti and Ross Cawthon, refuse to fade into memory and I miss them deeply. 

H\'{e}ctor Ram\'{i}rez contributed directly to this work and through shared struggle revealed himself a true friend. Chihway Chang showed unparalleled loyalty to this distant straggler. Kimmy Wu and Kassa Betre treated me gently all along the way.

This journey began with the education I received at the University of Pennsylvania, foremost through the grace of my undergraduate advisor Adam Lidz. The legion of professors there who endeavored to train me in this profession were led by Justin Khoury and Brig Williams. I thank also the City of Philadelphia, which though known for its fraternal love, adopted me as a son.  

In the now distant past, Ron Revere instilled in me enough enthusiasm for the natural world to last a lifetime. Before then, the people of France, Japan, and Quebec took in a stranger and welcomed me as their guest. And throughout my life, the enduring friendship of Lucas Leblanc has kept me steady.

These years in Chicago would have been hollow were it not for John-Henry Pezzuto and Madison Inman; Ruben Waldman, Elizabeth Ashley, and Ashley Guo; Judit Prat and Eric Oberla. Giulia Longhi treated me like family and made the city feel like home. Mr.\@ Brightside and Runa, cats, allowed me to pick them up. Cosmic Microwave Background, though just a starving waif in Jackson Park, managed to do the same for me. 

Ogura Satoko worked selflessly and tirelessly every single day for two years to help me achieve a dream, and Yiqi Yan and Shimeng Xu were the best companions in that quest that anyone could ask for.

India Weston supported me and cared for me boundlessly, though I could offer to her nothing in return.

Eric Passaglia and Abigail Friedman, and my siblings Abraham and Marta, know me well enough that no more need be written here. Martin and Alfreda Passaglia have made sure that some corner of this world is always kept warm for me. I walk in Abraham and Diana Friedman's footsteps, and I think of them every day. 

Finally I thank you the reader, for your sympathetic regard.

\preface

I develop the principles governing the production of our universe's primordial inhomogeneities during its early phase of inflation. As a guiding thread I ask what physics during inflation can lead to perturbations so large that they form black holes in sufficient abundance to be the dark matter. Chapter~\ref{chap:intro} presents the simplest paradigm for inflation, a single canonical field which slowly rolls, and shows that it cannot produce primordial black hole dark matter. This thesis then proceeds by gradually relieving the assumptions in that simple model. 

In Chapter~\ref{chap:eft}, abridged from Refs.~\cite{Passaglia:2018afq,Ramirez:2018dxe}, I use an effective field theory approach to generalize the canonical single-field model to single-clock inflation, inflation with only one dynamical degree of freedom. I present simple expressions for the power spectrum and bispectrum of the perturbations when slow roll is only transiently violated. I show through an example that they can be applied to the immense variety of models encompassed by the effective field theory. Despite this diversity, forming primordial black holes in single-clock inflation still requires violating slow roll.

In Chapter~\ref{chap:usr}, abridged from Ref.~\cite{Passaglia:2018ixg}, I present the prototypical single-field model which does produce primordial black holes by violating slow-roll, a model known as ultra-slow roll. Ultra-slow roll is an extreme scenario that also violates the assumption of an inflationary attractor solution which underlies single-clock inflation. Because of this, its non-Gaussianity violates a consistency relation and can amplify primordial black hole abundances, an effect I compute for both ultra-slow roll and its realistic implementations.

Finally, I establish in Chapter~\ref{chap:higgs}, abridged from Ref.~\cite{Passaglia:2019ueo}, the circumstances under which primordial black holes can be produced from the fluctuations of an extra spectator field during inflation. I focus on the Standard Model Higgs field and study its evolution from its early stochastic phase, through its roll down its large-field instability, all the way to its highly non-linear behavior during reheating.

Chapter~\ref{chap:conclusion} provides some brief concluding remarks.

\mainmatter

\chapter{Introduction}
\label{chap:intro}
\section{The Cosmic Microwave Puzzle}
\begin{figure}[t]
\begin{center}
\includegraphics[width=.65\linewidth]{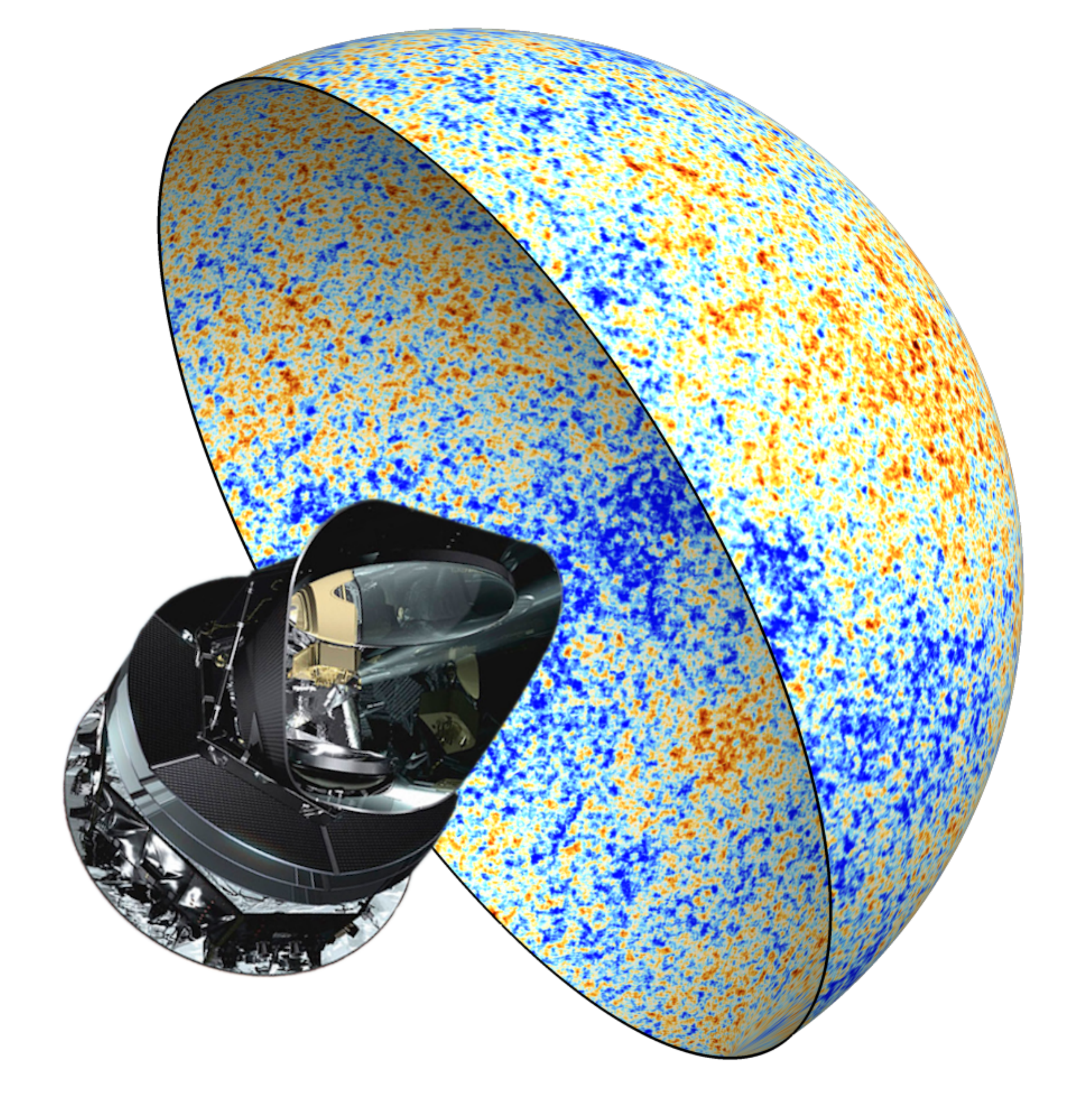}
\end{center}
\caption[The cosmic microwave background]{The {\it Planck} spacecraft is the most recent to have observed the microwave background, shown here over half the sky. It has detected superhorizon fluctuations in that map, sourced by the exponential expansion of our universe in its first instants, inflation. {\it Montage of European Space Agency images.}}
\label{fig:CMB}
\end{figure}

In 1964, Penzias and Wilson made the soothing discovery that we are bathed in a faint glow of microwave light \cite{Penzias:1965wn}. This radiation is made up of relic thermal photons, emitted everywhere in space when our universe was hotter and denser than it is today. Once our universe cooled enough for electrons and protons to form hydrogen, this light could travel largely unimpeded all the way until it reached us today, stretched and diluted by the expansion of the universe to a comfortable $2.7^{\circ} \Kelvin$. The discovery of the Cosmic Microwave Background (CMB) confirmed the Big Bang model and kickstarted cosmology as a physical science.

On further review though, the CMB exhibits the disconcerting feature that it is nearly the same temperature in all directions. Isotropy is uncomfortable because the light we receive from different directions was emitted from many different regions of space, and when we extrapolate those regions backwards in time, we find they have never been in causal contact with each other. Somehow our universe knows to be at nearly the same temperature everywhere.

Even more unsettling is the {\it nearly}. The COBE satellite in 1992 measured $1$ part in $10^5$ anisotropies in the temperature of the CMB \cite{Smoot:1992td}. On the one hand, we should be grateful for these small inhomogeneities in the universe at early times, because they seed the gravitational instabilities which lead to the formation of galaxies and cosmologists at late times. However, COBE found that the anisotropies are correlated across the many different causal horizons from which the CMB radiation was emitted -- regions of space which have seemingly never talked to each other somehow know to have precisely correlated perturbations.

The WMAP and Planck satellites (see Fig.~\ref{fig:CMB}), along with many ground- and ballon-based experiments, have confirmed and extended COBE's discoveries, measuring the anisotropies of the CMB temperature and polarization to exquisite precision and accuracy over a wide range of scales. In Fourier space, the spectrum of the primordial curvature perturbations $\zeta$ as a function of comoving scale $k$ has been measured to be
\begin{equation}
\label{eq:cmb_power}
\Delta^2_{\zeta} = A_s \left(\frac{k}{k_*}\right)^{n_s-1},
\end{equation}
with the amplitude $A_s \simeq (2.11 \pm .03) \times 10^{-9}$ and the tilt $n_s \simeq .966 \pm .004$ \cite{Aghanim:2018eyx}. $k_* = 0.05 \Mpc^{-1}$ is just a reference scale well measured by CMB experiments. A scale-invariant power spectrum would have $n_s = 1$, so beyond just the existence of superhorizon correlations we now have strong evidence that they are not exactly scale invariant.

The study of how this primordial curvature spectrum is transformed into the observed spectrum of temperature anisotropies has kept many people gainfully employed for decades now, and has enabled cosmologists to use the CMB as a powerful probe of our universe's contents and history. For example, study of the anisotropies has provided incontrovertible evidence for the existence of dark matter in our universe, to which we will return in \S\ref{sec:intro_pbhdm}. 

This thesis will focus on the physics in the very early universe which explains the puzzling superhorizon correlations. The answer falls under the paradigm of {\it cosmic inflation}, which explains how seemingly acausal regions communicated, and what they communicated about. Inflation posits new physics at the highest energy scales to explain the appearance of our universe on the largest length scales, and therefore represents the ultimate unification of particle physics and cosmology.

\section{Cosmic Inflation}

In an expanding universe the maximum comoving distance traveled by light since the beginning of time is the causal horizon
\begin{equation}
\label{eq:causalhorizon}
\eta = \int_0^{a} \frac{d \ln \tilde{a}}{\tilde{a} H},
\end{equation}
as a function of the scale factor $a$. $H$ is the Hubble rate, and with the Friedmann equations we can write the comoving Hubble radius $(a H)^{-1}$ as
\begin{equation}
\label{eq:comovinghubbleradius}
(a H)^{-1} \propto a^{\frac{1+3w}{2}},
\end{equation}
where $w$ is the equation of state of the universe. For a universe filled with radiation ($w=1/3$) or matter ($w=0$), the comoving Hubble radius is always increasing and the causal horizon is always dominated by the contribution from the last period of expansion.

We see correlations in the CMB emitted by two regions separated by a comoving distance $k^{-1}$ just today entering the comoving Hubble radius, and therefore $k \eta$ becoming larger than $1$ only recently. At decoupling they must have $k \eta \ll 1$ and were therefore not causally connected. This is the root of the {\it horizon problem} introduced in the previous section. 

Cosmic inflation posits that prior to the radiation dominated epoch in our universe there was a phase of exponential expansion: $H\sim$ constant, $w \sim -1$, $a \propto e^{H t}$. During such a phase the comoving Hubble radius \eqref{eq:comovinghubbleradius} shrinks and the causal horizon \eqref{eq:causalhorizon} is dominated by early time contributions. Regions which we would today calculate, neglecting inflation, to have never been in contact, could in fact have been contained within the same comoving Hubble radius early on. It is then not surprising that the radiation emitted by those regions has a correlated temperature.

We shall soon see that an exponential expansion phase not only allows for distant regions to have once been in contact, but also provides a mechanism to source the perturbations themselves. This is the remarkable success of inflation, and the reason it has been the favored paradigm since its development in the 1980s. 

In this introduction we cover the most basic model of inflation, canonical single-field slow roll. The rest of this thesis will then gradually relieve various assumptions of this model and search for the general principles that govern inflation and how these general principles translate to observations.

\subsection{Canonical single-field slow roll}

Scalar fields can easily accommodate the negative-pressure solutions needed to achieve $w\sim-1$ and exponential expansion. The only scalar field in the Standard Model is the Higgs, and we know enough about the Higgs to say that it is probably not the field driving inflation. We will therefore extend the Standard Model with a new scalar field $\phi$, minimally coupled to gravity and with a canonical kinetic term, so that our action now has a component
\begin{equation}
\label{eq:action}
S_\phi = \int d^4 x \sqrt{-g} \left[\frac{1}{2} g^{\mu \nu} \phi_{,\mu} \phi_{,\nu} - V(\phi) \right].
\end{equation}
The field $\phi (\vec{x}, t)$ satisfies the Klein-Gordon equation,
\begin{equation}
\label{eq:eom_intro}
\Box \phi = \frac{d V}{d\phi},
\end{equation}
which can be linearized into a background piece $\phi(t)$ and a perturbation $\delta\phi(\vec{x}, t)$.

The evolution of the background is then determined by the equation of motion
\begin{equation}
\ddot \phi + 3 H \dot \phi + \frac{d V}{d \phi} = 0,
\end{equation}
and the Friedmann equation
\begin{equation}
H^2 = \frac{1}{3} \left[ \frac{1}{2} \dot{\phi}^2 + V(\phi) \right].
\end{equation}
Overdots in this chapter denote coordinate time derivatives, and we set $\Mpl=1$ throughout. We see that so long as the inflaton's energy is dominated by a slowly varying potential, we can have an inflationary solution. We define a slow roll parameter
\begin{equation}
\epsilon_H \equiv -\frac{d \ln H}{d N} = \frac{1}{2}\frac{\dot\phi^2}{H^2},
\end{equation}
which should therefore be much less than unity. We will often use the $e$-folds of expansion $N \equiv \ln a$ as a time variable during inflation.

We need inflation to last many $e$-folds (at least $\sim60$ based on the horizon size today), and therefore we define a second slow-roll parameter,
\begin{equation}
\label{eq:intro_etaH}
\eta_H \equiv \frac{d\ln\epsilon_H }{d N},
\end{equation}
which should also be much less than $1$.

When these assumptions hold, the second-order equation of motion for $\phi$ can be approximated by the first-order equation
\begin{equation}
\label{eq:intro_attractor}
\dot \phi \simeq - \frac{1}{3 H}\frac{d V}{d \phi},
\end{equation}
with the velocity of the field a function only of the field's position along its potential. This is known as the slow-roll attractor solution.

\begin{figure}[t]
\begin{center}
\includegraphics[width=.65\linewidth]{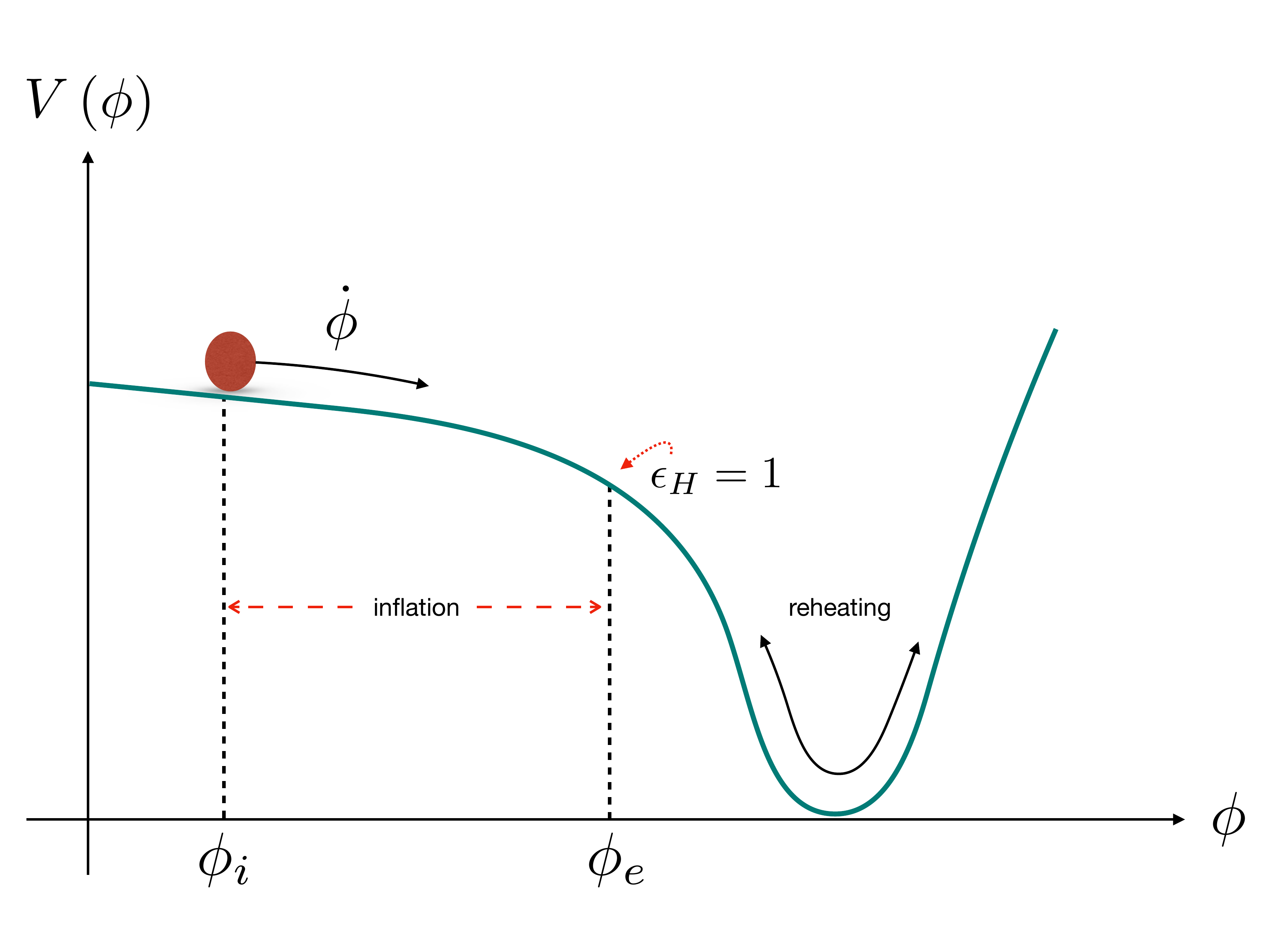}
\end{center}
\caption[Cartoon inflationary potential]{A typical inflationary potential. Observable scales cross the horizon early during inflation at $\phi_i$, and inflation proceeds until $\e \sim 1$ at $\phi_e$. Subsequently reheating occurs and the inflaton transfers its energy to Standard Model particles. {\it Courtesy of H\'{e}ctor Ram\'{i}rez.}}
\label{fig:potential}
\end{figure}

In Fig.~\ref{fig:potential}, we show a cartoon depiction of the inflationary potential convenient for visualization. Inflation occurs on a relatively flat region which supports slow-roll expansion. Observable scales might cross the shrinking comoving Hubble radius and leave causal contact at $\phi_i$, while inflation proceeds until $\e \sim 1$ at $\phi_e$.

We then enter a new phase, reheating, which results in the transfer of the inflaton's energy into a Standard Model plasma to begin our universe's radiation dominated phase. Reheating is the most poorly understood aspect of inflationary theory, but it can usually be glossed over because cosmologically relevant perturbations are much larger than the comoving Hubble radius at this time and thus reheating can only affect them in constrained ways.

So far we have managed to use single-field slow roll to explain why the CMB is isotropic: everything we see today was once in causal contact early in inflation. Now we will show that inflation can source the small and correlated anisotropies that we see in the sky.

\subsection{Inflationary perturbations}

Linearizing the Klein-Gordon equation \eqref{eq:eom_intro} yields the evolution equation for the Fourier mode $\delta \phi_k$ of the linear perturbation,
\begin{equation}
\ddot{\delta \phi}_k + 3 H \dot{\delta \phi}_k + \frac{k^2}{a^2} \delta \phi_k +  \frac{d^2 V}{d \phi^2} \delta \phi_k = 0.
\end{equation}
Here we have dropped perturbations to the metric, which are not important in slow roll in the spatially flat gauge.

We can scale out the expansion by using the Mukhanov variable $u_k \equiv a \delta \phi_k$ and, to lowest order in slow-roll, the equation of motion simplifies
\begin{equation}
\label{eq:intro_ms}
\frac{d^2 u_k}{d \eta^2} + \left(k^2 - \frac{2}{\widetilde{\eta}^2} \right) u_k = 0,
\end{equation}
where $\widetilde{\eta}$ is the positive decreasing conformal time remaining until the end of inflation,
\begin{equation}
\widetilde{\eta} \equiv \eta_{end} - \eta \sim \frac{1}{a H}.
\end{equation}
In the subhorizon regime $k \widetilde{\eta} \gg 1$, this equation looks like a harmonic oscillator which can be quantized by identifying $u_k$ as the wavefunction a quantum operator $\hat{u}_k$
\begin{equation}
\hat{u}_k = u_k \hat{a}_k + u_k^* {\hat{a}_k}^{\dagger},
\end{equation}
where $\hat{a}_k$ and $\hat{a}_k^{\dagger}$ are lowering and raising operators satisfying the usual commutation relation $[\hat{a}_k,\hat{a}_k^{\dagger}] = 1$. The wavefunction is normalized by imposing the canonical commutation relation $[\hat{u}_k, d \hat{u}_k /d \eta] = i$, yielding
\begin{equation}
u_k (k \widetilde{\eta} \rightarrow \infty) = \frac{1}{\sqrt{2 k}} e^{-ik \widetilde{\eta}}
\end{equation}
which is known as the Bunch-Davies initial condition. Eq.~\eqref{eq:intro_ms} classically evolves these subhorizon quantum zero-point modes through horizon crossing, and in slow-roll this has an analytic solution
\begin{equation}
u_k (\widetilde{\eta}) = \frac{1}{\sqrt{2 k}} \left(1-\frac{i}{k \widetilde{\eta}}\right) e^{-ik\widetilde{\eta}},
\end{equation}
leading to the late time ($k\widetilde{\eta} \rightarrow 0$) field fluctuation
\begin{equation}
\delta \phi_k \simeq \frac{i H}{\sqrt{2 k^3}}.
\end{equation}
This field fluctuation keeps evolving at late times as $H$ evolves, and therefore it is convenient to switch to a variable which is constant on superhorizon scales, like the curvature perturbation on uniform density slices $\zeta$\footnote{Chapters~\ref{chap:usr} and~\ref{chap:higgs} will cover in great detail scenarios in which $\zeta$ is not constant.},
\begin{equation}
\zeta = -\frac{\delta \rho}{\rho'} \simeq -\frac{\delta \phi}{\phi'},
\end{equation}
leading to a curvature power spectrum
\begin{equation}
\label{eq:intro_power}
\Delta^2_{\zeta} (k) = \frac{k^3 |\zeta_k|^2}{2 \pi^2} \simeq \left.\frac{H^2 }{8 \pi^2 \epsilon_H}\right\vert_{k \widetilde{\eta}=1},
\end{equation}
which should be evaluated at horizon crossing and is then constant until the mode reenters the horizon in the late universe.

This curvature power spectrum is precisely what is measured by the CMB in Eq.~\ref{eq:cmb_power}. We see that the near scale-invariance of the measured power spectrum is a reflection of near-constancy of $H$ during inflation. Moreover, the small red-tilt of measured power spectrum is a reflection of the fact that the slow-roll parameters are not exactly zero,
\begin{equation}
n_s - 1 = \frac{d \ln \Delta_\zeta^2 }{d \ln k} \sim -4 \epsilon_H - 2 \eta_H.
\end{equation}

Now that we have identified the observed curvature power spectrum with the two-point vacuum fluctuations of the scalar field, we expect that higher-point curvature correlators encode information about the interactions of that field. In the canonical slow-roll model we have constructed here, however, this a purely second-order effect, suppressed by additional factors of $\zeta \sim 10^{-5}$, and therefore unlikely to ever be observable.

However, in Chapter \ref{chap:eft} we will generalize the action \eqref{eq:action} by viewing $\phi$ as the Goldstone boson of broken time-translation symmetries during inflation, and we will drop the assumption of slow-roll. In this case, we will detail how to compute the three-point function and show that it is a powerful tool to distinguish the types of interactions felt by the inflaton. In Chapter \ref{chap:usr}, we will focus on a particular configuration, the squeezed-limit three-point function, which encodes the effect of a spatial dilation symmetry during inflation and which has important observational consequences.

\section{Primordial Black Hole Dark Matter}
\label{sec:intro_pbhdm}
	
	The superhorizon curvature perturbations which we have just computed eventually reenter the horizon, and if they are large enough they can then collapse to primordial black holes (PBHs). PBHs are interesting objects for many reasons, but in this thesis we focus on the fact that in some mass ranges they can be the dark matter (DM).

	The fractional energy-density in PBHs at formation time is related to the probability of the curvature lying above some threshold $\zeta_c$,
	\begin{align*}
	\beta \equiv \frac{\rho_{\rm PBH}}{\rho_{\rm tot}} &= 2 \int_{\zeta_c}^{\infty} d \zeta \frac{1}{\sqrt{2 \pi}\Delta_\zeta} e^{-\zeta^2/(2 \Delta_\zeta^2)}\\
	&\simeq \frac{\sqrt{2}}{{\pi}} \frac{\Delta_\zeta}{\zeta_c}e^{-\zeta_c^2/(2 \Delta_\zeta^2)}, \numberthis
	\end{align*}
	where in the last equality we assume PBH formation is a rare event and thus $\zeta_c \gg \Delta_\zeta$. The threshold can be estimated from simulations to be $\mathcal{O}(1)$, which means achieving a sizable abundance requires the power spectrum $\Delta_\zeta^2$ to grow relative to its small value at CMB scales \eqref{eq:cmb_power}. We therefore usually want to produce PBH DM on very small physical scales, so that there is more time during inflation for this to happen.

	However, the mass of a PBH formed by spherical collapse is equal to the mass contained in the horizon volume at collapse time,
	\begin{equation}
	M_H = \frac{4 \pi \rho}{3 H^3} = \frac{4 \pi}{H},
	\end{equation}
	and thus smaller scale perturbations lead to smaller mass PBHs. If the PBH mass is too small, the PBH DM will evaporate completely on cosmologically relevant timescales. The lightest PBHs which do not evaporate before matter-radiation equality have mass \cite{Motohashi:2017kbs}
	\begin{equation}
	M \sim 10^{-21} \Msun,
	\end{equation}
	which under some light assumptions can be shown to corresponds to physical scales which crossed the horizon
	\begin{equation}
	N_{\rm PBH} - N_{\rm CMB} \sim 40
	\end{equation} 
	$e$-folds after the CMB scales. To get the right DM abundance, the power spectrum at this scale has to be 
	\begin{equation}
	\Delta_\zeta (k_{\rm PBH}) \sim 10^{-2}.
	\end{equation}

	In order to successfully produce PBH DM using inflationary perturbations, we therefore require a $\sim 10^7$ enhancement of the power spectrum relative to its value on CMB scales \eqref{eq:cmb_power} within $\sim 40$ $e$-folds. 

	From the slow-roll expression for the power spectrum \eqref{eq:intro_power}, this implies that the fractional variation per $e$-fold of the first slow-roll $\epsilon_H$
	\begin{equation}
	\left\vert \frac{\Delta \ln \epsilon_H}{\Delta N} \right\vert \sim 0.4,
	\end{equation}
	meaning that there is necessarily an $\mathcal{O}(1)$ violation of the slow-roll condition \eqref{eq:intro_etaH} if PBHs are to make up the dark matter. 

	In fact the estimate here is a very conservative one, since such light PBHs would have observable levels of gamma ray Hawking radiation emission. PBH DM models therefore usually aim to produce PBHs in the `asteroid-mass' window around $10^{-17}\sim10^{-12} \Msun$ \cite{Montero-Camacho:2019jte}, and require a correspondingly larger slow-roll violation.

	Therefore we now have a no-go theorem (originally due to Ref.~\cite{Motohashi:2017kbs}) for primordial black hole production in canonical single-field slow-roll inflation. Any canonical single-field model which produces primordial black holes in sufficient abundance to be the dark matter must violate slow roll.

	The course of this thesis will now be to use primordial black holes as a guide through inflationary theory by gradually generalizing the assumptions made here.

\chapter{Single-Clock Inflation}
\label{chap:eft}

In this chapter, based on Refs.~\cite{Passaglia:2018afq,Ramirez:2018dxe}, we develop the unified effective field theory (EFT) of inflation, the most general form for the inflationary action which is consistent with unbroken spatial diffeomorphisms and a preferred temporal coordinate that represents the inflationary ``clock." 

This framework encompasses a wide variety of inflationary models, and we develop a complete formulation of the scalar power spectrum and bispectrum for the EFT in terms of a set of simple one-dimensional integrals which remain valid even when slow-roll is transiently violated. We show analytically that our expressions explicitly preserve the consistency relation between the power spectrum and the squeezed-limit bispectrum so long as the curvature perturbation is conserved outside the horizon. 

As an example application of our results, we compute the scalar power spectrum and bispectrum in a model in which potential-driven G-inflation at early times transitions to chaotic inflation at late times, showing that our expressions accurately track the power spectrum and bispectrum when conventional slow-roll approximations fail.

Despite the freedom inherent in the single-clock EFT, producing primordial black hole dark matter still generally requires violations of slow roll.

\section{Effective Field Theory}
	\label{sec:EFT}

	In this section we derive the action for scalar metric perturbations up to cubic order in the unified EFT of inflation. We begin in \S\ref{subsec:EFTLagrangian} by reviewing and generalizing the construction of the Lagrangian of the EFT of inflation, which we then expand to cubic order in scalar metric perturbations in \S\ref{subsec:EFTCubicAction}. We rewrite this action to make the squeezed-limit consistency relation manifest in \S\ref{subsec:CubicActionConsistency}. Finally, we study the structure of the EFT in the limits where it reduces to the Horndeski and beyond-Horndeski GLPV subclasses in \S\ref{subsec:HorndeskiGLPV}. 

	In general, we find that the cubic action for scalar perturbations can be written in terms of ten operators and manifestly leads to the squeezed-limit consistency relation during slow-roll. In the Horndeski and GLPV limits, six of the ten operators are present.

	\subsection{Unified Lagrangian}
		\label{subsec:EFTLagrangian}

		The unified EFT of inflation was presented in Ref.~\cite{Motohashi:2017gqb} with the complete set of quadratic operators that contribute  to theories where the metric perturbations obey a second-order equation in both
		time and space and temporal components of the metric remain non-dynamical. These restrictions ensure that the power spectra of scalar and tensor metric fluctuations 
		obey their usual form. We summarize here some of the essential features of that construction while extending it to include the complete set of cubic operators that contribute to the bispectrum.

		In the EFT construction, we seek the most general form for the action that is consistent with unbroken spatial diffeomorphisms and	a preferred temporal coordinate that represents the inflationary clock.  Using this preferred slicing, we decompose the metric into its $3+1$ Arnowitt-Deser-Misner () form
		\begin{equation}
		\label{eq:ADMMetric}
		\diff{s}^2 = -N^2 \diff{t}^2 + h_{ij} (\diff{x}^i + N^i \diff{t})(\diff{x}^j + N^j \diff{t}),
		\end{equation}
		with the lapse $N$, the shift $N^i$, and the spatial metric $h_{ij}$.  

		This metric and a unit timelike vector $n_\mu$ orthogonal to constant $t$ surfaces define the spatial tensors  that compose the EFT action. We construct an action invariant under spatial diffeomorphisms out of a general scalar function of these quantities
		\begin{equation}
		\label{eq:EFTAction} 
		S = \int \difffour{x} N \sqrt{h} \, L(N, {K^i}_j, {R^i}_j, t),
		\end{equation}
		in which $K_{\mu \nu} = n_{\mu;\nu} + n_\nu  n_{\mu;\beta} n^\beta$ is the extrinsic curvature, $R_{ij}$ is the three-dimensional Ricci tensor with trace $R ={R^i}_i$, and $h$ is the determinant of the three-dimensional metric $h_{ij}$. Semicolons here and throughout denote covariant derivatives with respect to the metric $g_{\mu \nu}$. Latin indices denote spatial coordinates, which are raised and lowered using $h_{ij}$. We use the shorthand summation convention
		\begin{equation}
		S_{i \ldots j}T_{i \ldots j}   \equiv \d^{i i'} \ldots \d^{j j'} S_{i \ldots j} T_{i' \ldots j'}, 
		\end{equation}
		for any two spatial tensors $S$ and $T$.
	
		We have not allowed additional spatial derivatives in  Eq.~\eqref{eq:EFTAction} since they lead to equations of motion that are	beyond second-order in spatial derivatives.  Thus we do not encompass the spatially covariant gravity \cite{Gao:2014soa,Gao:2014fra} or the Ho\v{r}ava-Lifshitz theories \cite{Horava:2009uw,Blas:2009yd,Blas:2010hb}. We 
		have also not allowed the lapse or shift to be dynamical, and thus we do not encompass the full set of degenerate higher order scalar tensor  (DHOST)\cite{BenAchour:2016fzp,Langlois:2017mxy} theories. For an extension of the EFT to such models, see Ref.~\cite{Motohashi:2020wxj}.

		Next we perturb the action \eqref{eq:EFTAction} around a spatially flat FLRW background, 
		\begin{equation}
		\left[N\right] = 1, \quad \left[N^i\right] = 0, \quad \left[h_{ij}\right] = a^2 \d_{ij},
		\end{equation}
		on which the extrinsic and intrinsic curvature are		
		\begin{equation}
		\left[{K^i}_j\right] = H {\d^i}_j, \quad \left[{R^i}_j\right] = 0,
		\end{equation}
		with $H \equiv d{\ln{a}}/dt$. Here and below the notation $\left[ \ldots \right]$ denotes evaluation on the background.

		In order to keep all terms that are at most cubic in metric perturbations, we expand the Lagrangian to cubic order in the ADM variables around the background. We define the Taylor coefficients
		\begin{align*}
		\label{eq:lagrangianpartials}
		\left[L\right] & =  \C, \numberthis\\
		\left[\frac{\partial L}{\partial X^i_{\,j}} \right] &=  \C_X {\delta_i^{j}}, \\ 
		\left[\frac{\partial^2 L}{\partial X^i_{\,j} \partial Y^k_{\hphantom{k}l}}\right] &=
		 \C_{XY} \delta_i^{\,j}  \delta_k^{\,l} + \frac{ \C_{\bar{X}\bar{Y}}}{2} (\delta^l_i \delta^j_k + \delta_{ik} \delta^{jl}), \\ 
		\left[\frac{\partial^3 L}{\partial X^i_{\,j} \partial Y^k_{\hphantom{k}l} \partial Z^m_{\,\,\, n}}\right] &= \C_{XYZ} \delta^j_i \delta^l_k \delta^n_m 
		\\&\quad+ \frac{\C_{\bar{X} \bar{Y} Z}}{2}  \delta^n_m (\delta^l_i \delta^j_k + \delta_{ik} \delta^{jl}) 
		\\&\quad+ \frac{\C_{X \bar{Y} \bar{Z}}}{2}  \delta^j_i (\delta^l_m \delta^n_k + \delta_{mk} \delta^{nl}) 
		\\&\quad+ \frac{\C_{\bar{X} Y \bar{Z}}}{2}  \delta^l_k (\delta^n_i \delta^j_m + \delta_{im} \delta^{jn})  
		\\&\quad+ \frac{\C_{\bar{X}\bar{Y}\bar{Z}}}{8} (\delta^j_k \delta^l_m \delta^n_i + \delta^j_k \delta^{l n} \delta_{m i} 
		\\&\quad+ \delta^{j l} \delta^n_k \delta_{m i} + \delta^{j l} \delta_{k m} \delta^n_i + \delta_{k m}  \delta^l_i \delta^{n j} 
		\\&\quad+ \delta^n_k \delta^j_m \delta^l_i  + \delta_{i k} \delta^{n j} \delta^l_m + \delta_{i k} \delta^j_m \delta^{n l}  ), 
		\end{align*}
		where $X, Y, Z \in \{N, K, R\}$ and the index structure is determined by the symmetry of the background. We treat scalars and traces with the same notation, so that the tensor ${N^i}_j = (N/3)\delta^i_j$. Thus $\C_{\bar{N}\bar{X}Y} = \C_{\bar{N}\bar{X}\bar{Y}}  = 0$ for any $X, Y$.  Otherwise, these coefficients are arbitrary functions of time which are invariant under subscript permutation in the EFT;
		they take different concrete forms in different specific inflationary models. Notationally, our $\C_{\bar{X}\bar{Y}}$ is equal to the $\tilde\C_{XY}$ of Ref.~\cite{Motohashi:2017gqb}. Up to cubic order we can write
		\begin{align*}
		\label{eq:cubicADMLagrangian}
			L =&\ \frac{1}{3!} \sum_{X,Y,Z} \left(\C_{XYZ} \d X \d Y \d Z + \C_{X\bar{Y}\bar{Z}} \d X \d {Y^i}_j \d {Z^j}_i \right. \\ 
			   &+ \C_{\bar{X}Y\bar{Z}} \d {X^i}_j \d Y \d {Z^j}_i  + \C_{\bar{X}\bar{Y}Z} \d {X^i}_j \d {Y^j}_i \d Z \\ 
			   &+ \left. \C_{\bar{X}\bar{Y}\bar{Z}} \d {X^i}_j \d {Y^j}_k \d {Z^k}_i \right) \\
			   &+ \frac{1}{2} \sum_{Y,Z} \left(\C_{YZ} \d Y \d Z + \C_{\bar{Y}\bar{Z}} \d {Y^i}_j \d {Z^j}_i \right) \\
			   &+ \C_N \d N + \C_R \d R + \frac{\C_N}{N} - \C_N,
			\numberthis
		\end{align*}
		with the sums running through all variable permutations with replacement. We have followed Ref.~\cite{Motohashi:2017gqb} in using integration by parts to eliminate the linear $\d K$ term up to a total derivative term as well as in using the background equation of motion to simplify some of the terms which are constant or linear in geometric quantities. 
		
		Finally, to ensure only second-order spatial derivatives in the equation of motion of perturbations we impose
		\begin{equation}
			\label{eq:QuadraticInheritance}
			\C_{\bar{K} \bar{K}} = - \C_{KK}, \qquad
			\C_{\bar{K}\bar{R}} = - 2 \C_{KR}, \qquad
			\C_{\bar{R}\bar{R}} = -\frac{8}{3} \C_{RR}.
		\end{equation}	
		This includes the Horndeski and GLPV classes.

	\subsection{Scalar Perturbations}
		\label{subsec:EFTCubicAction}

		We now restrict our attention to scalar metric perturbations and derive the quadratic and cubic actions for their dynamical
		field, the curvature perturbation. For scalar perturbations the ADM metric \eqref{eq:ADMMetric} takes the form		\begin{equation}
		N = 1+\d N, \quad N_i = \pa_i \psi, \quad h_{ij} = a^2e^{2\zeta} \d_{ij},
		\end{equation}
		where we have fixed the residual gauge freedom associated with spatial diffeomorphism invariance by taking a diagonal
		form for $h_{ij}$ \cite{Motohashi:2016prk}.   We call this choice unitary gauge.

		In unitary gauge, the perturbed geometric quantities are
		\begin{align*}
		\label{eq:ADMblocks}
		\d {K^i}_j =&\ \frac{1}{1+\d N}\Bigl[{\d^i}_j \left(\dot{\zeta}-H \d N\right)+ a^{-2} e^{-2\zeta} \left(\d^{ik} \pa_{k} \zeta \pa_{j} \psi \right. \\
		&+\d^{ik} \pa_{j} \zeta \pa_{k} \psi - \left. \d^{ik} \pa_k \pa_j \psi - {\d^i}_j \d^{ab} \pa_a \zeta \pa_b \psi\right) \Bigr],\\
		{\d R^i}_j =& -a^{-2} e^{-2\zeta} \left[\d^{ik} \pa_k \pa_j \zeta + {\d^{i}}_j \pa^2 \zeta + {\d^i}_j (\pa \zeta)^2 \right. - \left. \d^{ik}\pa_k \zeta \pa_j \zeta  \right].
		\numberthis
		\end{align*}
		Overdots in this chapter denote coordinate time derivatives, and here and throughout $(\pa \zeta)^2 \equiv \d^{a b} \pa_a \zeta \pa_b \zeta$ and $\pa^2 \zeta \equiv \d^{a b} \pa_a \pa_b \zeta$.
		Variation of the quadratic action with respect to the lapse and shift yields the Hamiltonian and momentum constraints
		\begin{equation}
		\label{eq:constraints}
		\d N = D_1 \dot{\zeta}, \quad \psi = D_2 \zeta + a^2 D_3 \chi,
		\end{equation}
		where $\chi$ is an auxiliary variable satisfying  $\pa^{2} \chi = \dot{\zeta}$ and the parameters $D_1$, $D_2$, and $D_3$ are
		\begin{align*}
		D_1 &= \frac{2 \C_{KK}}{2 H \C_{KK} - \C_{NK}}, \\
		D_2 &= \frac{4(\C_{NR} + \C_{R}-H \C_{KR})}{2 H \C_{KK} - \C_{NK}}, \\
		D_3 &= \frac{3 \C_{NK}^2 - 2 \C_{KK} (2 \C_{N} + \C_{NN})}{(2 H \C_{KK} - \C_{NK})^2}.
		\label{eq:Dn}
		\numberthis
		\end{align*}

		Since we are interested in the action to cubic order in perturbations, the lapse and shift should a priori be expanded beyond linear order. However, direct computation shows that the $\O{(\zeta^2)}$ lapse and shift parameters do not contribute to the cubic action. This is an example of the general result that the $\O{(\zeta^2)}$ lapse and shift parameters multiply the order $\O{(\zeta)}$ constraint equations and therefore do not contribute to the cubic action \cite{Maldacena:2002vr,Chen:2006nt,Pajer:2016ieg}.		
		
		After eliminating the lapse and shift, the quadratic action for the curvature $\zeta$ becomes
		\begin{equation}
		S_2 = \int \difffour{x} \ a^3 Q \left[ \dot{\zeta}^2 - \frac{c_s^2}{a^2} \left(\pa \zeta\right)^2 \right],
		\end{equation}
		in which $Q$ and $c_s^2$ are
		\begin{align*}
		Q &= \frac{\C_{KK} \left(2 \C_{KK} \left(2 \C_{N} + \C_{NN}\right)- 3 \C_{NK}^2\right)}{\left(2 H \C_{KK}-\C_{NK}\right)^2}, \numberthis\\
		c_s^2 &= \frac{2}{a Q} \biggl[\frac{\diff{}}{\diff{t}} \left(a \frac{2 \C_{KK} (\C_{NR} + \C_R)-\C_{KR} \C_{NK}}{2H \C_{KK}-\C_{NK}}\right) -  a \C_R\biggr].
		\end{align*}
		In terms of the $b_s$ parameter defined in Ref.~\cite{Motohashi:2017gqb}, $Q \equiv \e b_s / c_s^2$.
		The quadratic action provides the linearized equation of motion 
		\begin{equation}
		\label{eq:eom}
		\pa^2 \zeta = \frac{1}{a Q c_s^2} \frac{\diff{}}{\diff{t}} (a^3 Q \dot{\zeta}).
		\end{equation}

		We now plug in the perturbed geometric quantities \eqref{eq:ADMblocks} into the action \eqref{eq:EFTAction} with the Lagrangian \eqref{eq:cubicADMLagrangian}, eliminating the lapse and shift using the constraint equations \eqref{eq:constraints} and retaining terms up to cubic order in $\zeta$. We can also simplify the resulting action using integration by parts. Spatial boundary terms will not contribute to the in-in bispectrum, by momentum conservation, and will be omitted. Temporal boundary terms can contribute significantly and therefore must be retained \cite{Arroja:2011yj,Rigopoulos:2011eq}. 
				
		Finally, we can also use the linear equation of motion \eqref{eq:eom} to eliminate $\ddot\zeta$-type terms \cite{Burrage:2011hd,RenauxPetel:2011sb,Seery:2005gb}. The resulting cubic action is
		\begin{align*}
		\label{eq:cubiclagrangiandirect}
			S_3 =& \ S_3^{\text{Boundary}} + \int \diffcubed{x} \diff{t} \Bigl[a^3 F_1 \zeta  \dot\zeta {}^2 
			 + a F_2 \zeta  \left(\pa \zeta \right)^2 \numberthis \\
			 &+ a^3 \frac{F_3}{{H}} \dot\zeta {}^3
			 + a^3 F_4 \dot\zeta  \pa_a\zeta  \pa_a\chi 
			 + a^3 F_5 \pa^2\zeta  \left(\pa \chi \right)^2 \\
			 &+ \frac{F_6}{{H^3} a} \dot\zeta  \pa^2\zeta  \pa^2\zeta
			 + \frac{F_7}{{H^4} a^3} \left(\pa_a\pa_b\zeta \right)^2 \pa^2\zeta\\
			 &+ \frac{F_8}{{H^4} a^3} \pa^2\zeta  \pa^2\zeta  \pa^2\zeta 
			 + \frac{F_9}{{H^3} a} \pa^2\zeta \left(\pa_a\pa_b\zeta\right) \left(\pa_a\pa_b\chi\right) \Bigr],
		\end{align*}
		in which $F_1$ through $F_9$ are {dimensionless} time-dependent functions presented in Ref.~\cite{Passaglia:2018afq}. The temporal boundary terms are
		\begin{align*}
		\label{eq:boundaryaction}
			S_3^{\text{Boundary}} =& \int \diffcubed{x} \diff{t} \frac{\diff{}}{\diff{t}} \left[ a^3 G_1 \dot\zeta {}^3 
			 + a^3 G_2 \zeta  \dot\zeta {}^2 \right. \\
			 &+ a G_3 \zeta  \left(\pa \zeta \right)^2 
			 + a^3 G_4 \dot\zeta  \pa_a\zeta  \pa_a\chi  \\
			 &+ a G_5 \dot\zeta  \left(\pa \zeta \right)^2 
			 + \frac{G_6}{a} \left(\pa \zeta \right)^2 \pa^2\zeta\\
			 &+ a^3 G_7 \pa^2\zeta  \left(\pa \chi \right)^2
			 + a G_8 \pa_a\zeta  \pa_b\zeta  \pa_a\pa_b\chi\\  
			 &+ \left. a^3 G_9 \dot\zeta  \left(\pa_a\pa_b\chi \right)^2  \right],
	 	\numberthis
		\end{align*}
		in which $G_1$ through $G_9$ are time-dependent functions. 

		The $G_3$ and $G_6$ terms contain no time-derivatives of the fields and therefore do not contribute to
		bispectrum in the in-in formalism regardless of the behavior of their coefficients \cite{Burrage:2011hd,Adshead:2011bw}.

 		The remaining terms are suppressed relative to the usual $a^3 \zeta^2\dot\zeta$ boundary operator, which shall appear later in our construction, by the presence either of spatial derivatives, which yield relative factors of $k/aH \ll 1$, or by the presence of additional factors of $\dot{\zeta}$, which is suppressed outside the horizon. Therefore none of these terms contribute unless $G_n$ grows sufficiently quickly, so long as the boundary is taken when all modes are outside the horizon.

		We restrict our attention to scenarios which satisfy these mild conditions on the EFT parameters and therefore we hereafter discard $S_3^{\text{Boundary}}$ entirely.

	\subsection{Cubic Action and  Consistency Relation}
		\label{subsec:CubicActionConsistency}
	
		We can use to our advantage our ability to reorganize the cubic action using integration by parts and the equation of motion for $\zeta$ derived from the
		quadratic action. In particular, it is well known that in inflation with a single dynamical degree of freedom and a curvature perturbation which remains constant outside the horizon, the bispectrum in the squeezed limit should satisfy the consistency relation \cite{Cheung:2007sv,Creminelli:2004yq,Maldacena:2002vr,Creminelli:2011rh}
		\begin{equation}
		\lim_{k_S\rightarrow0} \frac{B_\zeta(k_S, k_L, k_L)}{P_\zeta(k_S) P_\zeta(k_L)} = -\frac{d \ln{ \Delta_\zeta^2(k_L)}}{d \ln{k_L}},
		\label{eq:ConsistencyRelation}
 		\end{equation}
		where $B_\zeta$ denotes the curvature bispectrum (see \S\ref{sec:GSR} for notation).  Here the power spectrum $P_\zeta$ is related to the dimensionless power spectrum $\Delta_\zeta^2$ by
		\begin{equation}
		\label{eq:SRDelta2}
		\frac{k^3}{2 \pi^2} P_\zeta \equiv \Delta_\zeta^2 
		\simeq \frac{H^2}{8\pi^2 Q c_s^3},
		\end{equation}
		where here and below $\simeq$ denotes a slow-roll relation.
		In the slow-roll approximation, the local slope of the power spectrum is nearly constant and is called the tilt 
		\begin{equation}
		\label{eq:tilt}
			\frac{d \ln \Delta_\zeta^2 }{d \ln k}  \simeq 	n_s - 1=  (-2 \e - q - 3 \sigma),
		\end{equation}
		where $q \equiv \dot{Q}/(HQ)$, $\sigma \equiv \dot{c}_s/(H c_s)$.		

		We expect the consistency relation to hold here, but at first glance -- or, in the language of \S\ref{sec:GSR}, when plugging in zeroth-order modefunctions --  the squeezed-contributing interactions $\zeta \dot{\zeta}^2$ and $\zeta (\pa \zeta)^2$ with their sources $F_1$ and $F_2$ are not obviously related to the tilt \eqref{eq:tilt}. We can rewrite these terms in such a way as to make the consistency relation manifest by generalizing the procedure in Refs.~\cite{Adshead:2013zfa,Creminelli:2011rh}.

		We first rewrite the squeezed-contributing action in terms of the quadratic Hamiltonian density
		\begin{equation}\Ha_2 = a^3 Q \left[ \dot{\zeta}^2 + \frac{c_s^2}{a^2} \left(\pa \zeta\right)^2 \right],\end{equation}
		and the quadratic Lagrangian density 
		\begin{equation}\L_2 = a^3 Q \left[ \dot{\zeta}^2 - \frac{c_s^2}{a^2} \left(\pa \zeta\right)^2 \right],\end{equation}
		such that
		\begin{align*}
		\label{eq:squeezeactionH2L2}
		S_{\text{squeezed}} =& \int \diffcubed{x} \diff{t} \frac{\zeta}{2 Q} \biggl[\left(\Ha_2 + \L_2\right) F_1 \\
							&+  \left(\Ha_2 - \L_2\right) \frac{F_2}{c_s^2} \biggr].
		\numberthis
		\end{align*}

		Next we note that several terms can be grouped into a vanishing boundary term. For a general function of time $F$,
		\begin{align*}
		\label{eq:squeezegrouprelation}
			\frac{1}{F} \frac{\diff{}}{\diff{t}} \left(\frac{F \zeta \Ha_2}{H}\right) =& \ \frac{\dot{\zeta}}{H} \L_2 - \zeta (\Ha_2 + 2 \L_2) - (q+\sigma) \zeta  \L_2 \\
			&+ \left(\frac{\dot{F}}{H F} + \e + \sigma\right) \zeta \Ha_2.
			\numberthis
		\end{align*}

		Ref.~\cite{Adshead:2013zfa} uses a similar relation with $F = 1/c_s^2$ to simplify the action in $k$-inflation. Here we generalize this grouping using
		\begin{equation}
		F = \frac{1}{2+q+\sigma} \left(2 - \frac{F_1}{2Q} + \frac{F_2}{2 c_s^2 Q}\right),
		\end{equation}
		such that the total $\zeta \L_2$ term on the right-hand side of Eq.~\eqref{eq:squeezegrouprelation} corresponds to the $\zeta \L_2$ term in Eq.~\eqref{eq:squeezeactionH2L2}, plus an additional factor of $2 \zeta \L_2$. 

		Making this substitution and using the specific functional forms of $F_1$ and $F_2$, we find a significant cancellation among the $\zeta \Ha_2$ terms which results in the squeezed action taking the form
		\begin{align*}
			S_{\text{squeezed}} =& \int \diff{t} \diffcubed{x} \left[\zeta (\Ha_2 + 2 \L_2) - \frac{F}{H} \dot{\zeta} \L_2 \right. + \left.\frac{\diff{}}{\diff{t}}\left(\frac{F}{H} \zeta  \Ha_2 \right) \right].
			\numberthis
		\end{align*}
		
		The boundary term here does not contribute to the bispectrum (see Ref.~\cite{Adshead:2013zfa}), and therefore we discard it. The $\dot{\zeta} \L_2$ term does not contribute to the squeezed limit. In order to make the consistency relation more manifest, we undo the grouping by using Eq.~\eqref{eq:squeezegrouprelation} with $F=1$. We also use
		\begin{equation}
		2 G \zeta \L_2 = \frac{\diff{}}{\diff{t}} (G a^3 Q \zeta^2 \dot{\zeta}) - \dot{G} a^3 Q \zeta^2 \dot{\zeta},
		\end{equation}
		which holds for all functions of time $G$, and in particular we use it with $G =  \e + 3\sigma/2 + q/2$.

		After these substitutions and including the terms in Eq.~\eqref{eq:cubiclagrangiandirect} that do not contribute to the squeezed limit, we obtain the cubic action for metric perturbations
		\begin{align*}
		\label{eq:cubicaction}
			S_3 =& \int \diffcubed{x} \diff{t} \biggl[a^3 Q \frac{\diff{}}{\diff{t}} \left(\e + \frac{3}{2}\sigma + \frac{q}{2}\right) \zeta^2 \dot{\zeta} 
		     \\&- \frac{\diff{}}{\diff{t}} \left[ a^3 Q \left(\e + \frac{3}{2}\sigma + \frac{q}{2}\right) \zeta^2 \dot{\zeta}\right] 
		     \\&+ (\sigma + \e) \zeta(\Ha_2 + 2 \L_2) + (1 - F) \frac{\dot{\zeta} \L_2}{H}
			 \\&+ a^3 \frac{F_3}{{H}} \dot\zeta {}^3
			 + a^3 F_4 \dot\zeta  \pa_a\zeta  \pa_a\chi 
			 \\&+ a^3 F_5 \pa^2\zeta  \left(\pa \chi \right)^2 
			 + \frac{F_6}{{H^3} a} \dot\zeta  \pa^2\zeta  \pa^2\zeta
			 \\&+ \frac{F_7}{{H^4} a^3} \pa^2\zeta \left(\pa_a\pa_b\zeta \right)^2
			    + \frac{F_8}{{H^4} a^3} \pa^2\zeta  \pa^2\zeta  \pa^2\zeta
			    \\&+ \frac{F_9}{{H^3} a} \pa^2\zeta (\pa_a\pa_b\zeta)(\pa_a\pa_b\chi) \biggr].
			 \numberthis
		\end{align*}
		
		Refs.~\cite{Adshead:2013zfa,Creminelli:2011rh} show explicitly in the context of more restricted inflationary models that the boundary term yields the slow-roll squeezed-limit consistency relation, while the first term on the first line contributes to the squeezed-limit at higher order in slow-roll, as does the first term on the third line (which can be seen by re-application of Eq.~\eqref{eq:squeezegrouprelation}). No other term contributes to the squeezed limit at lowest order in slow-roll, and therefore we can immediately see from Eq.~\eqref{eq:cubicaction} that the squeezed-limit consistency relation holds in the unified EFT of inflation during slow-roll. In \S\ref{sec:GSR}, we will show that the consistency relation holds even beyond slow-roll.

		While the cubic action \eqref{eq:cubicaction} ensures the consistency relation holds in slow-roll, no assumption of slow-roll has been made in its derivation. 

	\subsection{Horndeski and GLPV Subclasses}
		\label{subsec:HorndeskiGLPV}

		Though we write the EFT directly in terms of  the metric, the EFT can also be viewed as a four-dimensional scalar-tensor theory by transforming out of unitary gauge using the Stuckelburg trick \cite{Gao:2014soa,Cheung:2007st}. In this way, the EFT of inflation presented in \S\ref{subsec:EFTLagrangian} encompasses a large space of fully covariant models. In this section, we study the structure of the cubic action \eqref{eq:cubicaction} derived in \S\ref{subsec:CubicActionConsistency} in the Horndeski and GLPV model classes.

		The Horndeski and GLPV classes are constructed to avoid the Ostrogradsky instability \cite{Woodard:2015zca,Solomon:2017nlh}. The Horndeski class \cite{Horndeski:1974wa} is the most general 4-dimensional scalar-tensor theory with second-order equations of motion for the scalar field $\phi$. The Horndeski class can be broadened to include models which have higher than second-order equations of motion yet due to a degeneracy condition do not propagate an Ostrogradsky mode. This is  the beyond-Horndeski GLPV class \cite{Gleyzes:2014qga}, of which the Horndeski class is a subset. The GLPV class is an example of a DHOST theory \cite{BenAchour:2016fzp}. While the GLPV model can be represented with an action of the form \eqref{eq:EFTAction}, writing the other DHOST theories in our EFT would require generalizing Eq.~\eqref{eq:EFTAction} to include time derivatives of the lapse function \cite{Langlois:2017mxy}.

		The cubic action \eqref{eq:cubicaction} and the resultant bispectrum takes on a restricted form in the Horndeski and GLPV classes. This restriction follows from the ADM representation of the  action for Horndeski and GLPV models \cite{Kase:2014cwa},
		\begin{align*} 
			L =& \ A_2 + A_3 K 
			+ A_4 (K^2 - K^i_{\hphantom{i}j} K^j_{\hphantom{i}i})  + B_4  R
			\\
			& + A_5 ( K^3 - 3K K^i_{\hphantom{i}j} K^j_{\hphantom{i}i} + 2 K^i_{\hphantom{i}j} K^j_{\hphantom{i}k} K^k_{\hphantom{i}i} ) \\
			& + B_5 ( K^i_{\hphantom{i}j}R^j_{\hphantom{i}i}-\tfrac{1}{2}KR ).
			\numberthis
		\end{align*}
		Here  $A_n(X,\phi)$ and $B_n(X,\phi)$ are functions of the kinetic term $X= \nabla^\mu\phi \nabla_\mu\phi$ and field $\phi$.  In the unitary gauge of ADM,  $\phi\rightarrow\phi(t)$  and thus $X = -\dot{\phi}^2/ N^2$, so these quantities may also be considered
		as functions of $N$ and $t$. In the GLPV class, these functions are completely general, while in the Horndeski class they satisfy
		\begin{align*}
		A_4 &= 2 X B_{4,X} - B_4,\\
		A_5 &= - \frac{1}{3} X B_{5,X}.
		\numberthis
		\end{align*}
		
		We then take the appropriate partial derivatives in Eq.~\eqref{eq:cubicADMLagrangian} to get the various $\C$ variables in the Horndeski and GLPV theories. We find
		\begingroup
		\allowdisplaybreaks
		\begin{align*}
		 \C_{N} =&
		 - 2 X (A_{2,X}\!
		 + 3 A_{3,X} H
		 + 6 A_{4,X} H^2\!
		 + 6 A_{5,X} H^3 ), \\
		\C_{R}
		 =& \  B_{4}
		 - \frac{B_{5} H}{2},\\
		\C_{NN} =& \ 6 X \left(A_{2,X}
		 + 3 A_{3,X} H 
		 + 6 A_{4,X} H^2
		 + 6 A_{5,X} H^3\right) \\
		 &+ 4 X^2 \left(A_{2,XX}
		 + 3 A_{3,XX} H
		 + 6 A_{4,XX} H^2 \right. \\
		 &+ \left. 6 A_{5,XX} H^3\right), \\
		\C_{K} =& \  A_{3}
		 + 4 A_{4} H
		 + 6 A_{5} H^2, \\
		\C_{KK}
		 =& \ 2 (A_{4}
		 + 3 A_{5} H), \\
		\C_{NK}
		 =&  
		 - 2 \left(A_{3,X}
		 + 4 A_{4,X} H
		 + 6 A_{5,X} H^2\right) X ,\\
		\C_{NR}
		 =& 
		 - 2 \left(B_{4,X}
		 - \frac{B_{5,X} H}{2}\right) X ,\\
		\C_{KR}
		 =& - \frac{B_{5}}{2},\\
		\C_{NNN}
		 =&  
		 - 24 X \left(A_{2,X}
		 + 3 A_{3,X} H
		 + 6 A_{4,X} H^2 \right. \\
		 &+ \left. 6 A_{5,X} H^3\right)\\
		 &- 36 X^2 \left(A_{2,XX}
		 + 3 A_{3,XX} H
		 + 6 A_{4,XX} H^2 \right. \\
		 &+ \left. 6 A_{5,XX} H^3\right)\\
		 &- 8 X^3 \left(A_{2,XXX}
		 + 3 A_{3,XXX} H
		 + 6 A_{4,XXX} H^2 \right. \\
		 &+ \left. 6 A_{5,XXX} H^3\right), \\
		\C_{NNK}
		 =&  \ 6 X \left(A_{3,X}
		 + 4 A_{4,X} H
		 + 6 A_{5,X} H^2\right) \\
		 &+ 4 X^2 \left(A_{3,XX}
		 + 4 A_{4,XX} H
		 + 6 A_{5,XX} H^2\right),\\
		\C_{NNR}
		 =&  \ 6 X \left(B_{4,X}
		 - \frac{B_{5,X} H}{2}\right)  \\
		 &+ 4 X^2 \left(B_{4,XX}
		 - \frac{B_{5,XX} H}{2}\right) ,\\
		\C_{NKK}
		 =&
		 - 4 X (A_{4,X}
		 + 3 A_{5,X} H) ,\\
		\C_{N\bar{K}\bar{K}}
		 =& \ 
		  4 X (
		   A_{4,X}
		  +3 A_{5,X} H) ,\\
		 \C_{NKR} =& \ X B_{5,X}, \\
		 \C_{N\bar{K}\bar{R}} =&  - 2 X B_{5,X}, \\
		 \C_{KKK} =& \ 6 A_{5},\\
		 \C_{\bar{K}\bar{K}K} =& - 6 A_{5}, \\
		 \C_{\bar{K}\bar{K}\bar{K}} =&  \ 12 A_{5},
	  	\numberthis
		\end{align*}
		\endgroup
		in which ${}_{,X}\equiv d/dX$ and all other coefficients are either zero or determined by Eq.~\eqref{eq:QuadraticInheritance}. 
		
		Using these coefficients, one can show that in the Horndeski and GLPV cases $F_6$, $F_7$, $F_8$, and $F_9$ are identically zero using the expressions in Ref.~\cite{Passaglia:2018afq}. 
		Thus the cubic action reduces to
		\begin{align*}
		\label{eq:cubicactionHorndeski}
			S^{\text{GLPV}}_3 =& \int \diffcubed{x} \diff{t} \biggl[a^3 Q \frac{\diff{}}{\diff{t}} \left(\e + \frac{3}{2}\sigma + \frac{q}{2}\right) \zeta^2 \dot{\zeta} 
		     \\&- \frac{\diff{}}{\diff{t}} \left[ a^3 Q \left(\e + \frac{3}{2}\sigma + \frac{q}{2}\right) \zeta^2 \dot{\zeta}\right] 
		     \\&+ (\sigma + \e) \zeta(\Ha_2 + 2 \L_2) + (1 - F) \frac{\dot{\zeta} \L_2}{H}
			 \\&+ a^3 \frac{F_3}{{H}} \dot\zeta {}^3
			 + a^3 F_4 \dot\zeta  \pa_a\zeta  \pa_a\chi 
			 \\&+ a^3 F_5 \pa^2\zeta  \left(\pa \chi \right)^2 \biggr].
			 \numberthis
		\end{align*}

		It can also be shown that the $F_4$ and $F_5$ operators are suppressed by an additional factor of slow-roll parameters relative to the other operators. This result was shown for the Horndeski class in Ref.~\cite{DeFelice:2013ar}, and holds also for the GLPV class.

		This is the same form of the action as shown in Refs.~\cite{RenauxPetel:2011sb,DeFelice:2011uc,Gao:2011qe}, after undoing our grouping of the $F_1$ and $F_2$ terms. Our novel squeezed-action grouping of $F_1$ and $F_2$ also confirms the result of Ref.~\cite{DeFelice:2013ar} that the squeezed-limit consistency relation holds in slow-roll in Horndeski models
		and corroborates the result in Ref.~\cite{Fasiello:2014aqa} that GLPV leads to no new scalar bispectrum shapes relative to Horndeski. By writing it in this form we show that the squeezed-limit consistency relation holds in GLPV models in slow-roll.

\section{Power Spectrum and Bispectrum in Generalized Slow Roll}
	\label{sec:GSR}
	From our results so far, it is already clear that it is not much easier to produce PBH DM in the single-clock EFT than in canonical single-field. The slow-roll power spectrum \eqref{eq:SRDelta2} 		
	\begin{equation}
		\Delta_\zeta^2 	\simeq \frac{H^2}{8\pi^2 Q c_s^3},
	\end{equation}
	will still have to grow by $\sim10^7$ in $\sim40$ $e$-folds, as shown in Chapter~\ref{chap:intro}. Instead of implying a large drop in the background $\e$ as in canonical single field, this now implies a large change in $Q c_s^3$, 
	\begin{equation}
	\left\vert \frac{d \ln Q c_s^3}{d N} \right\vert \sim \mathcal{O}(1).
	\end{equation}
	Depending on how this change is achieved in terms of EFT parameters, it is possible that the background nonetheless proceeds in slow-roll. Regardless, slow-roll approximations for the perturbations must be violated.

	We now present in \S\ref{subsec:InInGSRFormalism} the generalized slow-roll and in-in formalisms, which we use to construct a complete integral formulation of the power spectrum and bispectrum resulting from the EFT action derived in \S\ref{sec:EFT}, valid when slow-roll is transiently violated. In \S\ref{subsec:squeeze}, we study the squeezed-limit of the bispectrum and show that the consistency relation holds beyond slow-roll. We relegate the explicit forms for the components of some expressions to the published version of this work, Ref.~\cite{Passaglia:2018afq}.

	\subsection{GSR and In-In Formalisms}
		\label{subsec:InInGSRFormalism}

		The tree-level three-point correlation function in the in-in formalism is given by \cite{Adshead:2013zfa,Maldacena:2002vr,Weinberg:2005vy,Adshead:2009cb}
		\begin{align*}
		\label{eq:inin}
		&\langle \hat{\zeta}_{\bf{k_1}}(t_*) \hat{\zeta}_{\bf{k_2}}(t_*) \hat{\zeta}_{\bf{k_3}}(t_*) \rangle =  2  \re{-i \int_{-\infty(1+i\epsilon)}^{t_*} \diff{t} \langle \hat{\zeta}^I_{\bf{k_1}}(t_*) \hat{\zeta}^I_{\bf{k_2}}(t_*) \hat{\zeta}^I_{\bf{k_3}}(t_*) H_{I}(t) \rangle}, \numberthis
		\end{align*} 
		with $H_I \simeq - \int \diffcubed{x} \L_3$ at cubic order \cite{Adshead:2008gk}. 

		The field operators $\hat{\zeta}^I$ are in the interaction picture, which means their corresponding modefunctions satisfy the free Hamiltonian's equation of motion \eqref{eq:eom}. $\hat{\zeta}_{\bf k}^I$ is the Fourier transform of the operator. 
		We define the corresponding modefunctions $\zeta_k(t)$ as
		\begin{equation}
		\hat{\zeta}^I_{\bf{k}}(t) = \zeta_k(t) \hat a({\bf k}) + \zeta_k^* \hat a^\dag(-{\bf k}) ,
		\end{equation}
		where the annihilation and creation operators satisfy
		\begin{equation}
		[\hat a({\bf k}), \hat a^\dag({\bf k}')] = (2\pi)^3 \delta({\bf k}-{\bf k}')
		\end{equation}
		as usual.    Using these relations  the power spectrum can be evaluated from the modefunctions at a time $t_*$
		taken to be after all the relevant modes have left the horizon 
			\begin{align}
		\langle \hat{\zeta}^I_{\textbf{k}}(t_*)\hat{\zeta}^I_{\textbf{k}'}(t_*)\rangle & = 
		 (2\pi)^3\delta^{3}(\textbf{k}+\textbf{k}') |\zeta_k(t_*)|^2  
		\nonumber \\ &\equiv  (2\pi)^3\delta^{3}(\textbf{k}+\textbf{k}') P_\zeta (k). 
		 \end{align}

		Translational and rotational invariance requires that the three-point correlators be encapsulated in the bispectrum $B_\zeta$ as
		\begin{equation}
		\langle \hat{\zeta}_{\bf{k_1}} \hat{\zeta}_{\bf{k_2}} \hat{\zeta}_{\bf{k_3}}\rangle
		= (2\pi)^3 \delta^{3}({\bf{k_1}}+{\bf{k_2}}+{\bf{k_3}}) B_{\zeta}(k_1,k_2,k_3), 
		\end{equation}
		in which we have suppressed the evaluation at $t_*$. The dimensionless parameter conventionally constrained by experiment is
		\begin{equation}
		f_{\textrm{NL}} (k_1, k_2, k_3) \equiv \frac{5}{6} \frac{B_\zeta(k_1, k_2, k_3)}{P_\zeta(k_1) P_\zeta(k_2) + \text{perm.}} \ \ .
		\end{equation}
		Here and throughout `$+\text{perm.}$' denotes the two additional cyclic permutations of indices.

		In order to evaluate the in-in integral \eqref{eq:inin} and compute $B_{\zeta}(k_1,k_2,k_3)$, we need to solve the equation of motion \eqref{eq:eom} for the interaction picture modefunctions $\zeta_k(t)$.
		However, beyond the slow-roll approximation, there is no general analytic solution to the equation of motion. The generalized slow-roll approach is to solve the equation of motion iteratively \cite{Adshead:2013zfa,Stewart:2001cd,Choe:2004zg,Dvorkin:2009ne,Hu:2011vr,Kadota:2005hv}.
		It is convenient to express the modefunction in dimensionless form as
		\begin{equation}
		y \equiv \sqrt{\frac{k^3}{2\pi^2} } \frac{f}{x} \zeta_k,
		\end{equation}
		where
		\begin{equation}
		f \equiv 2 \pi a s \sqrt{2 Q c_s},
		\end{equation}
		$x= k s$ and the sound horizon 
		\begin{equation}
		s \equiv \int_a^{a_\text{end}} \frac{ \diff{\tilde{a}}}{\tilde a} \frac{c_s}{\tilde{a} H},
		\end{equation}
		with $a_{\rm end}$ denoting the end of inflation.
		
		The formal solution to Eq.~\eqref{eq:eom} is 
		\begin{equation}
		\label{eq:modefunctionformal}
		y(x) = y_0 (x) - \int_x^\infty \frac{\diff{\tilde x}}{\tilde x^2} g(\ln \tilde{s}) y(\tilde x) \im{y_0^*(\tilde x) y_0(x)},
		\end{equation}
		in which  $\tilde x \equiv k \tilde{s}$, $g(\ln s) \equiv (f'' - 3f')/f$	and $' \equiv d/d \ln s$.
		The zeroth order solution with Bunch-Davies initial conditions is 
		\begin{equation}
		y_0 (x) = \left(1+\frac{i}{x}\right) e^{i x}.
		\end{equation}
		The first-order solution is obtained by plugging in the zeroth-order solution into the right hand side of Eq.~\eqref{eq:modefunctionformal}.

		Every order in the GSR hierarchy of solutions is suppressed relative to the previous order by the $g$ factor, whose time integral is assumed to be small but whose value can evolve and become transiently large unlike in the slow-roll approximation -- we call such a case ``slow-roll suppressed". When operators in the cubic action are
		also slow-roll suppressed, as is the case for the $\zeta^2 \dot{\zeta}$ and $\zeta(\Ha_2 + 2 \L_2)$ terms, it suffices to use the zeroth-order solution for the modefunctions in computing the bispectrum to first-order in slow-roll parameters. Operators with general EFT coefficients, however, are not necessarily slow-roll suppressed and therefore the first-order modefunction solution must be used in order to maintain a consistent first-order solution.  

		In the GSR formalism, the power spectrum to first-order in slow-roll parameters is \cite{Miranda:2012rm,Motohashi:2017gqb}
		\begin{equation}
		\label{eq:GSRpower}
		\ln \Delta_\zeta^{2} = G(\ln s_{*}) + \int_{s_{*}}^\infty {\frac{\diff s}{s}} W(ks) G'(\ln s),
			\end{equation}
		with the power spectrum window function
		\begin{equation}
		W(u) = {\frac{3 \sin(2 u)}{2 u^3}} - {\frac{3 \cos (2 u)}{u^2}} - {\frac{3 \sin(2 u)}{2 u}},
		\end{equation}
		the power spectrum source
		\begin{equation}
		G = - 2 \ln f    + \frac{2}{3} (\ln f )',
		\end{equation}
		and $s_*$ the sound horizon corresponding to the superhorizon time $t_*$.

		The first-order bispectrum result follows the same schematic form as the first-order power spectrum result: a windowed integral over a source. Each operator $i$ in the cubic action contributes a set of sources and windows to the bispectrum which are indexed by $j$ according to their asymptotic scalings at $x\ll1$ and $x\gg 1$. Thus we denote these sources and windows as $S_{ij}$, $W_{ij}$. 

		At zeroth-order in GSR modefunctions, the bispectrum integrals depend only on the triangle perimeter $K \equiv k_1+k_2+k_3$, and all shape dependence is held outside the integrand by corresponding $k$-weights  $T_{ij}$. The integrals take the form
		\begin{align*}
		\label{eq:GSRIntegration}
		 I_{ij}(K) = S&_{ij}(\ln s_*) W_{ij}(K s_*) + \!\int_{s_*}^\infty \!\frac{\diff s}{s}  S_{ij}'(\ln s) W_{ij}(K s).\numberthis
		\end{align*}
		
		At first-order in the GSR modefunctions, each operator yields a shape-dependent boundary contribution resulting from the removal of certain nested integrals using integration by parts \cite{Adshead:2013zfa}. These contributions are of the form $\left[T_{i B} I_{i B} (2 k_3) + \text{perm.}\right]$.
		Together, the perimeter-dependent and shape-dependent integrals enable computation of the complete bispectrum of the effective field theory of inflation to first-order in slow-roll parameters,
		\begin{align*}
		\label{eq:GSRBi}
		B_{\zeta}&(k_1,k_2,k_3) =  \frac{(2 \pi)^4}{4} \frac{\Delta_\zeta(k_1) \Delta_\zeta(k_2) \Delta_\zeta(k_3)}{k_1^2 k_2^2 k_3^2} \numberthis\\ &\times \Big\{  \sum_{ij} T_{ij} I_{ij}(K) +
		 \sum_{i=2}^9 \left[T_{iB} I_{iB}(2 k_3) +{\rm perm.}\right] \Big\}.
		\end{align*}
		
		We provide the sources, windows, and $k$-weights that each operator in the cubic action contributes to this expression in the full published version of this work, Ref.~\cite{Passaglia:2018afq}. In Tab.~\ref{tab:OperatorSummary}, we give summary information for each operator. The following section focuses on establishing the squeezed-limit consistency relation beyond slow-roll from these results.

		\begin{table}[tb]
		\begin{center}
		\begin{tabular}{ l  c c  c c }
		$i$  &  Operator &  Source & Squeezed & GLPV \\
		 \hline
		 &&&&\\[-10pt]
		0 \ {} & $\zeta^2 \dot \zeta$ & $ \dfrac{2 \e + 3 \sigma + q }{2 f}$ & yes &  Supp. \\[10pt]
		1 & $\zeta({\Ha}_2+2 {\L}_2)$  & $\dfrac{\sigma+\e}{f}$ & yes &  Supp.  \\[10pt]
		2 & $\dot \zeta {\cal L}_2$  & $ \dfrac{c_s}{a H s} \dfrac{F-1}{f}$ & no & Free \\[10pt]
		3 & $\dot \zeta^3$ & $- \dfrac{1}{Q} \dfrac{c_s}{a {H} s} \dfrac{F_3}{f}$ & no &  Free \\[10pt]
		4 & $ \dot\zeta (\partial\zeta)  \pa \chi  $ & $-\dfrac{1}{2Q}\dfrac{F_4}{f}$ & no & Supp. \\[10pt]
		5 & $ \pa^2 \zeta (\pa \chi)^2  $ & $\dfrac{1}{Q}\dfrac{F_5}{f}$ & no & Supp. \\[10pt]
		6 & $\dot{\zeta} \pa^2 \zeta \pa^2 \zeta$ & $ { \dfrac{1}{Q c_s^4}  \left( \dfrac{c_s}{a Hs} \right)^3}
		\dfrac{F_6}{f}$ & no & -- \\[10pt]
		7 & $(\pa_a \pa_b \zeta)^2 \pa^2 \zeta$ & 
		$
		{ \dfrac{1}{Q c_s^6}  \left( \dfrac{c_s}{a Hs} \right)^4}
		\dfrac{F_7}{f}$ & no & -- \\[10pt]
		8 & $(\pa^2 \zeta) (\pa^2 \zeta) (\pa^2 \zeta)$ & ${ \dfrac{1}{Q c_s^6}  \left( \dfrac{c_s}{a Hs} \right)^4}
\dfrac{F_8}{f}$ & no & -- \\[10pt]
		9 & $(\pa^2 \zeta) (\pa_a \pa_b \zeta) (\pa_a \pa_b \chi)$ & ${ \dfrac{1}{Q c_s^4}  \left( \dfrac{c_s}{a Hs} \right)^3}
\dfrac{F_9}{f}$ & no & -- \\[10pt]
		\end{tabular}
		\end{center}
		\caption[GSR bispectrum operators and sources]{GSR bispectrum operators, sources, whether they contribute to the squeezed-limit, and their status in the GLPV class and its subset the Horndeski class. ``Supp.'' denotes an operator which is slow-roll suppressed, while ``Free'' an operator which is not. The $i=6$, $i=7$, $i=8$, and $i=9$ operators are identically zero in the GLPV and Horndeski classes.}
		\label{tab:OperatorSummary}
		\end{table}

	\subsection{Consistency Relation}
		\label{subsec:squeeze}

		In \S\ref{subsec:CubicActionConsistency}, we argued from the cubic action that the squeezed-limit consistency relation \eqref{eq:ConsistencyRelation} holds during slow-roll inflation. Now that we have the complete integral forms of the bispectrum to first-order in slow-roll parameters, we can examine the squeezed-limit consistency relation in more detail, in particular focusing on its form beyond slow-roll.

		We first confirm our expectation from \S\ref{subsec:CubicActionConsistency} that only the
		 $i=0$ and $i=1$ operators  contribute in the squeezed-limit. In the published version of this work, Ref.~\cite{Passaglia:2018afq}, we show from the full expressions for GSR sources, windows, and weights that in the squeezed-limit $x_L / x_S \gg 1$, $x_S \ll 1$, we have 
		\begin{equation}
		\sum_{i=2}^{9} \Bigl[ \sum_j T_{ij} I_{ij} + \left[T_{iB} I_{iB}(2 k_3) + \text{perm.}\right] \Bigr] = 0,
		\end{equation}
		and therefore the operators $i=2$ to $9$ have no net squeezed contribution.

		As for the $i=0$ and $i=1$ operators, only  $I_{01}$, $I_{02}$, $I_{11}$, and $I_{12}$ contribute to squeezed triangles as $k_L/k_S$.   We can then generalize a calculation from Ref.~\cite{Adshead:2013zfa} to show that the consistency relation holds even beyond slow-roll. The GSR expression for the squeezed bispectrum is
		\begin{align*}
		\frac{12}{5} f^{\text{squeezed}}_{\text{NL}} = \lim_{k_S \rightarrow 0} &\frac{1}{\Delta (k_S)} \Bigl[-2 I_{01}(2 k_L)  + 4 I_{02}(2 k_L) + 2 I_{11}(2 k_L) - 2 I_{12}(2 k_L) \Bigr]. \numberthis
		\end{align*}

		To leading order we can substitute $1/\Delta(k_S) \rightarrow f_*$, resulting in
		\begin{align*}
		\label{eq:GSRconsistency}
		\frac{12}{5} f_{\text{NL}}
		\approx&
		-2 \frac{f'}{f}\Big|_{s_*} 
		+  f_* \int_{s_*}^\infty \frac{d s}{s}   \bigg[ \left( \frac{\e}{f} \right)' W_{\epsilon}( k_L s) \numberthis \\ &  \qquad 
		+
		\left( \frac{\sigma}{f} \right)' W_{\sigma}( k_L s) 
		+
		\left( \frac{q}{f} \right)' W_{q}( k_L s) \bigg], 
		\end{align*}
		where
		\begin{align*}
		 W_{\epsilon}(x) =& \ \frac{1}{x}  \sin(2 x),\\
		 W_{\sigma}(x) =& \ \frac{2}{x}  \sin(2 x) - \cos (2 x),\\
		 W_{q}(x) =& \ \frac{1}{x}  \sin(2 x) - \cos (2 x),
		 \numberthis
		 \end{align*}
		and we have evaluated the boundary term during slow-roll as
		\begin{equation}
		(2 \e + 3 \sigma + q)|_{s_*} \simeq -2 \frac{f'}{f}\Big|_{s_*}.
		\end{equation}

		We need to compare this GSR expression for the squeezed bispectrum to the GSR expression for the tilt of the power spectrum \cite{Adshead:2012xz},
		\begin{align*}
		\label{eq:slope}
		\frac{ d \ln \Delta_\zeta^2}{d \ln k}\Big|_{k_L}  &= \int_{s_*}^\infty \frac{ds}{s} W'(k_L s) G'(\ln s)
		\numberthis\\
		&=
		2 \frac{f'}{f}\Big|_{s_*} + \int_{s_*}^\infty \frac{ds}{s} \left( \frac{f'}{f} \right)' W_{n}(k_L s),
		\end{align*}
		where
		\begin{equation}
		W_{n}(x) =-2 \cos(2 x) + \frac{2}{x} \sin(2x).
		\end{equation}

		We see immediately from comparing the boundary terms in Eqs.~\eqref{eq:GSRconsistency} and \eqref{eq:slope} that the squeezed limit consistency relation holds in slow-roll. The integral contributions become significant during slow-roll violations. For a sharp feature at $k_L s \gg 1$, the parameters with the highest numbers of derivatives dominate and
		\begin{equation}
		\label{eq:sharpstepappx}
		\left( \frac{f'}{f} \right)'  \approx \frac{f''}{f} \approx - \frac{f_*}{2} \left( \frac{\sigma + q}{f} \right)',
		\end{equation}
		which, when combined with the windows in the desired limit, establishes consistency beyond slow-roll between Eqs.~\eqref{eq:GSRconsistency} and \eqref{eq:slope}.

		We have made two assumptions in deriving the consistency relation beyond slow-roll. First, we have assumed that the net change in the power spectrum between two different scales
		is slow-roll suppressed and thus that we can send $1/ \Delta (k_S)$ to $f_*$.  Implicitly this requires that 
		any slow-roll violation is highly transient so that the integrated effect of transient violations remains small.   Therefore
		second, we assume that the sources of slow-roll violation are sharp in their temporal structure using Eq.~\eqref{eq:sharpstepappx}. The inflationary model we consider in the following section can violate these approximations by allowing large changes in the power spectrum outside the well observed regime. Nonetheless, we expect that the consistency relation when computed exactly holds in general as long as $\zeta$ freezes out after
		horizon crossing.		
		
\section{Worked Example: Transient G-Inflation}
	\label{sec:GInflation}
	In this section, we illustrate the
	calculation of  the scalar bispectrum in our general formalism for the unified EFT of inflation with a specific  model with cubic galileon interactions
	  in which slow-roll is transiently violated. This model is not constructed to form PBH DM, which we shall instead do in the following chapter, but rather to achieve a specific tensor-to-scalar ratio and resolve certain reheating instabilities.  We briefly review this transient G-inflation model and its power spectrum in \S\ref{subsec:GInfModel} and present its bispectrum in \S\ref{subsec:GInfResults}.

	\subsection{Scalar Power Spectra}
		\label{subsec:GInfModel}
		The transient G-inflation model is presented in detail along with its scalar and tensor power spectra in Ref.~\cite{Ramirez:2018dxe}. We briefly review it here.

		We assume that the Lagrangian density takes the form 
		\begin{equation}
		\L = -X/2 - V(\phi) + f_3(\phi) \frac{X}{2} \Box \phi + \frac{R}{2},
		\end{equation}
 		with the chaotic inflation potential $V(\phi) = m^2 \phi^2/2$. In Ref.~\cite{Ohashi:2012wf}, this model is considered with a constant $f_3 = -M^{-3}$. The constant $f_3$ model suffers from two problems: for the measured value of the scalar tilt $n_s$, it predicts too large a tensor-to-scalar ratio $r$; and for some values of $m$ and $M$ the inflaton has a gradient instability $c_s^2 < 0$ during reheating whose resolution would lie beyond the scope of the perturbative EFT.

		Transient G-inflation shuts off the G-inflation term before the end of inflation by using a $\tanh$ step-like feature in $f_3$
		\begin{equation}
		f_3 (\phi) = -M^{-3} \left[ 1 + \tanh\left(\frac{\phi-\phi_r}{d}\right)\right].
		\end{equation}
		
\begin{figure}[t]
		\begin{center}
		\includegraphics[width=.7\linewidth]{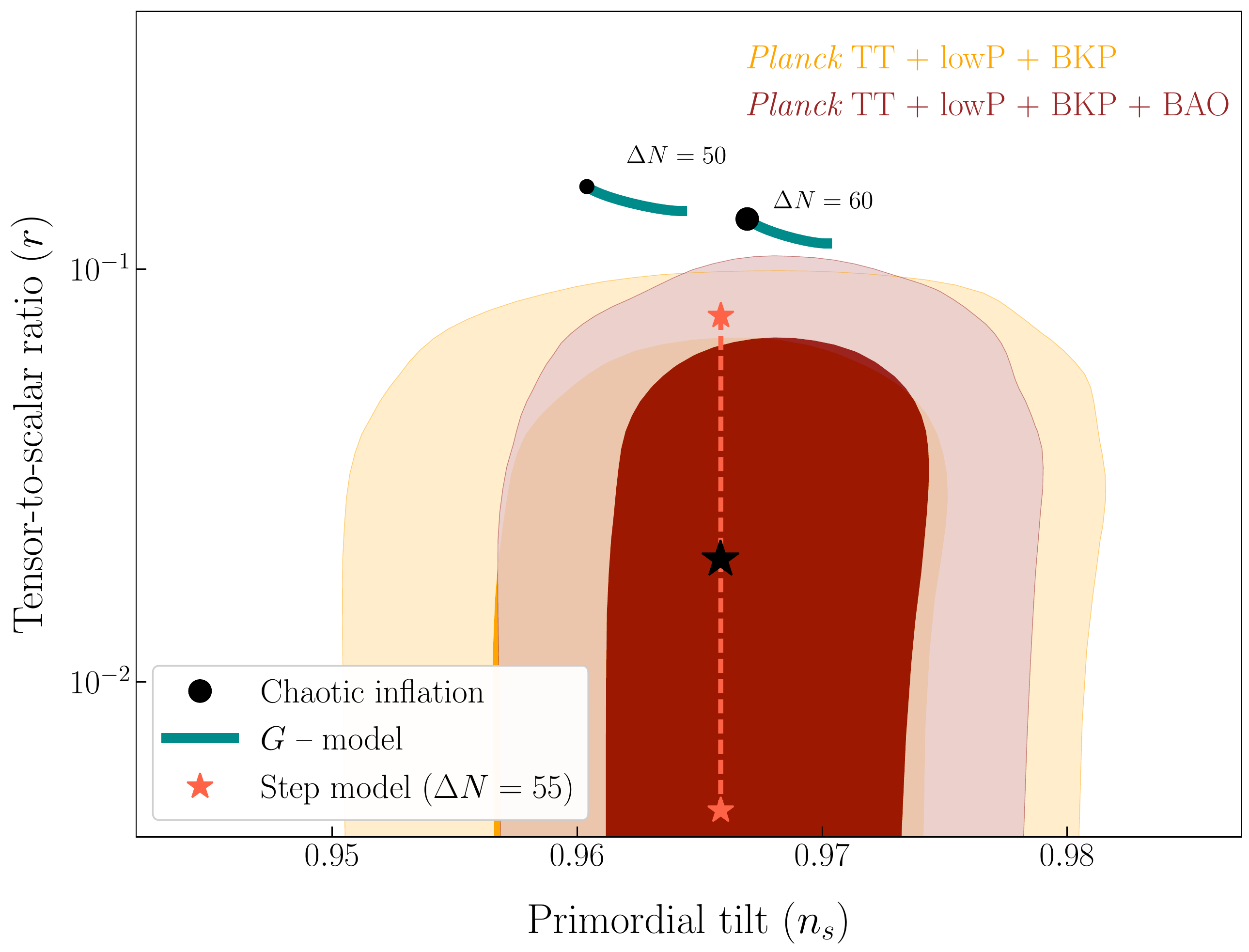}
		\end{center}
		\caption[Tensor-to-scalar ratio and spectral tilt: theory and experiment]{{\it Planck} and BICEP-Keck constraints in the $(n_s,r)$ plane along with the range of predictions from canonical chaotic inflation (black points), G-inflation (green lines), and the transient G-inflation `step' model (orange line and stars). The black star is the `large-step' model of this work.}
		\label{fig:nsr}
		\end{figure}

		Prior to the step, the inflaton is in a G-inflation regime, while after the step, the inflaton follows the slow-roll attractor solution of chaotic inflation. Because the $f_3 X \Box \phi$ term in the Lagrangian becomes negligible after the step, the gradient instability at the end of inflation is avoided. By having the transition start just as the CMB scale exits the horizon, the tilt $n_s$ is decoupled from the tensor-to-scalar ratio $r$ and therefore the model can be consistent with observations, as seen in Fig.~\ref{fig:nsr}.

		We consider two parameter sets for the transient G-inflation model, a `large-step' model and a `small-step' model.
		The large-step model is the fiducial model of Ref.~\cite{Ramirez:2018dxe}. The inflaton mass scale $m = 2.58 \times 10^{-6}$ is chosen to satisfy the {\it Planck} 2015 TT+lowP power spectrum amplitude. The Galileon mass scale $M=1.303 \times 10^{-4}$ suppresses the tensor amplitude relative to the scalar amplitude when the CMB mode $k_{\text{CMB}} = 0.05$\,Mpc$^{-1}$ exits the horizon $55$ $e$-folds before the end of inflation. The remaining parameters $\phi_r=13.87$ and $d=0.086$ control the step and are chosen such that the tilt and running satisfy observational constraints.

		The small-step model is chosen by the same procedure, save for the parameter $M$ which is selected for a larger tensor amplitude. The other parameters are adjusted to keep the tilt and amplitude of the power spectrum fixed. The resultant parameter set is $\{m, M, \phi_r, d\} = \{6.50 \times 10^{-6},\ 48.25 \times 10^{-4},\ 14.67,\ 0.021\}$. In this model, the power spectrum evolution before and after the step is much smaller as inflation is never in a fully G-inflation dominated phase, and thus we are closer to the regime of validity of the argument in \S\ref{subsec:squeeze}.


		\begin{figure}[t]
		\begin{center}
		\includegraphics[width=.65\linewidth]{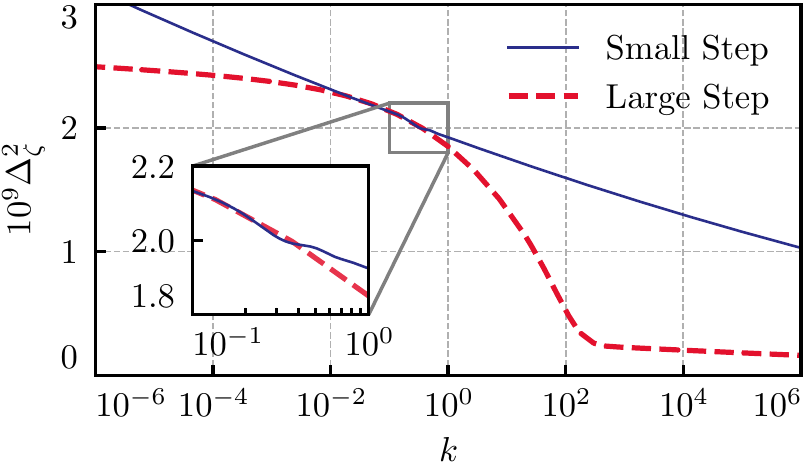}
		\end{center}
		\caption[GSR power spectra]{The GSR power spectra for the transient G-inflation models we consider.  In the small step model, the transition has a small amplitude but rapid variation 
		whereas in the large step model, it has a large amplitude and slow variation. For comparison to exact solutions of the Mukhanov-Sasaki equation, see Ref.~\cite{Ramirez:2018dxe}.}
		\label{fig:powerSpectra}
		\end{figure}

		In both models, slow-roll is transiently violated as the inflaton traverses the step, and thus the GSR formalism should be used in place of the traditional slow-roll approach for power spectrum and bispectrum observables. We show the GSR power spectra for these models in Fig.~\ref{fig:powerSpectra}.  In the small-step model, the deviations from scale invariance 
		are small in amplitude but rapidly varying in $k$ (see inset).   In the large-step model, they
		are large in amplitude but smoother in scale.    We shall see next that these properties 
		also apply to the bispectrum.

	\subsection{GSR Bispectrum}
		\label{subsec:GInfResults}

		We now compute the bispectrum for the transient G-inflation models of \S\ref{subsec:GInfModel} using the GSR formulas from \S\ref{sec:GSR}.
		
		We begin by computing the squeezed bispectrum, where the consistency relation allows us to check our computations by comparing the bispectrum result in the squeezed limit to the slope of the GSR power spectrum using Eq.~\eqref{eq:ConsistencyRelation}. We choose to fix the ratio $k_S / k_L = 10^{-2}$. From the analytic analysis in \S\ref{subsec:squeeze}, we know that the only operators which contribute to the squeezed limit are the $i=0$ and $i=1$ operators, and their sources are manifestly related to the local slope of the power spectrum. Thus we expect these operators to enforce the consistency relation. 

		\begin{figure}[t]
		\begin{center}
		\includegraphics[width=.65\linewidth]{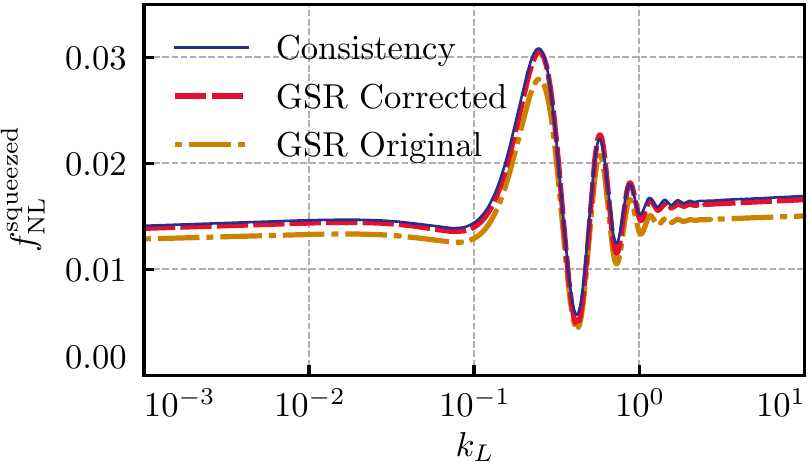}
		\end{center}
		\caption[Squeezed bispectrum in small-step G-inflation]{Squeezed bispectrum for small-step transient G-inflation. We see excellent agreement between the GSR bispectrum and the consistency relation curve, though with a slight amplitude error. By applying a simple correction to account for modefunction evolution outside the horizon, we can eliminate this error completely.}
		\label{fig:shortSqueezed}
		\end{figure}

		The accuracy of the GSR approximation in the squeezed-limit for the small-step case is shown in Fig~\ref{fig:shortSqueezed}. The GSR bispectrum result closely tracks the consistency relation result before, during, and after the step in the power spectrum. Slow-roll violations during the transition appear as sharp features in the sources which, when integrated against the windows, induce oscillatory features in the squeezed bispectrum and in the tilt of the power spectrum. 
		
		While the  GSR bispectrum calculation and the power spectrum based consistency relation expectation agree on the period and phase of these features, there is a small amplitude difference between the curves before, during, and after the transition. This error occurs because the
		bispectrum and power spectrum are calculated to first-order in slow-roll suppressed
		quantities.  In particular the
		 consistency relation check of \S\ref{subsec:squeeze} ignores corrections due to the
		 evolution in $f$ which would be picked up in the next order of the GSR iteration. 
		  Since there is some slow-roll suppressed evolution in $f$ between the epochs when
		   $k_S$ and $k_L$ freeze out, or equivalently in the power spectra at the two scales,
		   a correspondingly small error is induced in the bispectrum. 
		   
		   In this case, where the 
		   change in the  power spectrum  between $k_S$ and $k_L$ is insignificant, this error is minor. Nonetheless, in the upcoming large-step example the power spectrum will significantly evolve across freeze-out epochs and this error will become large. In Refs.~\cite{Adshead:2013zfa,Adshead:2012xz}, it is shown that next-order terms in the GSR hierarchy provide a correction factor
		\begin{equation}
		\label{eq:gsr_squeeze_corr}
		R_0 = 1 + \frac{n_s-1}{2} \ln \left(\frac{k_S}{k_L}\right),
		\end{equation}
		assuming that the squeezed bispectrum integrals receive most of their contributions
		at horizon crossing for $k_L$.  

		This correction multiplies the zeroth-order bispectrum contributions from the $i=0$ and $i=1$ terms and corrects for the leading-order integrated evolution of $f$. Since in the following example the power spectrum evolution will be large, we generalize this correction to the non-leading integrated evolution of $f$ by choosing.
		\begin{align}\label{eq:gsr_squeeze_corr_resum}
		R = \Delta(k_S)/\Delta(k_L).
		\end{align}
		
		We show in Fig.~\ref{fig:shortSqueezed} that this correction eliminates the small amplitude error, improving the consistency between the squeezed bispectrum and the derivative of the
		 power spectrum. This correction does not impact triangle shapes where all three modes are comparable in scale. For a formulation of GSR which avoids this type of error by maintaining order-by-order modefunction freeze-out, see Ref.~\cite{Miranda:2015cea}.

		\begin{figure}[t]
		\begin{center}
		\includegraphics[width=.65\linewidth]{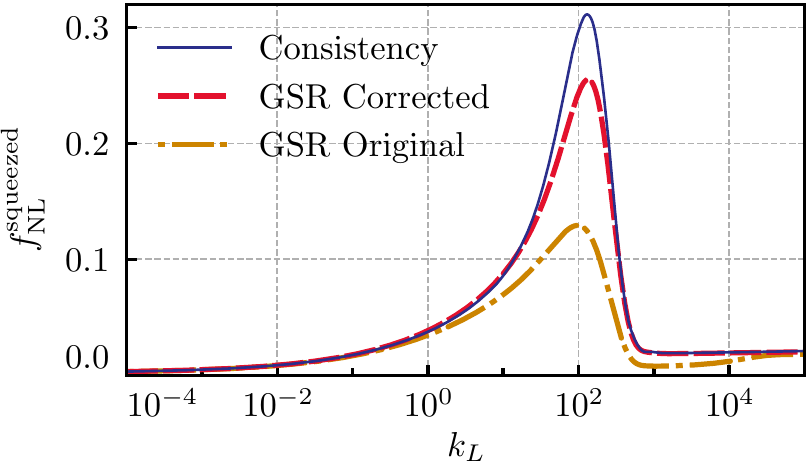}
		\end{center}
		\caption[Squeezed bispectrum in large-step G-inflation]{Squeezed bispectrum for large-step transient G-inflation. Despite the large evolution of the power spectrum in this model, the corrected GSR bispectrum tracks closely the consistency relation. The discrepancy between the corrected bispectrum result and the consistency relation at the peak indicates that the next-order term in the GSR hierarchy becomes important there.}
		\label{fig:wideSqueezed}
		\end{figure}

		We show in Fig.~\ref{fig:wideSqueezed} the squeezed bispectrum for the large-step model. In the large-step model, the $i=0$ and $i=1$ sources are much wider than in the small-step case and thus the bispectrum appears as a single peak rather than an oscillatory function. In addition, for this choice of model parameters the power spectrum evolution is large and thus the GSR squeezed bispectrum makes a significant error across the step. Nonetheless, correcting the bispectrum for the integrated evolution $f$ between $k_S$ and $k_L$ with Eq.~\eqref{eq:gsr_squeeze_corr_resum} succeeds in explaining most of this discrepancy.
		
		The residual errors in Fig.~\ref{fig:wideSqueezed} at the peak of the squeezed bispectrum can be understood as a reflection of other iterative corrections in the GSR hierarchy, modes which converge
		only slowly in this large-step case. 
		These terms are associated
		with the dynamics of the $k_L$ modes and similar corrections are required for
		the power spectrum
		as well.  
		 In fact, it is explicitly shown in Ref.~\cite{Ramirez:2018dxe} 
		that the $g$ terms in the power spectrum expansion reach order unity during the
		transition, which explains why  higher order GSR contributions are necessary to ensure the consistency relation holds at the bispectrum peak.

		\begin{figure}[t]
		\begin{center}
		\includegraphics[width=.65\linewidth]{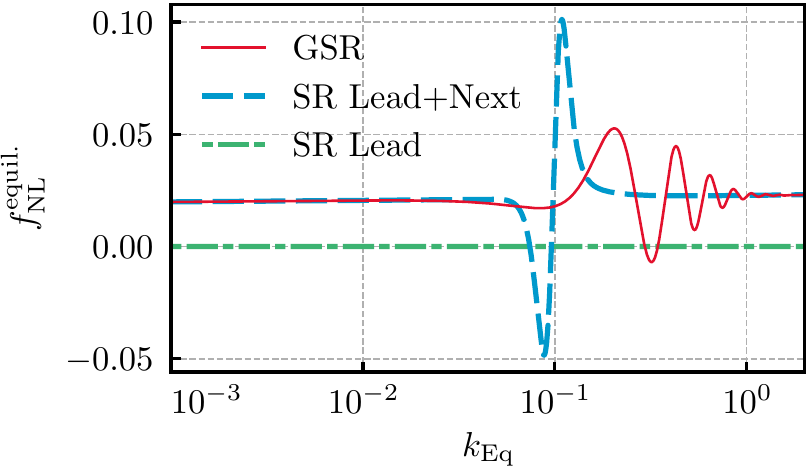}
		\end{center}
		\caption[Equilateral bispectrum in small-step G-inflation]{Equilateral bispectrum for small-step transient G-inflation. For this set of parameters, $|f_3| \ll 1$ and thus the leading-order slow-roll contribution Eq.~\eqref{eq:GInfEquilBispectrumSR} remains nearly $0$. The next-to-leading order SR contribution, Eq.~(100) of Ref.~\cite{DeFelice:2013ar}, dominates, and agrees with the GSR computation before and after the step. During the transition, the SR hierarchy is violated and the SR expression fails to accurately track the GSR bispectrum.}
		\label{fig:shortEquil}
		\end{figure}

		\begin{figure}[t]
		\begin{center}
		\includegraphics[width=.65\linewidth]{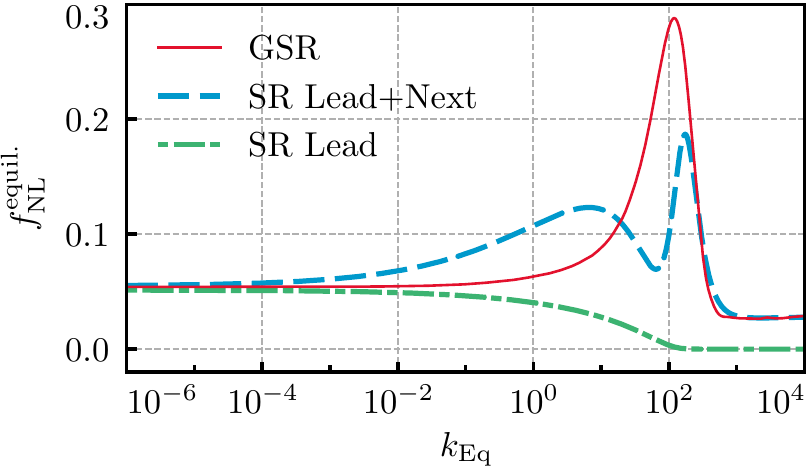}
		\end{center}
		\caption[Equilateral bispectrum in large-step G-inflation]{Equilateral bispectrum for large-step transient G-inflation. In this case the leading-order SR contribution  Eq.~\eqref{eq:GInfEquilBispectrumSR} dominates prior to the transition where $|f_3| \gg 1$. After the transition, the leading-order SR contribution goes to zero while the next-to-leading order terms in Eq.~(100) of Ref.~\cite{DeFelice:2013ar} come to dominate. The GSR result again agrees with the SR results before and after the transition, while the SR result shows an erroneous double-peak feature during the transition. Despite the enhancement due to the slow-roll violation, $|f_{\rm NL}^{\rm equil.} | <1$  at all times.}
		\label{fig:wideEquil}
		\end{figure}

	We next turn to the equilateral bispectrum. Only the $i=2$ and $i=3$ operators yield contributions which are not slow-roll suppressed  (see Tab.~\ref{tab:OperatorSummary}). In the 
	slow-roll approximation one would take their sources to be constant in Eq.~\eqref{eq:GSRBi}
	and obtain 
		\begin{equation}
		\label{eq:EquilBispectrumSR}
		f_{\mathrm{NL}}^{\mathrm{lead;\ equil.}} \simeq \frac{35}{108} (1 - F)_{\textrm{SR}} + \frac{5}{81} \left(\frac{F_3}{Q}\right)_{\textrm{SR}},
		\end{equation}
in which the $``\textrm{SR}"$ subscript denotes that the functions should be expanded to zeroth order in slow-roll. This can be shown to agree analytically with the result for the leading-order equilateral bispectrum in the literature for Horndeski models, Eq.~(97) of Ref.~\cite{DeFelice:2013ar}. In the specific case of transient G-inflation, Eq.~\eqref{eq:EquilBispectrumSR} takes the form

		\begin{align*}
		\label{eq:GInfEquilBispectrumSR}
		&f_{\mathrm{NL}}^{\mathrm{SR, equil.}} \simeq \frac{5 f_3^2 H^2 \dot\phi {}^2 \left(17
		 + 94 f_3 H \dot\phi 
		 - 17 f_{3,\phi} \dot\phi^2\right)}{81 \left(1
		 + 4 f_3 H \dot\phi 
		 - f_{3,\phi} \dot\phi^2\right)^2 \left(1
		 + 6 f_3 H \dot\phi 
		 - f_{3,\phi} \dot\phi^2\right)}, \numberthis
		\end{align*}
		in which ${}_{,\phi}\equiv d/d\phi$.
		
		When $|f_3|$ is large, as in pure G-inflation, the leading-order equilateral bispectrum dominates over slow-roll suppressed terms and leads to a larger bispectrum than in canonical inflation.
		However, when $|f_3|$ is small, as occurs in the small step model and after the transition in the wide step model, the leading order contribution to the equilateral bispectrum is subdominant to the slow-roll suppressed contributions from the $i=0$ and $i=1$ operators. For this case, Ref.~\cite{DeFelice:2013ar} computes a next-to-leading order contribution to the bispectrum, which results from considering the contributions from slow-roll suppressed operators, the next-order in slow-roll contributions from the $i=2$ and $i=3$ operators, as well as SR corrections to the modefunctions.

		In Fig.~\ref{fig:shortEquil} and Fig.~\ref{fig:wideEquil}, we compare the total equilateral bispectrum in GSR with the leading-order slow-roll expression \eqref{eq:GInfEquilBispectrumSR} as well as Eqs.~(97) and (100) of Ref.~\cite{DeFelice:2013ar}, formulas which include the next-to-leading order contributions.
		
		In the small-step case, inflation before and after the transition is nearly canonical and thus the equilateral bispectrum is dominated by the $i=0$ operator. At the transition the $i= 1$, $i=2$, and $i= 3$ operators  contribute, while the $i= 4$ and $i= 5$ operators remain subdominant throughout. As expected, the leading-order slow-roll bispectrum is subdominant throughout while the slow-roll formula including the  next-to-leading order contributions agrees well with GSR before and after the transition. However, during the transition it  displays radically different behavior from the GSR curve and fails to reproduce the oscillatory equilateral bispectrum resulting from the sharp sources.
		
	In the large-step case, inflation before the step is in a G-inflation dominated phase. During this phase, the $i=2$ and $i=3$ operators dominate the equilateral bispectrum. In the G-inflation dominated limit, $f_3 \rightarrow -\infty$, the leading-order contribution in slow-roll to the equilateral bispectrum (\ref{eq:GInfEquilBispectrumSR}) approaches $235/3888 \sim 0.06$. The slow-roll suppressed contribution only yields a small adjustment to this value. This is significantly smaller than might be expected from the $k$-inflation scaling, for example in DBI inflation $f_{\mathrm{NL}}^{\textrm{equil.}} \simeq \frac{35}{108} (1-1/{c_s^2})$, which with the G-inflation $c_s^2 \simeq 2/3$ yields $f_{\mathrm{NL}}^{\textrm{equil.}} \simeq -0.16$.
	Note also the difference in sign.

 After the transition, $f_3 \rightarrow 0$ and the leading-order contribution goes to zero while next-to-leading order contributions become important. Once more, while the leading-order and next-to-leading order SR formulas can accurately track the GSR bispectrum when the usual slow-roll hierarchy is maintained, they fail during the transition when
 this hierarchy is violated. In particular, the next-to-leading order SR formula predicts an erroneous double peak structure in the equilateral bispectrum.

\section{Discussion}
	\label{sec:Discussion}

	In this chapter, we developed an effective field theory approach for the study of the power spectrum and bispectrum in single-clock inflation beyond the usual slow-roll approximation. 
	This approach begins with the most general action which breaks temporal diffeomorphisms but preserves spatial diffeomorphisms. 
	In addition we require that the scalar degree of freedom obeys a standard dispersion relation at leading order so that power spectra behave in the usual way. 

	Our approach of studying the action directly in unitary gauge yields a wider set of terms in the action than explicitly considered in previous work \cite{Cheung:2007st,Baumann:2011su,Senatore:2009gt,Bartolo:2010di}, and in particular our action encompasses the Horndeski \cite{Horndeski:1974wa} and GLPV \cite{Gleyzes:2014qga} classes.

	From this starting point we derive the quadratic and cubic actions for scalar curvature perturbations, making use of integration by parts and the equation of motion while discarding boundary terms which are suppressed outside the horizon. 
	By appropriately grouping the operators, we isolate the ones that contribute in the squeezed limit and highlight the consistency relation between the power spectrum and the squeezed bispectrum. 
	The resultant cubic action contains ten operators, of which six are present in the Horndeski and GLPV classes, and of these six operators four are slow-roll suppressed.

	We then compute the power spectrum and the tree-level bispectrum contribution from each operator using the in-in and GSR formalisms which are valid beyond the slow-roll limit.
	Our GSR results enable computation of the power spectrum and any bispectrum configuration
	for all the operators in our action from a set of simple one-dimensional integrals.
	
	In particular the GSR expressions confirm that the
	 consistency relation holds not just in the slow-roll approximation but also in the case of rapidly
	 varying sources.  
	This result extends works which show that the consistency relation explicitly holds in slow-roll, for specific models, or for certain subclasses of EFT operators \cite{Cheung:2007sv,Bartolo:2013exa,DeFelice:2013ar,Adshead:2013zfa}.

	As an explicit example, we compute the power spectrum and bispectrum for a specific inflationary model in the Horndeski class in which slow-roll is transiently violated, the transient G-inflation model \cite{Ramirez:2018dxe}. For this model, our first-order GSR results for the equilateral bispectrum show qualitatively different behavior from the slow-roll results in the literature during the slow-roll violating phase. This model also highlights corrections for squeezed configurations
	from non-leading GSR terms which can be important in models in which the power spectrum deviates dramatically from scale-invariance between freeze-out epochs. 
	
	The large number of time-dependent coefficients in the EFT of inflation allows a rich range of behavior of perturbations beyond slow-roll. By condensing this large family of coefficients  into a small number of integrals, we 
	have provided the tools with which the bispectrum for a very general class of inflation models
	can be easily studied. 

\chapter{Single-Field Beyond Single-Clock}
\label{chap:usr}

The squeezed-limit consistency relation we emphasized in Chapter~\ref{chap:eft} came about because the single-clock background is an attractor, so long-wavelength perturbations appear to short-wavelength modes as a simple rescaling of the background. In this chapter, based on Ref.~\cite{Passaglia:2018ixg}, we will first show that the non-Gaussianity produced by this coordinate shift has no effect on PBH abundances as measured in local coordinates, and therefore does not alleviate the no-go theorem for PBH DM in single-field slow-roll that we showed in Chapter \ref{chap:intro}.

We will then present a single-field model which does produce PBH DM, by violating the slow-roll approximation in a phase known as {ultra-slow-roll (USR)}~\cite{Kinney:2005vj}. In USR, the field velocity is no longer uniquely determined by the field position and the background is no longer an attractor -- USR is therefore a single field model which is not single-clock, and it violates the squeezed-limit consistency relation.

We will therefore study the effect on PBH abundances of USR's enhanced squeezed non-Gaussianity, and show that it does not have a qualitatively important effect on PBH abundances.

Finally, implementations of USR inflation which are consistent with CMB measurements must have the USR phase be transient. We explore in depth the phenomenology of such transient USR models, such as inflection-point inflation, to show that their squeezed non-Gaussianity is highly suppressed unless they satisfy certain specific conditions which we detail.

\section{Primordial Black Holes and Ultra-Slow Roll}
\subsection{No Go for Single Field Slow Roll}
	\label{sec:nogo}

	Following Ref.~\cite{Motohashi:2017kbs}, we showed in Chapter~\ref{chap:intro} that in canonical single field inflation the comoving curvature power spectrum
	\begin{equation}
	\Delta_\curv^2  \equiv \frac{k^3}{2 \pi^2} P_\curv 
	\end{equation}
	must reach at least
	\begin{equation}
	\label{eq:PBHcriterion}
	\Delta^2_\curv \sim 10^{-2}
	\end{equation}
	within $\sim42$ $e$-folds from the epoch when CMB scales exited the horizon, at which $\Delta_\curv^2 \simeq
	10^{-9}$, for the dark matter to be entirely composed of PBHs. 
	In slow roll, the power spectrum satisfies
	\begin{equation}
	\label{eq:Delta2SR}
	\Delta^2_\curv\simeq
	\frac{H^2}{8 \pi^2 \e}.
	\end{equation}
	Therefore such an enhancement of $\Delta^2_\curv$ requires a slow-roll violation of at least $\eta \equiv d \ln \e / dN \sim  1$ after horizon exit of the CMB modes but well before the end of inflation.
	
	In this section we update this slow-roll no-go theorem to include local non-Gaussianity which modulates
	short-wavelength power in a long-wavelength mode. In particular 
	since the formation of a PBH depends on the density fluctuation averaged on the horizon
	scale at 
	reentry of the perturbations, horizon scale power that is modulated by superhorizon wavelength fluctuations can
	in principle enhance formation.  We study whether such a modulation can make it possible to produce a 
	substantial fraction of the dark matter in PBHs with slow-roll inflation.

	In the presence of a long-wavelength fluctuation $\curv_{\Long}$, low pass filtered for 
	 comoving wavenumbers $k\le k_L$, the power spectrum at $\kShort \gg \kLong$ becomes position dependent
	\begin{equation}
	\label{eq:modulation}
	P_\curv(\kShort, x) = P_\curv(\kShort) \left[ 1 + \frac{d \ln P_\curv(\kShort)}{d \curv_{\Long}} \curv_{\Long} (x) \right].
	\end{equation}
	 By multiplying by and averaging over the long-wavelength mode, 
	\begin{eqnarray}
	\langle \curv_{\Long}(x) P_\curv(\kShort, x) \rangle_{\curv_{\Long}} &\simeq &  \int\frac{d^3 k_L}{(2\pi)^3}
	B_\zeta(\kLong,\kShort,\kShort),
	\end{eqnarray}
	we can relate the power spectrum response to the curvature bispectrum
	 $B_\zeta$,
	 \begin{equation}
	\label{eq:response}
	\frac{d \ln P_\curv(\kShort)}{d \curv_{\Long}} \simeq \frac{	B_\zeta(\kLong,\kShort,\kShort)} {P_\curv(\kShort) P_\curv(\kLong)} \simeq \frac{12}{5} \fNL(\Squeezedks).
	\end{equation}
	Here $\fNL$ is the standard dimensionless non-Gaussianity parameter
	\begin{equation}
	\label{eq:fNLdefinition}
	\fNL (k_1, k_2, k_3) \equiv \frac{5}{6} \frac{B_\curv(k_1, k_2, k_3)}{P_\curv(k_1) P_\curv(k_2) + \text{perm.}} ,\ \ 
	\end{equation}
	in which `$+\text{ perm.}$' denotes the two additional cyclic permutations of indices and the
	approximation \eqref{eq:response} assumes the squeezed limit $\kShort \gg \kLong$.

	\begin{figure*}[t]
	\includegraphics[width=\linewidth]{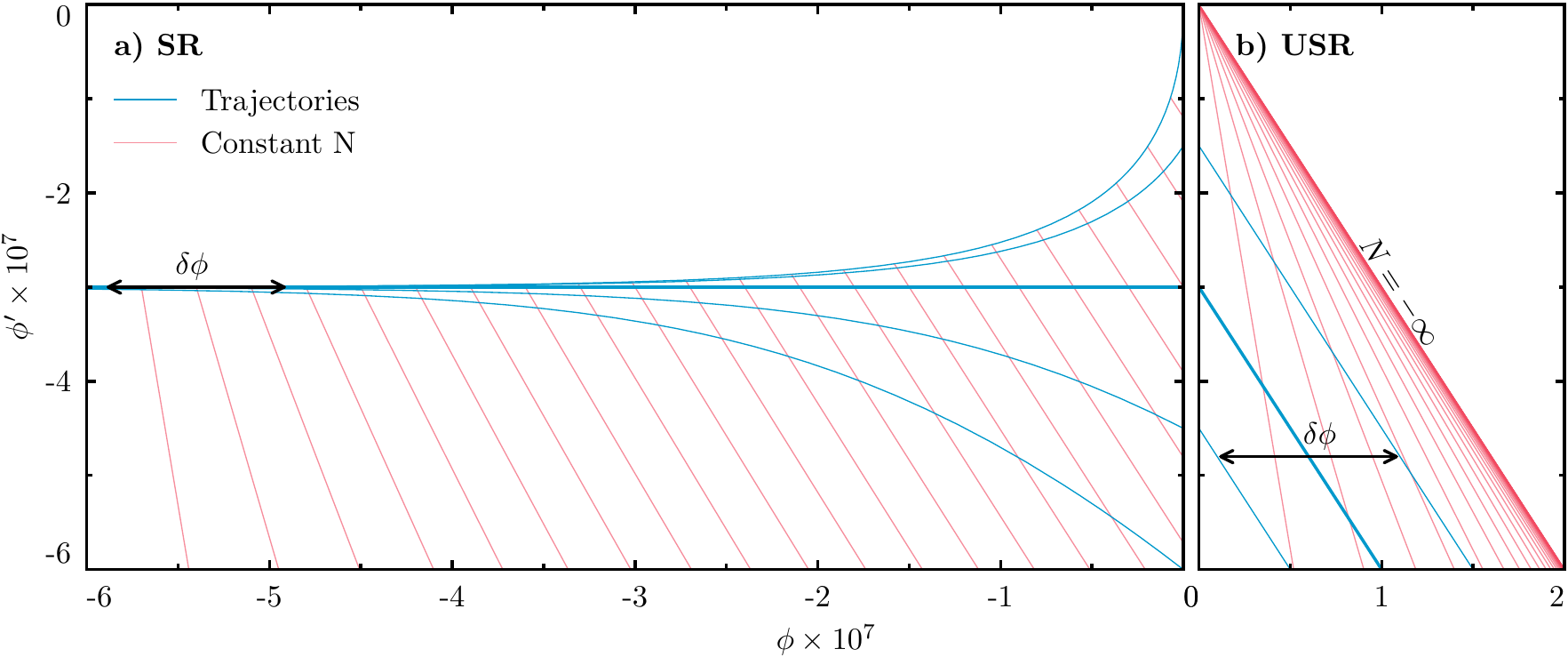}
	\caption[Phase spaces of slow-roll and ultra-slow roll]{Phase space diagram for a) slow roll (SR) and b) ultra-slow roll (USR). Shown are
	background trajectories (blue lines), lines of constant $e$-folds (red lines) to the end of inflation (left edge of panels), and field fluctuations $\delta \phi$ (arrows).
	SR trajectories converge to the attractor for different initial kinetic energies at $\phi=0$.  SR field fluctuations
	$\delta \phi$  follow the  attractor trajectory and can be absorbed into a change in  $e$-folds leaving a change in the relationship between local and global coordinates, but no local imprint on observables  once clocks are synchronized to a fixed field value at the end of inflation. 	
	USR field fluctuations  can still be absorbed into a local background but no longer the background
	of the unperturbed universe (thick blue line).   Since different USR trajectories experience  different numbers of $e$-folds to the end,
	the power spectrum becomes position dependent, with $\fNL$ reflecting the $e$-folding asymmetry between
	positive and negative $\delta\phi$ or $\partial^2 N/\partial \phi^2$. 
		} 
	\label{fig:phasespaceSR+USR}
	\end{figure*}\clearpage

	In single-field inflation, $\fNL(\Squeezedks)$ has a constrained form when $\zeta$ is conserved outside the horizon. The curvature perturbation is equivalent to a field fluctuation in spatially flat
	gauge $\zeta=-\delta\phi/\phi'$, with primes denoting derivatives with respect to 
	$e$-folds  $'=d/dN$ here and throughout.  Therefore for a constant $\zeta$, the field fluctuation evolves according to 
	\begin{equation}
	 \delta\phi' = \frac{\phi''}{\phi'}\delta\phi,
	 \label{eq:attractor}
	\end{equation}
	and the phase-space trajectory of the long-wavelength field perturbation follows that of the background
	itself.  Short-wavelength modes evolving in a long-wavelength perturbation then also follow the phase-space trajectory of the background, with the only difference being the local $e$-folds which determines the relationship between physical and comoving wavenumber (see Fig~\ref{fig:phasespaceSR+USR}\hyperref[fig:phasespaceSR+USR]{a}). 

	Single-field inflation on the slow-roll attractor \eqref{eq:attractor} therefore satisfies the consistency relation \cite{Maldacena:2002vr}
	\begin{equation}
	\lim_{\kLong/\kShort\rightarrow0}\frac{12}{5} \fNL (\Squeezedks) = -\frac{d \ln{ \Delta_\curv^2(\kShort)}}{d \ln{\kShort}}.
	\label{eq:ConsistencyRelationfNL}
	\end{equation}

	This implies a modulation of the small-scale power spectrum due to the long-wavelength mode according to Eqs.~\eqref{eq:modulation} and~\eqref{eq:response} as
	\begin{equation}
	\label{eq:GlobalDilation}
	P_\curv(\kShort, x) = P_\curv(\kShort) \left[ 1 -\frac{d \ln{ \Delta_\curv^2(\kShort)}}{d \ln{\kShort}} \curv_{\Long} (x) \right].
	\end{equation}
	This modulation is zero at the scale where the power spectrum peaks and corresponds to a dilation of scales rather than an amplitude enhancement.  
	
	In general the physical effect of a  dilation of scales is to change the mass scale of PBHs rather than enhance their abundance.  However in the slow-roll case, there is actually a change in neither abundance nor mass scale.
	Though the dilation \eqref{eq:GlobalDilation} does occur in global comoving coordinates, in single-field inflation a freely-falling
	observer will not see this dilation locally. 

	For a given perturbed metric, 	the standard Fermi normal coordinates (FNC)~\cite{Manasse:1963} can be constructed with respect to  a central timelike geodesic of a comoving observer~\cite{Senatore:2012ya,Senatore:2012wy}, such that $g_{\mu\nu}^{\rm FNC} \simeq \eta_{\mu\nu}$ up to tidal corrections.
	In order to absorb the effects of superhorizon perturbations out to the horizon scale of a local observer, as required for PBH calculations, we utilize  conformal Fermi normal coordinates (\fncb{}) \cite{Pajer:2013ana}.  \fncb{} are
	constructed such that $g_{\mu\nu}^{\overline {\rm FNC}} \simeq a^2 \eta_{\mu\nu}$, i.e.~a conformally flat, locally 
	Friedmann-Lema\^itre-Robertson-Walker (FLRW) form where the global scale factor $a$ of the background
	universe is evaluated
	at the proper time of the central observer.
	
	As shown in Ref.~\cite{Pajer:2013ana}, for single-field slow-roll inflation the bispectrum in \fncb{}
	is related to the comoving-gauge bispectrum by an additional term proportional to the tilt of the power spectrum as 
	\begin{align*}
	\label{eq:Bzetatrans}
	\lim_{\kLong/\kShort\rightarrow0} B_{\bar\curv}(\Squeezedks) =\:& P_\curv(\kLong) P_\curv(\kShort) \frac{d\ln \Delta^2_\curv(\kShort)}{d\ln \kShort} \\
	& + B_{\curv}(\Squeezedks),
	\numberthis
	\end{align*}
	where barred symbols denote quantities in the \fncb{} frame. This additional term neatly cancels the comoving-gauge squeezed bispectrum from the consistency relation~\eqref{eq:ConsistencyRelationfNL} 
	and thus in single-field slow-roll inflation
	\begin{equation}
	\label{eq:BzetabarCancellation}
	\lim_{\kLong/\kShort\rightarrow0}  B_{\bar\curv}(\Squeezedks) = 0.
	\end{equation}
	There is therefore no modulation of the power spectrum in \fncb{} 
	\begin{equation}
	\label{eq:PzetabarModulation}
	P_{\bar\curv}(\kShort, x) = P_{\bar\curv}(\kShort),
	\end{equation}
	and the small-scale power spectrum in \fncb{} does not depend on the value of the long-wavelength perturbation. All local observers therefore see the same small-scale power spectrum regardless of their position in the long-wavelength mode. 
	
	Physically, the cancellation in Eq.~\eqref{eq:BzetabarCancellation} occurs because the bispectrum from the consistency relation encodes the effect on small-wavelength modes of evolving in a separate universe with a background evolution defined by the long-wavelength mode. Once the long-wavelength mode is frozen, this effect is just to change coordinates in the separate universe relative to global coordinates. When making local observations, an observer knows nothing of the global coordinates and instead makes measurements in coordinates corresponding to the separate universe. The formation of
	PBHs is a local process and so their properties also do not depend on their position in the long-wavelength
	mode.

	This lack of local modulation can also be understood from the phase-space diagram Fig.~\ref{fig:phasespaceSR+USR}\hyperref[fig:phasespaceSR+USR]{a}. 
	Relative to the end of inflation 
	at a fixed field value, perturbed trajectories in slow roll are indistinguishable from the background trajectory and thus observers making measurements relative to the end of inflation
	cannot from any local measurement decide whether they inhabit different regions of a long-wavelength
	curvature perturbation.
	
	This leads us to our first conclusion:  squeezed non-Gaussianity cannot produce PBHs as a significant fraction of the dark matter in canonical single-field slow-roll inflation.  For such PBHs to form in canonical
	single-field inflation, the slow-roll approximation must be violated, at least transiently, to either produce large Gaussian
	or non-Gaussian fluctuations. In this sense, the slow-roll no-go theorem shown in Ref.~\cite{Motohashi:2017kbs} is robust and does not change.

	Models that evade this no-go result typically have a period when the inflaton rolls on a very
	flat potential where Hubble friction is insufficient to keep the inflation on the slow-roll attractor.  
	The ultra-slow-roll model, where the inflaton potential is perfectly flat, provides the prototypical example for
	such studies as we shall  see next.

\subsection{Ultra-Slow-Roll Inflation}
	\label{sec:usr}

	Ultra-slow roll~\cite{Kinney:2005vj} is a model of single-field inflation which greatly enhances the scalar  power spectrum while also breaking the single-field consistency relation \eqref{eq:ConsistencyRelationfNL}  for the squeezed bispectrum by violating the attractor condition (\ref{eq:attractor})~\cite{Namjoo:2012aa,Martin:2012pe}. It is therefore possible to spatially modulate the local power in small scale
	density fluctuations relevant for PBHs with long-wavelength modes.
	In this section we examine whether this non-Gaussian modulation can significantly enhance the PBH abundance in ultra-slow roll.

	USR is characterized by a potential which is sufficiently flat before its end, which we denote with $\phi=0$, that the Klein-Gordon equation takes the form 
	\begin{equation}
	\ddot\phi \simeq - 3 H \dot\phi,
	\label{eq:KGUSR}
	\end{equation}
	where here and throughout overdots denote derivatives with respect to the coordinate time $t$. If the
	potential energy dominates then $H\simeq \text{const.}$ and	Eq.~\eqref{eq:KGUSR} then implies $\phi''\simeq -3\phi'$ and hence $\phi'\simeq -3\phi+\text{const.}$, defining a family of trajectories in the phase-space diagram, as depicted by the blue trajectories in Fig.~\ref{fig:phasespaceSR+USR}\hyperref[fig:phasespaceSR+USR]{b}.  
	Therefore, the phase-space trajectory of the background evolution depends on the initial kinetic energy and does not exhibit attractor behavior.

	For an exactly flat potential at $\phi>0$, an inflaton with insufficient initial kinetic energy will not cross the plateau to 
	reach $\phi=0$, neglecting stochastic effects.
	In Fig.~\ref{fig:phasespaceSR+USR}\hyperref[fig:phasespaceSR+USR]{b} we focus on classical trajectories that can reach $\phi=0$ within finite $e$-folds, and hence the upper right triangle region is inaccessible.

	The solution to Eq.~\eqref{eq:KGUSR} is $\dot\phi \propto a^{-3}$ and so $\e\propto a^{-6}$ and $\eta = -6$.   
	Since the analytic solution of the Mukhanov-Sasaki equation for $\curv$ in the superhorizon limit is given by
	\begin{equation}
	\label{eq:shsol-s} 
	\curv \simeq c_1 + c_2 \int \frac{\diff{t}}{a^3 \e} , 
	\end{equation}
	with integration constants $c_1$ and $c_2$, it is dominated by the second mode which 
	grows in USR since $(a^3 \e)^{-1} \propto a^{3}$ rather than decays as it does in slow roll.
	With $H\simeq \text{const.}$, Eq.~\eqref{eq:shsol-s} gives $\zeta\propto a^{3}$ and hence in the spatially flat gauge $\delta\phi=-\zeta\phi'=\text{const.}$, implying that $\delta\phi'=0$, unlike the case of the slow-roll attractor \eqref{eq:attractor}.

	The power spectrum in this model depends on the value of $\e$ at the end of USR, 
	\begin{equation}
	\label{eq:Delta2USR}
	\Delta^2_\curv\simeq
	\dfrac{H^2}{8 \pi^2 \epsilon_{H, \text{end}}},
	\end{equation}
	and thus can be very large if $\epsilon_{H,\End} \ll 1$. 
 	
 	One can employ a gauge transformation from spatially flat gauge to comoving gauge to show that the squeezed-limit non-Gaussianity takes the form (see App.~B of Ref.~\cite{Passaglia:2018ixg})
	\begin{equation}
	\label{eq:USRfNL}
	\lim_{\kLong/\kShort\rightarrow0}\frac{12}{5} \fNL (\Squeezedks) = 6.
	\end{equation}
	Since the USR power spectrum is scale invariant, the large value of $\fNL$ in USR violates
	the consistency relation. 

	The physical origin of this large value for $\fNL$ can  be seen from the phase-space diagram Fig.~\ref{fig:phasespaceSR+USR}\hyperref[fig:phasespaceSR+USR]{b}.
	Due to the initial kinetic energy dependence of the background evolution, a USR perturbation cannot be mapped into a change in the background clock along the same phase-space trajectory. Instead, long-wavelength perturbations $\delta\phi$ carry no corresponding $\delta\phi'$ and so shift the USR trajectory to one with a different relationship between $\phi$ and $\phi'$.   On this shifted trajectory, the short-wavelength power spectrum attains a different value at the end of USR.  More generally, if a local measurement is sensitive to $\phi'$ at the end of inflation, as in the case of $\Delta^2_\curv(\kShort)$, then different observers will produce different measurements depending on their position in the long-wavelength mode. 
	
	This graphical representation of $\fNL$ can be turned into a computational method through
	the so-called $\delta N$ formalism  \cite{Starobinsky:1985aa,Salopek:1990jq,Sasaki:1995aw,Sugiyama:2012tj}. When the expansion shear for a local observer is negligible, as it is in USR above the horizon, the nonlinear evolution of the curvature fluctuation
	follows the evolution of local $e$-folds.    On spatially flat hypersurfaces, the field fluctuation can
	be absorbed into a new conformally flat FLRW background on scales much shorter than the
	wavelength and so the local $e$-folds may be calculated from the Friedmann equation of a separate
	universe.  The position-dependent power spectrum is therefore the second order change in $e$-folds due to 
	a short-wavelength $\delta \phi_S$ on top of a long-wavelength $\delta \phi_L$.  Since in USR these perturbations
	leave $\phi'$ unchanged, the non-Gaussianity parameter can be computed from the $e$-folds as a function of 
	phase-space position of the background $N(\phi,\phi')$  as
	\begin{equation}
	\label{eq:deltaN}
	\frac{12}{5} \fNL = 2 \frac{\partial^2 N}{\partial \phi^2} \left/ \left(\frac{\partial N}{\partial \phi}\right)^2 \right. ,
	\end{equation}
	at fixed $\phi'$. 	
	
	The consequence of this formula can be visualized through Fig.~\ref{fig:phasespaceSR+USR}\hyperref[fig:phasespaceSR+USR]{b}
	 as the effect of perturbations on phase-space trajectories. Around a chosen background trajectory, the long-wavelength perturbation is reabsorbed into a new background, a horizontal shift to a new trajectory. Short-wavelength perturbations living in this new background induce a second shift in the trajectory, hence the second derivative.  Visually, the fact that for the same amplitude of field fluctuation $|\delta\phi|$, a positive fluctuation intersects more surfaces of constant $N$ than
	a negative fluctuation indicates a large $\fNL$.
	Refs.~\cite{Namjoo:2012aa,Chen:2013eea,Cai:2017bxr,Pattison:2017mbe} follow this approach to analytically 
	compute its value in complete agreement with the in-in approach or the gauge-transformation approach. We shall again exploit the $\delta N$ formalism in \S\ref{sec:transientUSR}.

	Despite the violation in the consistency relation, the coordinate transformation for the bispectrum Eq.~\eqref{eq:Bzetatrans} still holds and the transformation from global comoving coordinates to \fncb{} leads to the same additional tilt-dependent term in the bispectrum as in the canonical case so long as the transformation to \fncb{} is performed when modes are frozen outside the horizon after the end of inflation.\footnote{
	\fncb{} can still be established during the USR phase but are more closely related
	to spatially flat gauge than comoving gauge in temporal synchronization (see also App. B of Ref.\cite{Passaglia:2018ixg}).  In spatially flat gauge, a superhorizon field fluctuation $\delta\phi$ can 
	be absorbed into a new, nearly conformally flat FLRW background, as we  exploit with the 
	$\delta N$ formalism.}
	After this time, the construction follows Ref.~\cite{Pajer:2013ana} exactly. This procedure of transforming coordinate systems after inflation is followed for slow-roll inflation in Ref.~\cite{Cabass:2016cgp} to compute the next-to-leading order term in the bispectrum transformation. Practically, it corresponds to the clock-synchronization condition that all local observers make their measurements at fixed proper time after the end of inflation.
	 
	 Given the scale invariance of the spectrum, the tilt-dependent transformation from comoving gauge to \fncb{} 
	leaves neither an enhancement of the local power in the long-wavelength mode nor 
	a modulation of the mass of the PBHs.  On the other hand, since the transformation term no longer cancels with the comoving-gauge $\fNL$ 
	itself, a large value of the latter can in principle enhance PBH formation locally. 

	If $\fNL(\Squeezedks)$ is described by the USR result Eq.~\eqref{eq:USRfNL}, then the local power spectrum can be enhanced by a factor $12/5\times \fNL \times \curv_{\Long} = 6 \times \curv_\Long$. Therefore the non-Gaussian response enhances the local power spectrum by an order unity quantity unless the long-wavelength mode is large, i.e.\ 
	\begin{equation}
	\curv_\Long \gtrsim 10^{-1}.
	\end{equation}
	However, the scale invariance of USR would then imply
	\begin{equation}
	\Delta^2_\Short = \Delta^2_\Long \sim \langle \curv_\Long^2(x)\rangle \gtrsim 10^{-2},
	\end{equation}
	which satisfies the criterion Eq.~\eqref{eq:PBHcriterion} for PBH formation, and therefore PBHs would already be produced at scale $\kShort$ even before accounting for the non-Gaussian response. Note that the conversion from $\Delta^2_\curv$ to spatial variance involves a summation over $\kLong$ and
	gives a logarithmic factor which depends on the total $e$-folds of USR.   In a realistic model this logarithmic 
	factor must be finite 
	so as to also satisfy constraints from the CMB.   

	This result is the second main conclusion of this work: in a USR model which does not produce a significant PBH abundance under the Gaussian approximation, the squeezed non-Gaussian response enhances the local power spectrum by at most
	\begin{equation}
	 \frac{\Delta P_\curv}{P_\curv} \lesssim 1,
	\end{equation}
	and therefore the squeezed non-Gaussian response does not qualitatively change Gaussian conclusions.  
	Of course as they originate from rare fluctuations, PBHs can change in their abundance but these changes can 
	be reabsorbed into model parameters that make no more than an order unity change in the power spectrum.
	In particular squeezed non-Gaussianity cannot make a model that falls far short of making PBHs the dark matter under
	the Gaussian assumption into one that does.

	Since inflation has to end and observational constraints should be satisfied on CMB scales, the simple picture presented here must be modified to account for transitions into and out of USR.   In \S\ref{sec:transientUSR} we shall explore whether even this level of enhancement still holds in such models of transient USR inflation.

\section{Transient Ultra-Slow Roll and non-Gaussianity}
	\label{sec:transientUSR}
	
	In addition to a graceful exit problem, USR inflation is incompatible with the measured tilt of the CMB power spectrum \cite{Aghanim:2018eyx} and is in tension with constraints on local non-Gaussianities in the CMB \cite{Ade:2015ava}, and therefore any USR phase must begin after CMB modes exit the horizon and must take care not to grow those modes after horizon exit.

	One model proposed in the literature for PBH production with a transient USR phase is inflection-point inflation \cite{Motohashi:2017kbs,Garcia-Bellido:2017mdw}. In \S\ref{subsec:inflection}, we show that the transition out of USR in inflection-point inflation induces
	\begin{equation}
	\fNL(\Squeezedks) \ll 1,
	\end{equation}
	and therefore non-Gaussianities do not enable  PBHs to be the dark matter in inflection-point inflation.

	This numerical result can be understood from Ref.~\cite{Cai:2017bxr}'s analytic study of infinitely sharp potential transitions between USR and SR, which we review briefly in \S\ref{subsec:infinite}.  Transitions where the inflaton velocity monotonically decreases to reach an attractor solution lead to squeezed non-Gaussianity that is proportional to the potential slow-roll parameters on the attractor. Conversely, transitions where the inflaton instantly goes from having too much kinetic energy for the potential it evolves on to suddenly having insufficient kinetic energy for a now much steeper potential conserve the USR non-Gaussianity.  We call the latter transitions
	 \textbf{large}, which we will define specifically below [see Eq.~\eqref{eq:CaiCriterion}].  

	In \S\ref{subsec:flat}, we generalize the analysis of Ref.~\cite{Cai:2017bxr} to potentials which do not have an infinitely sharp break, and in particular we study how quickly the inflaton must traverse the potential feature to reproduce the USR result. We show that to conserve the USR result Eq.~\eqref{eq:USRfNL} the transition must be \textbf{fast} in that it completes in a small fraction of an $e$-fold [see Eq.~\eqref{eq:dN}].   

	We conclude that the large $\fNL$ of USR will only be preserved if the transition to SR is both
	large and fast. For all other cases,  the enhancement to the local power spectrum
		\begin{equation}
	\frac{\Delta P_\curv}{P_\curv} \ll 1,
	\end{equation}
	and so squeezed non-Gaussianity in transient USR does not generally affect the conclusions on PBH formation.

	\subsection{Slow--Small Transition: Inflection-Point Inflation}
		\label{subsec:inflection}

		Inflection-point inflation is characterized by a potential which supports a slow-roll phase when CMB scales exit the horizon followed by a slow-roll violation and subsequent ultra-slow-roll phase which enhances the power spectrum at small scales. This USR phase is generally unstable and lasts just a few $e$-folds before the inflaton loses enough kinetic energy to lock onto the attractor solution of the potential and slow-roll inflation resumes \cite{Pattison:2018bct}.	We call this transition \textbf{slow} because the inflaton kinetic energy decreases monotonically to the slow-roll value, and \textbf{small} because the potential slow-roll parameters on the attractor are comparable to the kinetic energy at the end of the USR phase.

		We consider an inflection potential of the form explored in Ref.~\cite{Motohashi:2017kbs} following Ref.~\cite{Garcia-Bellido:2017mdw},
		\begin{equation} 
			V(\phi) = \frac{\lambda v^4}{12} \frac{x^2 (6 - 4 a x + 3 x^2)}{(1 + b x^2)^2},
			\label{eq:GBInflectionPotential}
		\end{equation}
		where $x=\phi/v$. We study this model with the parameters 
		\begin{equation}
		\left\{a, b-1, \lambda, v\right\} = \left\{3/2,\ 4 \times 10^{-5} ,\ 7 \times 10^{-8},\ 0.658 \right\}.
		\end{equation} In terms of the auxiliary variables of Refs.~\cite{Motohashi:2017kbs,Garcia-Bellido:2017mdw}, this model has 
		\begin{equation}\left\{\beta,\ \Delta N_{\SR} \right\}=\left\{4\times10^{-5},\ 125\right\}.\end{equation}
	
		These parameters are finely tuned to significantly suppress $\e$ after the CMB scale $k_0 = 0.05\ \text{Mpc}^{-1}$ exits the horizon $55$ $e$-folds before the end of inflation while also preventing the inflection point from trapping the inflaton for too many $e$-folds. Nonetheless our qualitative results for the non-Gaussianity are not sensitive to the specific functional form of the potential nor to the parameter set above.  

		Note that even with fine-tuning, this model does not fit observational constraints from the CMB	(e.g.,~\cite{Aghanim:2018eyx}) because the power spectrum is too red (scalar slope $n_s = 0.91$) due to the proximity of the inflection point to CMB scales.	This additional red tilt implies a larger value of $\epsilon$ at CMB scales and hence a larger relative suppression of
		$\epsilon$ and growth of the power spectrum during the USR phase.   Without this enhancement, the
		inflection model falls far short of forming PBHs as the dark matter~\cite{Motohashi:2017kbs} and so we 
		choose these parameters to study whether models on the threshold of forming sufficient PBHs for Gaussian 
		fluctuations can be made to do so through non-Gaussianity in the model.

		\begin{figure}[t]
		\centering
		\includegraphics[width=.65\linewidth]{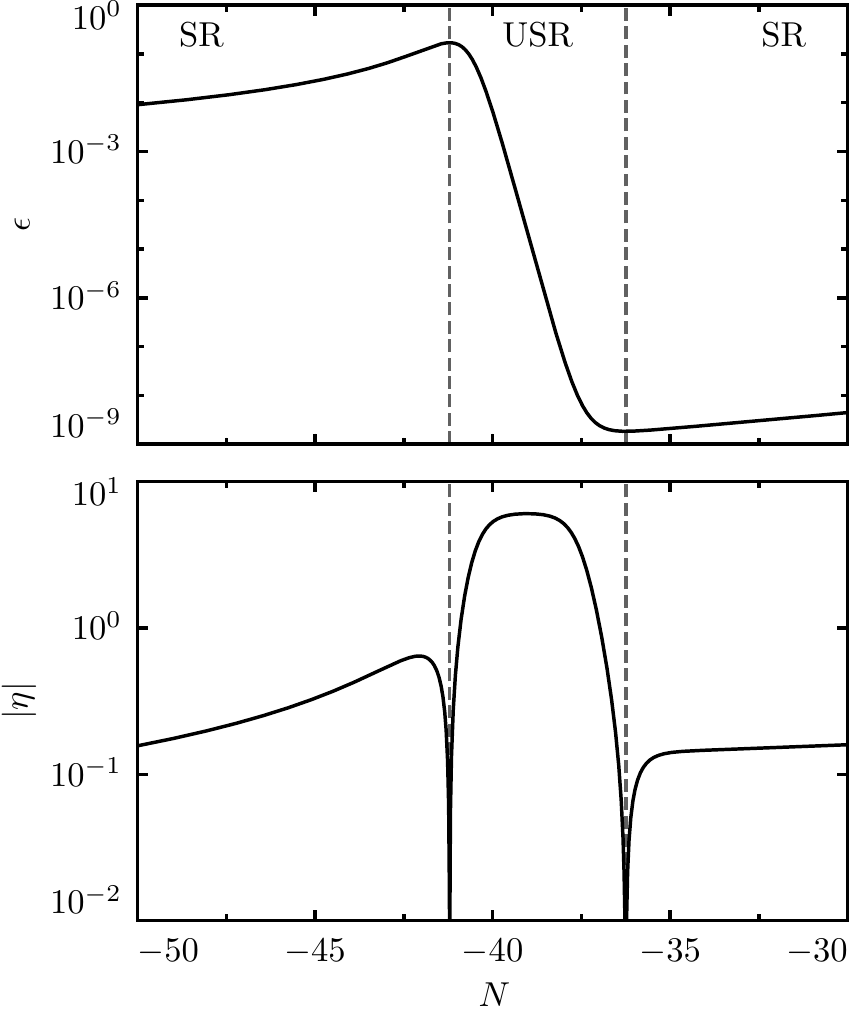}
		\caption[Inflection-point background]{Inflection-point model background parameters $\e \equiv - d\ln H/d N$ and $\eta \equiv d\ln\e / d N$. $\eta$ experiences two zero-crossings as $\epsilon$ reaches critical points entering and exiting the USR period, which we use to delineate the USR phase from the SR phases.
				Here the transient period lasts for $\sim5$ $e$-folds, but only achieves $\eta \simeq -6$ for 
		a shorter period. }
		\label{fig:GBBackground}
		\end{figure}

		Along the inflaton trajectory, the potential~\eqref{eq:GBInflectionPotential} has a single inflection point, where $d^2V/d\phi^2 = 0$ is satisfied, between two close points where $dV/d\phi = 0$.  In this region the slope of the potential is tiny, and hence the USR condition $|dV/d\phi| \ll |\phi'| H^2$ is satisfied briefly, after which slow roll quickly resumes.
		The evolution of the slow-roll parameters $\e$ and $\eta$ in this model is shown in Fig.~\ref{fig:GBBackground}. The model exhibits a transient period where $\e \propto a^{-6}$ and thus the USR result $\eta\simeq-6$ is temporarily achieved. 

		\begin{figure}[t]
		\centering
		\includegraphics[width=.65\linewidth]{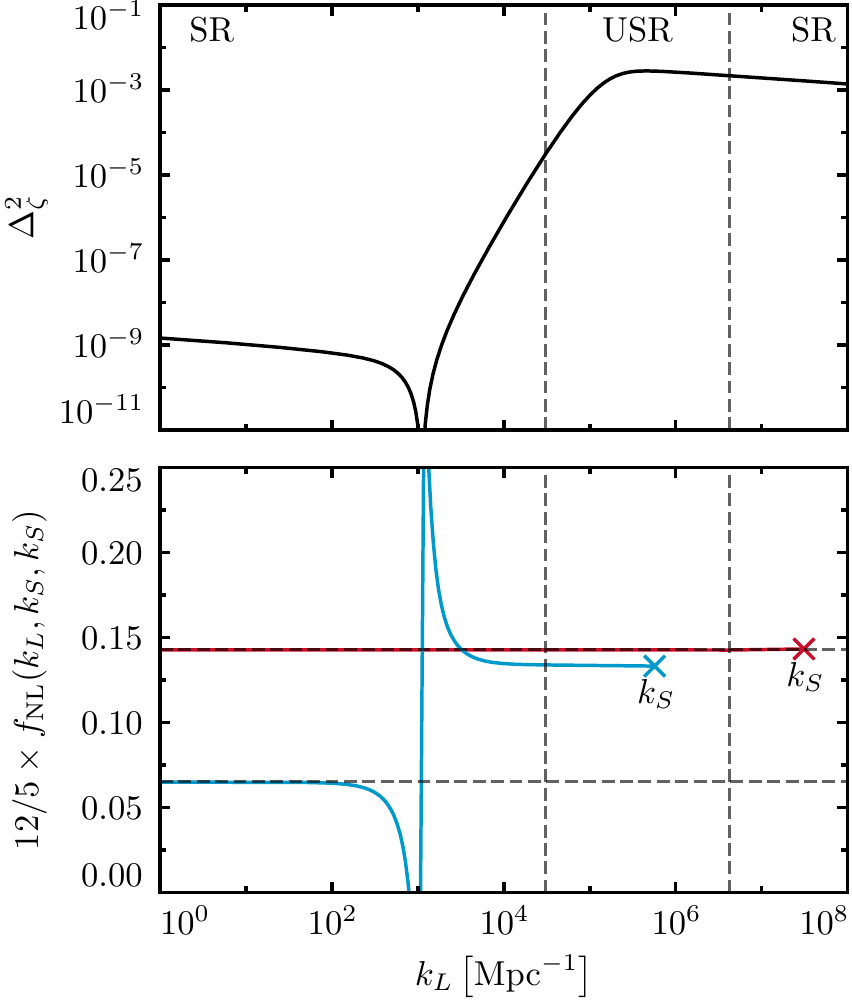}
		\caption[Inflection-point power spectrum and non-Gaussianity]{Inflection-point model power spectrum $\Delta^2_\curv$ and  non-Gaussianity parameter $\fNL(\Squeezedks)$. The vertical dashed lines delineate modes which cross the horizon during the SR and USR periods (see Fig.~\ref{fig:GBBackground}). 
		For $\fNL$, $\times$'s denote values for $\kShort>\kLong$ with blue a mode that exits the horizon during USR and red after USR. 
		The horizontal dashed lines denote the consistency relation expectation for the two modes and the spike 
		in $\fNL$ reflects the near zero in $\Delta_\zeta^2$ rather than a large bispectrum
		 (see \S\ref{subsec:inflection} for further discussion).
		 }
		\label{fig:GBPowerSpecAndfNL}
		\end{figure}\clearpage

		The upper panel of Fig.~\ref{fig:GBPowerSpecAndfNL} shows the power spectrum $\Delta^2_\curv$ produced by inflection-point inflation with the potential~\eqref{eq:GBInflectionPotential}, computed by numerically solving the Mukhanov-Sasaki equation,
		\begin{equation}
		\label{eq:MukhanovSasaki}
		\frac{1}{a^2\epsilon}\frac{d}{ds} \left( a^2 \epsilon \frac{ d \curv_k}{d s} \right)  + k^2 \curv_k = 0,
		\end{equation}
		where  $s \equiv \int_{t}^{t_{\rm end}} dt / a$, with  Bunch-Davies initial conditions at $ks\gg 1$ of the form
		\begin{equation}
		\label{eq:BunchDavies}
		\curv_k^0 = \frac{1}{2 a\sqrt{ k \epsilon}} \left(1 + \frac{i}{k s}\right) e^{i k s}.
		\end{equation}

		Modes which exit the horizon well before the USR phase do not grow outside the horizon and their power spectrum satisfies the slow-roll result \eqref{eq:Delta2SR}. Modes which exit the horizon shortly before the USR phase do, however, grow outside the horizon leading to the rise before peak power in the USR phase.

		This behavior can be understood in more detail from the exact, but formal, solution of the Mukhanov-Sasaki equation	\eqref{eq:MukhanovSasaki} \cite{Hu:2016wfa}
		\begin{equation}
		\label{eqn:Rprimesoln}
		\curv'= -\frac{1}{a^3 \e H}\biggl[  \int \frac{d a}{a} a^3 \left(\frac{k}{aH}\right)^2 (\e H) {\curv} +{\rm const.}\biggr]. 
		\end{equation}
		In the SR phase, $\e$ is roughly constant and the growing integrand provides the leading contribution 
		\begin{equation}
		\label{eqn:RprimesolnSR}
		\curv'\simeq -\left(\frac{k}{aH}\right)^2  {\curv} ,  \qquad {\rm (SR)},
		\end{equation}
		and hence the curvature perturbation $\curv \propto e^{\frac{1}{2}\left(\frac{k}{aH}\right)^2}$ freezes out to a constant for $k/(aH)\ll 1$ as in Eq.~\eqref{eq:shsol-s}.
		On the other hand, in the USR phase, since $\zeta \propto a^3$ outside the horizon from \eqref{eq:shsol-s}, it immediately holds that
		\begin{equation} 
		\label{eqn:RprimesolnUSR} \curv' \propto a^3,  \qquad {\rm (USR)},
		\end{equation}
		outside the horizon. 
		One can also see that \eqref{eqn:RprimesolnUSR} is consistent with \eqref{eqn:Rprimesoln} as follows.  
		With $\zeta \propto a^3$, the integral in \eqref{eqn:Rprimesoln} acts as 
		$\simeq \int d \ln a \, a^{-2}$, which is dominated by early times and hence converges to a constant, whereas
		the prefactor grows as $\propto a^3$, resulting in \eqref{eqn:RprimesolnUSR}.
		Thus for a mode which spends $N_{\SR}$ $e$-folds outside the horizon in slow roll, it takes $N_{\USR} = 2/3 \times N_{\SR}$ $e$-folds of USR inflation to raise ${\curv}'$ back to order unity. Therefore at a fixed duration  $N_{\USR}$ of USR inflation, modes which exit the horizon more than $3/2 \times N_{\USR}$ $e$-folds before USR remain constant while modes which exit within $3/2 \times N_{\USR}$ grow outside the horizon. 
		After the USR phase, modes freeze in and the smooth change in the slope of the potential assures
		a slow increase in $\epsilon$ and a smooth transition of the power spectrum to the final SR phase.

		The power spectrum shown in the upper panel of Fig.~\ref{fig:GBPowerSpecAndfNL} exhibits a near-zero minimum $\Delta^2_\curv \sim 8 \times 10^{-16}$.
		Similar behavior occurs in other models in which the growing mode overtakes the constant mode (see, e.g., Refs.~\cite{Cicoli:2018asa,Ozsoy:2018flq,Byrnes:2018txb} and \S\ref{subsec:flat}). 
		This phenomenon can also be understood in detail from the formal solution Eq.~\eqref{eqn:Rprimesoln}, in which it can be seen that in slow roll the superhorizon mode approaches its slow-roll freezeout value with decreasing amplitude, i.e.\ with
		\begin{equation}
		\label{eq:FreezeOutArg}
		\Arg\left[\frac{\curv'}{\curv}\right] = \pi  + \O\left(\frac{k}{a H}\right) , \qquad {\rm (SR)},
		\end{equation}
		where the order of the correction follows from using the approximate SR form \eqref{eq:BunchDavies}
		in \eqref{eqn:Rprimesoln}.

		While $\zeta\propto a^3$ in the USR superhorizon limit, at the onset of USR, the curvature perturbation must reach this limit from the SR side.     Let $a=a_*$ at the onset of the USR phase, then the curvature
		evolves as
		\begin{equation}
		\curv' = \curv' \Big\vert_{a_*}  \times \left(\frac{a}{a_{*}}\right)^3 , \qquad {\rm (USR)},
		\end{equation}
		with the boundary condition $\curv' \vert_{a_*}$ given approximately by the SR solution for a smooth transition. Given the relative sign in the leading order SR expression \eqref{eqn:RprimesolnSR}, this represents an increase in the decay rate of $|\zeta|$ and thus before modes can grow as $\zeta \propto a^3$ in USR they must reverse sign. 

		There is a mode which experiences just enough evolution outside the horizon by the end of the USR phase to go from its freezeout value to near-zero.   The corresponding value of the power spectrum at the minimum is determined by the small out-of-phase component, i.e.\ how close Eq.~\eqref{eq:FreezeOutArg} is to $\pi$ and therefore how far outside the horizon this mode is when USR begins. Thus the longer the USR phase is, the deeper the minimum is.

		Modes which exit after this minimum are dominated by their superhorizon growth, and as modes exit the horizon closer to the USR phase they grow for a longer period and thus the power spectrum grows with increasing $k$. While a prolonged USR phase leads to a constant  $\Delta^2_\curv$ for modes which exit the horizon during USR [cf. Eq.~\eqref{eq:Delta2USR}], the inflection model touches the $\eta\simeq-6$ phase only briefly and the power spectrum therefore exhibits a peak  $\Delta^2_{\curv} (\kPeak) = 2.8 \times 10^{-3}$. This peak falls a factor of a few short of the value $\Delta^2_\curv\sim10^{-2}$ required for PBHs to form all the dark matter (see \S\ref{sec:nogo} and note that a model with the right tilt at CMB scales must fall much further short of this requirement \cite{Motohashi:2017kbs}). After the USR phase, the model returns to the slow-roll attractor and $\Delta^2_\curv$ is once more described by Eq.~\eqref{eq:Delta2SR}.

		It is now interesting to ask whether the power spectrum of the upper panel of Fig.~\ref{fig:GBPowerSpecAndfNL} can be locally enhanced by a factor of a few to exceed the threshold~\eqref{eq:PBHcriterion} for	PBH dark matter. According to Eqs.~\eqref{eq:modulation} and \eqref{eq:response}, for the power spectrum at a short-wavelength scale $\kShort$ to be significantly enhanced, we require a large long-wavelength perturbation $\curv_{\Long}$ and a large correlation $\fNL$. 

		In the lower panel of Fig.~\ref{fig:GBPowerSpecAndfNL}, we plot $\fNL(\Squeezedks)$ as a function of the long-wavelength mode $\kLong$ for two different values of the short-wavelength mode $\kShort$. The red upper curve shows $\fNL$ for a short-wavelength mode which exits the horizon after the end of USR, while the blue lower curve shows $\fNL$ for a short-wavelength mode which exits the horizon during USR. The upper and lower horizontal dashed lines show the consistency relation expectation $\fNL$ in the limit $\kLong/\kShort \rightarrow 0$.

		The numerically computed bispectrum for a short-wavelength mode which exits the horizon after USR, the red upper curve, agrees with the consistency relation. In other words, the short-wavelength perturbation $\curv_{\Short}$ retains no memory that, while it was inside the horizon, the long-wavelength perturbation $\curv_{\Long}$ outside the horizon grew in USR. This is because $\fNL(\Squeezedks)$ is set when $\kShort$ exits the horizon and the modes $\curv_{\Long}$ are already frozen at this time. Because the transformation of the bispectrum to \fncb{}   involves a subtraction of the consistency relation component, Eq.~\eqref{eq:Bzetatrans}, we conclude that short-wavelength modes which exit after the USR phase show no response to long-wavelength modes in local coordinates and therefore no enhancement of local PBH abundance.

		For a short-wavelength mode which exits the horizon during USR, the blue curve, the above logic does not hold. The numerically computed bispectrum $\fNL(\Squeezedks)$ does not agree with the consistency relation when $\curv_{\Long}$ grows outside the horizon. For such triangles, $12/5 \times \fNL \simeq 0.13$ while the consistency relation predicts $12/5 \times \fNL \simeq 0.065$.  
		
		Conversely, for the frozen $\curv_\Long$ modes that correspond to modes that exited the horizon well before USR, the consistency relation for $\fNL$ does hold. This is a successful test of our numerical computation, since in this limit the long-wavelength mode remains constant outside the horizon and just shifts the local coordinates for the small-wavelength mode along the background trajectory.

		Fig.~\ref{fig:GBPowerSpecAndfNL} also shows that when $k_\Short$ exits the horizon during USR, the near-zero of $\Delta^2_\curv$ induces a feature on $\fNL(\Squeezedks)$. This is due to the division by the power spectrum in the definition of $\fNL$, Eq.~\eqref{eq:fNLdefinition}. In particular, when $k_\Short$ exits the horizon $\curv_\Long$ has not yet reached its final (tiny) value set at the end of USR and thus a non-zero bispectrum $B_\curv(\Squeezedks)$ can be obtained. After the end of USR, $\curv_\Long$ is very small and thus $\fNL$ is amplified. The physical effect of this feature is negligible since, to obtain the power spectrum response, $\fNL$ should be multiplied by $\curv_\Long$, which has a minimum at this feature.

		More generally, the USR phase does enhance $\fNL$ relative to the consistency relation value. Hence the non-Gaussianity in \fncb{}  , $\bar{f}_{\rm NL}$, is non-zero. There is therefore an enhancement of the PBH abundances due to squeezed non-Gaussianity, which is not the case in single-field inflation on the attractor. However, for the $k_\Short$ shown in Fig.~\ref{fig:GBPowerSpecAndfNL} in blue, both $\fNL$  and $d \ln \Delta_\curv^2 / d \ln k$ are so small that, once multiplied by $\curv_\Long \ll 1$, the position-dependent effect on $\curv_\Short$ is insignificant.

		Quantitatively, we can summarize the PBH abundance enhancement in this model by choosing $\kShort$ and $\kLong$ in the USR phase, where $\fNL$ is nearly constant.  In particular, to eliminate the tilt-dependent coordinate effects on the abundance and to maximize the Gaussian part of the power spectrum, we can choose $\kShort=\kLong=\kPeak$. This triangle is not squeezed but since in USR $\fNL$ is the same for all triangle shapes, this triangle does serve as a summary statistic for local non-Gaussianity in inflection-point inflation.

		Doing so, we compute numerically that $12/5 \times \fNL(\kPeak,\kPeak,\kPeak) = 0.13$.  To obtain the response, we set $\curv_{\Long}$ to the peak value $\curv^{\rm RMS}(\kPeak) = \sqrt{\Delta^2_\curv (\kPeak)} \simeq 0.05$. Squeezed non-Gaussianity can therefore enhance the local power spectrum by a factor of at most $\sim 0.006$ and so in  inflection-point inflation, its ability 
		to enhance the local power spectrum  is negligible,
		\begin{equation}
		\frac{\Delta P_\curv}{P_\curv} \ll 1.
		\end{equation}

		In this model, we do not recover the USR squeezed limit result $12/5 \times \fNL \simeq 6$ and therefore do not enhance the small-scale power spectrum by an order-unity quantity. This is a reflection of the analytic result of Ref.~\cite{Cai:2017bxr} that transitions from a USR phase to a SR phase which are monotonically decreasing in the field velocity suppress the USR non-Gaussianity, and similar results were found numerically in Ref.~\cite{Atal:2018neu}.

		In the following sections, we will show that this suppression of non-Gaussianity is generic to transition models, except for the special case where the transition is both fast and large.

	\subsection{Infinitely Fast Transitions}
		\label{subsec:infinite}

		Infinitely fast transitions from USR to SR were considered in Refs.~\cite{Cai:2017bxr} and \cite{Pattison:2018bct},	and Ref.~\cite{Cai:2017bxr} established analytically that the final level of non-Gaussianity is sensitive to the way USR is exited.

		Specifically Ref.~\cite{Cai:2017bxr} considered the case where a pure ultra-slow-roll potential is joined to a slow-roll potential $V_{\rm SR}$ at a field position which we label $\phi_2$ for ease of generalization later:
		\begin{equation}
		\label{eq:inffast}
		V(\phi) = 
		\begin{cases}
		V_{\rm SR}(\phi_2) , &\phi > \phi_2 \quad {\rm (USR)} \\ 
		V_{\rm SR}(\phi) , &  \phi \le \phi_2 \quad {\rm (SR)}
		\end{cases}
		\end{equation}
		and hence the potential has an infinitely sharp discontinuity in slope at $\phi_2$. We call this an infinitely fast transition from USR to SR because the inflaton rolls over this discontinuity instantaneously.  $V_{\rm SR}(\phi)$ can be characterized in general by the potential slow-roll parameters for $\phi \le \phi_2$
		\begin{equation}
		\epsilon_V \equiv 
		\frac{1}{2} \left( \frac{1}{V}  \frac{d V}{d\phi}\right)^2, 
		\quad \eta_V\equiv \frac{1}{V} \frac{d^2V}{d\phi^2}.
		\end{equation}
		The transition can be characterized by the {strictly positive} amplitude parameter\footnote{The $h$ defined in \eqref{eq:h} is equivalent to Ref.~\cite{Cai:2017bxr}'s $-h/6$.}
		\begin{equation}
		\label{eq:h}
		h \equiv \sqrt{\frac{\epsilon_V (\phi_2^-)}{\epsilon(\phi_2^+)}},
		\end{equation}
		where
		\begin{equation}\label{eq:phidefNarrow}\phi_2^{\pm} \equiv \lim_{\omega\rightarrow0} \phi_2 \pm \omega,\end{equation}
		which is a ratio between the potential slow-roll parameter at the beginning of the SR phase and the Hubble slow-roll parameter $\epsilon$ at the end of the USR phase. 

		If $h=1$, the kinetic energy at the end of USR is just enough to keep the field on the attractor of the SR phase.
		The $h \ll 1$ limit therefore corresponds to the small transition, a monotonic transition from USR to the SR attractor 
		where the inflaton continues to slow down before hitting the attractor and hence the power spectrum
		continues to evolve.  
		Conversely for $h\gg 1$, Ref.~\cite{Cai:2017bxr} showed that the perturbations freeze out at $N(\phi_2)$.
		We call this a {\bf large} transition because the inflaton instantly goes from having too much kinetic energy for the potential it evolves on to suddenly having insufficient kinetic energy for a now much steeper potential
		\begin{equation}
		\label{eq:CaiCriterion} h\gg 1 \implies
		\epsilon_V
		\gg  \epsilon = \frac{1}{2} \phi'^2 \ , 
		\end{equation}		
		
		Since perturbations do not freezeout immediately for a finite value of $h$ the final level of non-Gaussianity is  not given by Eq.~\eqref{eq:USRfNL} but rather can be shown analytically to be \cite{Cai:2017bxr}
		\begin{equation} 
		\label{eq:CaifNL}
		\lim_{\kLong/\kShort\rightarrow0} \frac{12}{5} \fNL(\Squeezedks) = 2 \frac{h ( 3 h + \eta_V)}{(h+1)^2},
		\end{equation}
		for scales $\kLong$, $\kShort$ which cross the horizon during USR. 	Eq.~\eqref{eq:CaifNL} yields the USR result Eq.~\eqref{eq:USRfNL} only in the limit $h\gg1$, and thus for infinitely fast transitions the USR non-Gaussianity is conserved only when the transition is large.
		
		The enhancement of the local power spectrum is suppressed for small transitions, but we shall next see
		that it is also suppressed if the transition is not sufficiently fast.   
		Therefore, the transition needs to be large and fast to recover the USR non-Gaussianity.  
		In contrast, the inflection model of the previous
		section is an example where the transition is both small and slow.

	\subsection{Fast/Slow--Large/Small Transitions}
		\label{subsec:flat}

		In order to study in more detail the phenomenology of transient USR inflationary phases beyond the slow-small transition of \S\ref{subsec:inflection} and the infinitely fast limit of \S\ref{subsec:infinite}, we construct a toy inflationary model which begins in SR, enters a USR phase, and then transitions back to SR. We implement this with a potential 
		where the slope of an otherwise linear potential makes two transitions across adjustable
		widths in field space
		\begin{align*}
			\frac{d V}{d\phi}(\phi) &= \frac{\beta}{2} \left[1 
		 + \tanh \left(\frac{\phi 
		 - \phi_1}{\delta_1}\right) \right] + \frac{\gamma}{2} \left[ 1 
		 +  \tanh \left( \frac{\phi_2-\phi}{\delta_2}\right)\right]	,\numberthis
		\end{align*}
		and hence
		\begin{align*} \label{eq:tanhpotential}
			V(\phi) &= V_0 + \frac{\beta}{2} \left[\phi 
		 + \delta_1 \log \left\{ \cosh \left(\frac{\phi 
		 - \phi_1}{\delta_1}\right)\right\} \right]\\
		  &\quad+ \frac{\gamma}{2} \left[ \phi 
		 - \delta_2 \log \left\{ \cosh \left( \frac{\phi_2-\phi}{\delta_2}\right)\right\}\right]. \numberthis
		\end{align*}
		This potential describes three phases with finite transitions, which is a natural generalization of the model~\eqref{eq:inffast} with two phases with instant transition considered in \S\ref{subsec:infinite}.
		The model parameters $\left\{\delta_1,\ \delta_2\right\}$ and $\left\{\phi_1,\ \phi_2\right\}$ determine the widths and positions of two transitions, respectively.
		The limit $\delta_1,\delta_2\to 0$ amounts to instant transitions, where the potential is composed of
		a flat plateau of amplitude $V_0$ for $\phi_2<\phi<\phi_1$ in between two linear pieces of slope $\left\{\beta,\ \gamma\right\}$, which we set positive. 
		By modifying these parameters, we can set the duration of the USR phase as well as the circumstances of its beginning and end. 
		By constructing the transition in $dV/d\phi(\phi)$ rather than in $V(\phi)$ directly, 
		we trivially obtain	a monotonic potential where the field always rolls downhill.

		Since this is a toy model that we introduce to illustrate the fast/slow  and large/small distinction, we do
		not attempt to accurately fit measurements at CMB scales or to appropriately end inflation. Once all modes which we are interested in have frozen out in the latter slow-roll phase, we end inflation by hand. By adjusting the potential before and after the plateau, one could turn this toy model into a model which can fit CMB observations and produce PBHs while ending inflation gracefully without changing the conclusions we draw below.

		For $\phi>\phi_1$ the potential has a positive slope $\simeq \beta$ and the inflaton follows the slow-roll attractor.
		In order for the inflaton to leave the slow-roll attractor and enter a USR phase, the transition must be sufficiently sharp that the inflaton enters the flat region of the potential with excess kinetic energy. Thus we fix the entry parameters $\{\beta,\ \phi_1,\ \delta_1\} = \{10^{-14},\ 0,\ 10^{-2}\}$ to guarantee such a transition. By having inflation start on the slow-roll attractor, we are freed from having to specify initial conditions during USR.
		
		The region $\phi_2<\phi<\phi_1$ marks the USR phase where the potential is approximately flat.
		We fix the amplitude of the potential in the flat plateau $V_0= 2 \times 10^{-14}$, chosen to ensure that USR modes in our fiducial model are still perturbative, i.e.~$\Delta^2_\zeta \lesssim 1$ for the durations of USR we consider here.

		Finally, for $\phi<\phi_2$ the potential has a positive tilt $\simeq \gamma$ and the inflaton returns to the slow-roll attractor.
		Among the remaining parameters $\{\phi_2,\ \delta_2,\ \gamma\}$, $\phi_2$ determines the duration of the USR period, and $\{\delta_2,\gamma\}$ set the circumstances of the exit from USR. $\phi_2$ in particular must be very finely tuned to allow several $e$-folds of USR inflation while still reaching the transition point in a reasonable amount of time. The instant transition of \S\ref{subsec:infinite} corresponds to taking $\delta_2\to0$ and to focusing on the inflaton behavior around $\phi = \phi_2$.	

		We generalize Ref.~\cite{Cai:2017bxr}'s analysis to transitions of finite width between the flat and slow-roll potentials by allowing $\delta_2 \neq 0$. We start by generalizing  the definition for the start and end of the transition, Eq.~\eqref{eq:phidefNarrow}. We choose the end of the transition $\phi_2^-$ from the potential through
		\begin{equation}
		\phi_2^- \equiv \phi_2 - 2 \delta_2.
		\end{equation}
		The beginning of the transition, $\phi_2^+$, is not simply $\phi_2+2\delta_2$ since the USR phase persists while $\epsilon_V \ll \epsilon$. Instead, we choose to define the beginning of the transition through the deviation from the USR analytic solution, 
		\begin{equation}
		1-\frac{\phi'_{\USR}}{\phi'}\bigg\vert_{\phi_2^+}  = 0.05 \times \left(1-\frac{\phi'_{\USR}}{\phi'}\bigg\vert_{\phi_2^-}\right),
		\end{equation}
		where $\phi'_{\USR}$ is the analytic solution for the field velocity in USR, and $\phi'$ is the actual field velocity, which is evaluated numerically.
		By computing the field velocity deviation relative to the change at the end of the transition $\phi_2^-$ we guarantee that $\phi_2^+$ can be defined even for small and fast transitions.
				
		In other words, $\phi_2^-$ is roughly where the potential completes its transition, and $\phi_2^+$ is roughly where the field velocity begins to leave the USR solution. While the specific criteria chosen here are arbitrary, they are useful for classifying transition regimes and in the $\delta_2 \rightarrow 0$ limit the choices here return the limit Eq.~\eqref{eq:phidefNarrow} up to percent-level factors. 

		From these definitions for $\phi_2^+$ and $\phi_2^-$ we compute $h$ by evaluation of Eq.~\eqref{eq:h} and we quantify the duration of the transition from USR inflation to the beginning of the relaxation process
		\begin{equation}
		\label{eq:dN}
		d_N \equiv N(\phi_2^-) - N(\phi_2^+).
		\end{equation}

		\begin{figure}[t]
		\centering
		\includegraphics[width=.65\linewidth]{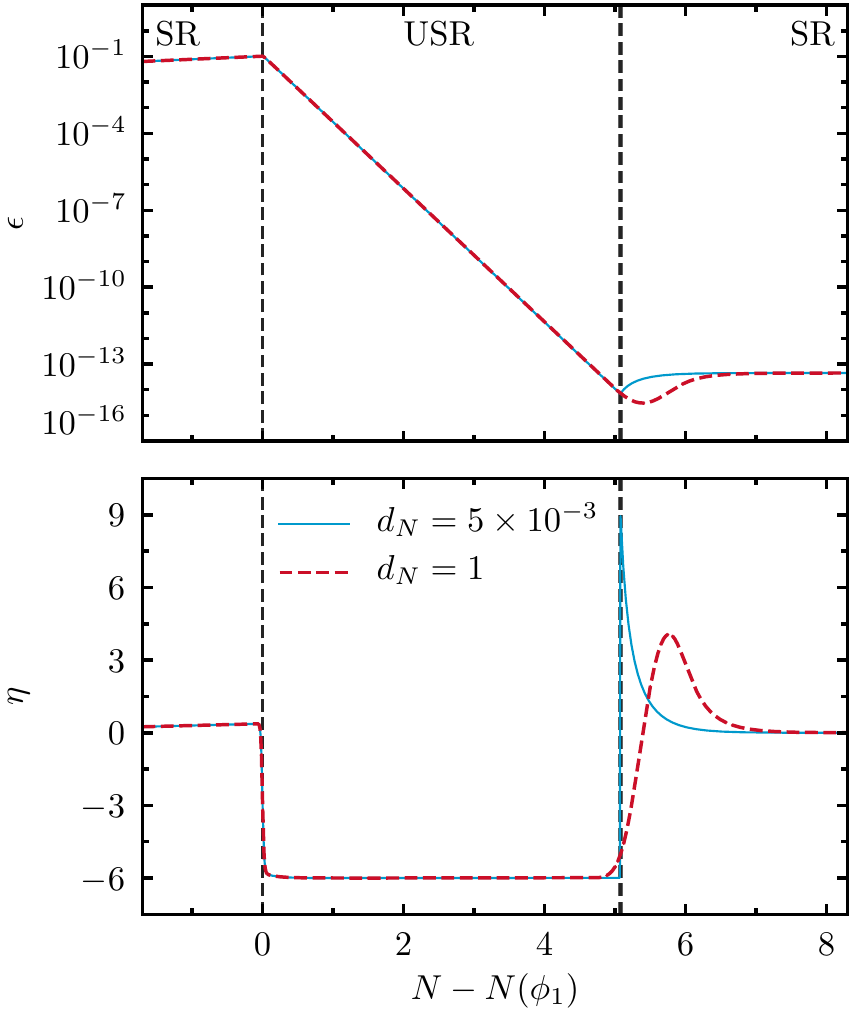}
		\caption[Ultra-slow-roll transition models]{Transition model background parameters $\e \equiv - d\ln H/d N$ and $\eta \equiv d\ln\e / d N$ for a fast vs.~slow transition, denoted by the $e$-fold width $d_N$, with a fixed large transition ($h=2.5$). The vertical lines mark $N(\phi_1)$ and $N(\phi_2^+)$.  The specific parameter choices used for these models are described in \S\ref{subsec:flat}. 
		}
		\label{fig:USRBackground}
		\end{figure}

		The situation of \S\ref{subsec:infinite} corresponds to the limit $d_N \rightarrow 0$ and we now generalize this result by exploring the impact of the duration $d_N$ on the resultant non-Gaussianity. 
		In Fig.~\ref{fig:USRBackground}, we show the background parameters $\e$ and $\eta$ for two models with a large transition $h$ = 2.5, one fast ($d_N = 5\times 10^{-3}$) and the other slow  ($d_N = 1$). For these models $\gamma$ is fixed at $6 \times 10^{-21}$, while the fast transition has $\left\{\phi_2 ,\ \delta_2\right\} = \left\{-0.1580281699,\ 2.12 \times 10^{-10} \right\}$ and the slow transition has $\left\{\phi_2 ,\ \delta_2\right\} = \left\{-0.1580282187,\ 3.6\times10^{-8} \right\}$.  Notice the amount of fine-tuning in $\phi_2$ required to achieve subpercent-level control of the transition amplitude and duration.

		Defining $n \equiv N - N(\phi_2^-)$ as the positive increasing number of $e$-folds elapsed since the potential transition, in the fast transition limit $d_N\rightarrow0$ the solution for $\eta$ after the transition point behaves according to the analytic result \cite{Cai:2017bxr} 
				\begin{equation}
		\label{eq:AnalyticEta}
		\eta(n > 0)=\frac{6 (h-1)}{1+h(e^{3 n}-1)}~.
		\end{equation}
		This is reflected in the behavior of the blue solid curve in Fig.~\ref{fig:USRBackground}, which behaves as Eq.~\eqref{eq:AnalyticEta} up to a ${\sim0.3\%}$ difference in the $h$ parameter as defined here compared to the $h$ parameter in the exact $\delta_2 \rightarrow 0$ limit. 
		
		On the other hand, in the case where $d_N$ is large, the red dashed curve of Fig.~\ref{fig:USRBackground}, the numerical solution for $\eta$ deviates significantly from this analytic form. This can be understood by Taylor expanding Eq.~\eqref{eq:AnalyticEta} around the transition point $n=0$,
		\begin{equation}
		\eta (n > 0) =  6(h-1) \left( 1 - 3 h n \right)   + \O\left(n^2\right),
		\end{equation}
		from which we can see that after $\phi_2^-$, $\eta$ evolves on a timescale $n \sim  h^{-1}$. Thus if the transition timescale $d_N$ is larger than this timescale, the evolution of $\eta$ will differ from the analytic solution \eqref{eq:AnalyticEta}.

		The behavior of $\eta$ is important because it controls the freezeout of perturbations through  the evolution of $\epsilon$ in Eq.~\eqref{eqn:Rprimesoln}, and it comes directly into the source of squeezed non-Gaussianity in the in-in formalism, where the cubic Lagrangian relevant in the squeezed limit is
		\begin{align*}
		 \label{eq:cubicLagrangianSqueezed}
		 \L_3 \equiv& \ a^3 \e 
		 \frac{\diff{}}{\diff{t}} \left(\e + \frac{\eta}{2} \right) \curv^2 \dot{\curv}
	     - \frac{\diff{}}{\diff{t}} \left[ a^3 \e \left(\e + \frac{\eta}{2} \right) \curv^2 \dot{\curv}\right] \\
	     &+ \e \curv(\Ha_2 + 2 \L_2) \\
	     &- \frac{\diff{}}{\diff{t}} \left[ \frac{a^3 \e}{H} \curv \dot{\curv}^2 + a^3 \frac{\e^2}{2 H} \dot\curv  \pa_a\curv  \pa_a(\partial^{-2} \dot\curv) \right].
		 \numberthis
		\end{align*}
		$\Ha_2$ here is the quadratic Hamiltonian density
		\begin{equation}\Ha_2 = a^3 \e \left[ \dot{\curv}^2 + \frac{1}{a^2} \left(\pa \curv\right)^2 \right].
		\end{equation}

		Therefore, the timescale $d_N$ plays an important role in 
		changing the non-Gaussianity produced in USR.

		\begin{figure}[t]
		\centering
		\includegraphics[width=.65\linewidth]{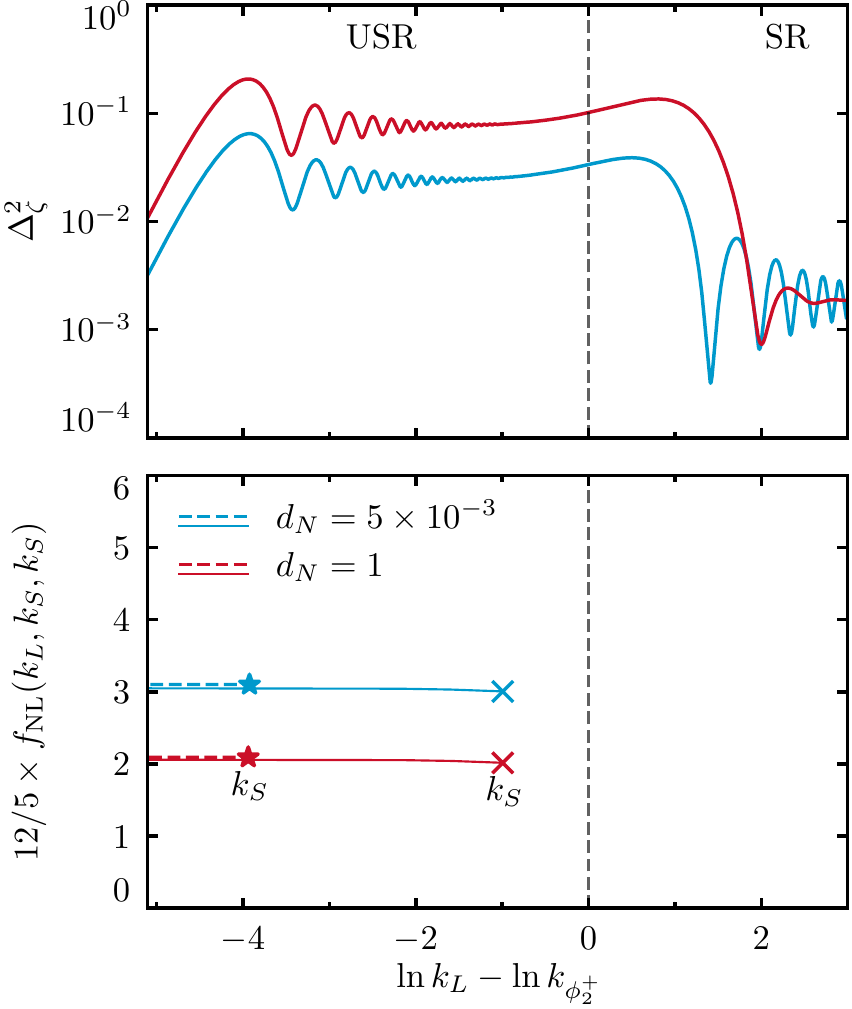}
		\caption[Power spectrum and non-Gaussianity of {\it Large} transitions]{Transition model power spectrum $\Delta^2_\curv$ and non-Gaussianity parameter $\fNL(\Squeezedks)$ for the large-fast (blue) and large-slow (red) transition models of Fig.~\ref{fig:USRBackground}.   The left edge 
		corresponds roughly to a mode which exits the horizon at the beginning of USR. Conventions for displaying $\fNL$ are the same as in Fig.~\ref{fig:GBPowerSpecAndfNL} except both  $\kShort$ values for each model cross the horizon before the end of USR (vertical line, $k_{\phi_2^+}$) (see \S\ref{subsec:flat} for further discussion).}
		\label{fig:USRPowerSpecAndfNL}
		\end{figure}\clearpage

		The upper panel of Fig.~\ref{fig:USRPowerSpecAndfNL} shows the power spectra for the same large-fast and large-slow models as Fig.~\ref{fig:USRBackground}. Once more, the red upper curve shows the slow transition, while the blue lower curve shows the fast transition. The vertical dashed line shows a mode which exits the horizon at the onset of the transition $N(\phi_2^+)$.

		The power spectra in these models show many of the same features as the inflection-point power spectrum, Fig.~\ref{fig:GBPowerSpecAndfNL}, and therefore we focus on the USR and transition regions of the plot, where unlike the inflection model these models have power spectrum plateaus for modes which exit during USR. This plateau is modulated by small oscillations sourced by the first feature in the potential. The USR to SR transition feature in the potential also induces power spectrum oscillations, with the fast transition model having more oscillations due to the sharper source. The slow transition model has a larger power spectrum than the fast transition as $\e$ reaches a lower level in this model (see Fig.~\ref{fig:USRBackground}) and thus the modes grow more.

		The lower panel of Fig.~\ref{fig:USRPowerSpecAndfNL} shows the non-Gaussianity for these models for two different values of the short-wavelength mode, one marked with a cross and the other a star and both exiting during the USR phase, as a function of the long-wavelength mode, curves with correspondingly solid and dashed lines. All triangles yield the same value of $\fNL$ when the legs exit during the USR phase up to corrections of order $\kShort/k_{\phi_2^+}$, consistent with the result in the inflection model \S\ref{subsec:inflection} and the exact USR result of \S\ref{sec:usr}. However, for neither model does the level of the non-Gaussianity agree with the analytic result for USR Eq.~\eqref{eq:USRfNL}, $5 \fNL /12= 6$. This is the result of Ref.~\cite{Cai:2017bxr}, that the residual level of $\fNL$ depends on the value of the transition parameter $h$, and in particular the fast transition model yields the result expected from Eq.~\eqref{eq:CaifNL} for a transition with $h=2.5$, $\fNL \simeq 3.1$.
		However, the slow model has the same $h=2.5$ as the fast model, yet a smaller value of $\fNL$. This is due to the slow nature of the transition in the $d_N=1$ model. Fast transitions yield Eq.~\eqref{eq:CaifNL}, while slow transitions suppress the non-Gaussianity.

		We model this effect with the ansatz that a given transition length $d_N$ sets an upper bound to the transition amplitude, independent of $h$. We can define an effective transition amplitude
		\begin{equation}
		\label{eq:ansatz}
		h_{\rm eff} \equiv \left[ \left(\frac{1.5}{d_N}\right)^{-3}+ h^{-3} \right]^{-1/3},
		\end{equation}
		where the exponent serves merely to interpolate between the two limits and the factor of $1.5$ comes from calibrating the results to the form of \eqref{eq:CaifNL}
		\begin{equation}
		 \frac{12}{5} \fNL(h_{\rm eff}) =  2 \frac{h_{\rm eff} ( 3 h_{\rm eff} + \eta_V)}{(h_{\rm eff}+1)^2},
		 \label{eq:fNLhEff}
		\end{equation}
		where in this toy model with a linear SR potential we set $\eta_V=0$.			

		\begin{figure}[t]
		\centering
		\includegraphics[width=.65\linewidth]{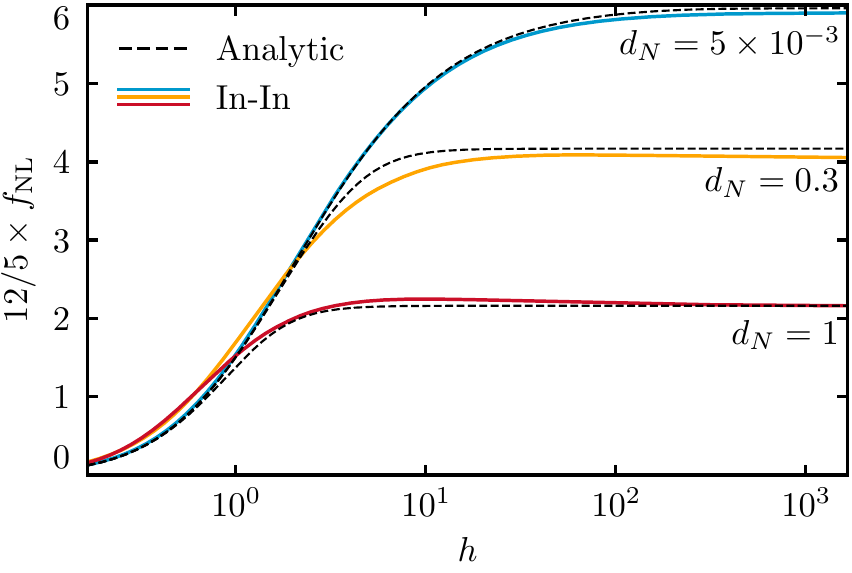}
		\caption[Non-Gaussianity for {\it Large} and {\it Small} transitions]{Transition model non-Gaussianity parameter $\fNL$ as a function of amplitude $h$ for various $e$-fold widths $d_N$ with both $k_\Long$ and $k_\Short$ exiting the horizon during USR.  Colored lines show the numerical in-in computation and dashed black lines show the calibrated analytic prediction from Eqs.~\eqref{eq:ansatz}~and~\eqref{eq:fNLhEff}.   
		To reach the USR result $12\fNL /5 = 6$, the transition must be large $h\gg 1$ and
		fast $d_N \ll 1$ (see \S\ref{subsec:flat} for further discussion). 
		}
		\label{fig:fNLvsh}
		\end{figure}

		\begin{figure*}[ht]
		\includegraphics[width=\linewidth]{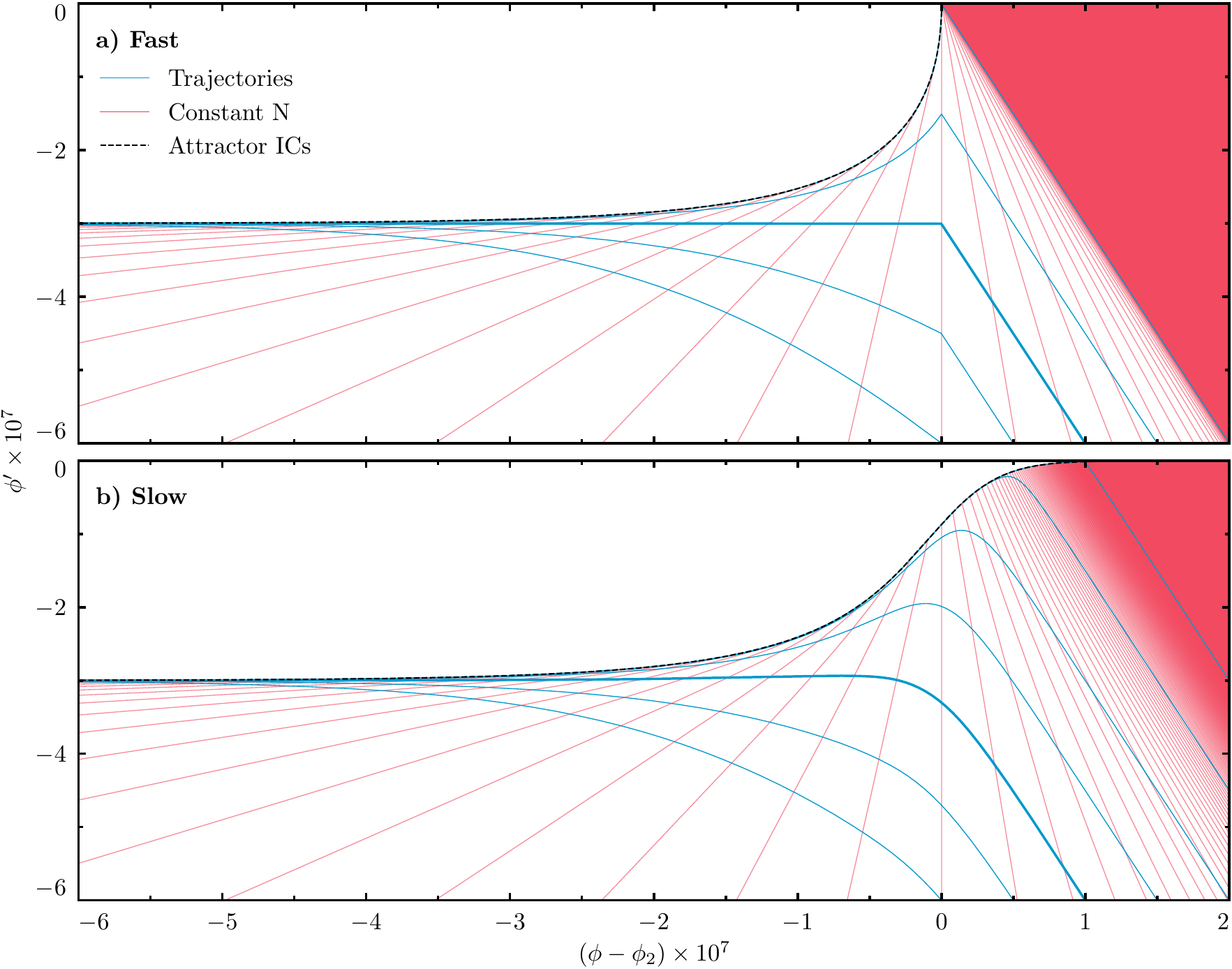}
		\caption[Phase space of ultra-slow-roll transitions]{ Phase-space diagram for the USR to SR transition models with model trajectories (blue lines) and
		constant $e$-fold surfaces (red lines) relative to the transition at $\phi_2$. The top panel a) has $\delta_2 = 2.12\times10^{-10}$, such that most trajectories correspond to  fast transitions, while the bottom panel b) has $\delta_2 = 3.6\times10^{-8}$,
		such that most trajectories correspond to  slow transitions.   In each panel, higher trajectories represent larger transitions, with  $h=1$ as the large-small dividing line (thick blue).  
			 Constant $N$ surfaces become space-filling in the top right corners. 
			 A large $\fNL$ requires a fast-large transition as can be visualized by $\delta N$, the change
			 in $e$-folds given a shift in the initial field $\delta \phi_i$ that takes the local background to
			 a new trajectory (see \S\ref{subsec:flat} for further discussion).}\label{fig:phasespaceTransition}
		\end{figure*}\clearpage

		In Fig.~\ref{fig:fNLvsh}, we compare the ansatz \eqref{eq:ansatz} coupled with the analytic formula  \eqref{eq:fNLhEff} (black dashed curves) to the full numerical in-in computation of the bispectrum (colored solid curves) for modes that exit the horizon during USR
		as a function of the transition amplitude $h$ for three different transition speeds $d_N$.  Every point along these lines corresponds to a different set of parameters for our toy model. 

		The analytic formula \eqref{eq:ansatz} agrees well with the in-in computation for all values of $h$ and $d_N$. For small-slow transitions (small $h$ and large $d_N$), where $\fNL$ is small and becomes proportional to the potential slow-roll parameters on the attractor, there is a slight difference between the numerical and analytic $\fNL$. This we attribute to small differences in the value of $h$ as defined for the infinitely fast transitions and as defined for slow transitions since the errors  decrease for smaller values of $d_N$.  For large transitions $h\gg1$, where $\fNL$ is largest, the analytic ansatz produces a slight overestimate of $\fNL$ as the transitions become faster.  This is due to nonlinearities between the true $h_{\rm eff}$ and $1/d_N$ in the large $h$ limit which our ansatz does not model.
		Since these differences are minor,
		we conclude from the analytic formula that to produce a large level of non-Gaussianity after the transition from USR to SR requires $h_{\rm eff} \gg 1$ and thus the transition must be large, ${h \gg 1 }$, and fast, ${d_N \ll 1}$.

		Just as in the SR (\S\ref{sec:nogo}) and exact USR case (\S\ref{sec:usr}), the effect on $\fNL$ of the transition from USR to SR can be understood visually from the way phase-space trajectories intersect constant $N$
		surfaces. Fig.~\ref{fig:phasespaceTransition}\hyperref[fig:phasespaceTransition]{a} shows the phase-space 
		trajectories (blue lines) and constant $N$ surfaces (red lines) for a narrow $\delta_2 = 2.12 \times 10^{-10}$ and therefore faster transitions, such that the fast model of Fig.~\ref{fig:USRBackground} and Fig.~\ref{fig:USRPowerSpecAndfNL} corresponds to a trajectory in this space. Fig.~\ref{fig:phasespaceTransition}\hyperref[fig:phasespaceTransition]{b} shows the phase space for a wider $\delta_2 = 3.6 \times 10^{-8}$ and therefore slower transitions, and the slow model of Fig.~\ref{fig:USRBackground} and Fig.~\ref{fig:USRPowerSpecAndfNL} evolves through this space.

		Trajectories near the top of each panel have the inflaton speed up after the transition and hence
		have a large $h$, with the large-small dividing line of $h=1$ denoted by thick blue lines.
		Notice also that the
		union of the panels of Fig.~\ref{fig:phasespaceSR+USR} gives the limit of infinitely fast transitions,
		with the exception that here constant $N$ surfaces are plotted relative to the transition feature
		$N(\phi_2)$ rather than the end of inflation $N=0$.  Trajectories are evenly spaced in $\phi$ at the point where they cross the bottom edge at $\phi' =-6$ in a range that reflects a reasonable amount of USR $e$-folds as we describe next.

		Due to the smooth nature of the potential \eqref{eq:tanhpotential}, for any finite $\phi>\phi_2$ the potential slope $dV/d\phi$ has a finite positive value. Thus, unlike in the exact USR case (Fig.~\ref{fig:phasespaceSR+USR}\hyperref[fig:phasespaceSR+USR]{b}) or the infinitely fast case, all trajectories 
		with any finite $\phi'$ for $\phi>\phi_2$ will eventually cross $\phi_2$.   

		Black dashed curves in Fig.~\ref{fig:phasespaceTransition} depict the envelope of such trajectories, neglecting stochastic effects, and correspond to initial conditions on the attractor on the very nearly flat potential.  Consequently, the constant $e$-fold surfaces become increasingly tightly packed and eventually space-filling, in contrast to the empty upper right triangle in Fig.~\ref{fig:phasespaceSR+USR}\hyperref[fig:phasespaceSR+USR]{b}.  We choose not to continue showing trajectories which take such large numbers of $e$-folds to traverse the nearly flat plateau of the potential.

		By the same $\delta N$ arguments of \S\ref{sec:usr} we can immediately see from these phase spaces why a large ($h \gg 1$) transition is necessary to conserve the USR non-Gaussianity. Here $\delta N$ refers to the change in the total number
		of $e$-folds elapsed to a fixed field position on the SR side for a shift in the initial field
		 position $\delta \phi_i$ on the USR side which then shifts the whole trajectory. Note that
		 $\delta N$ combines the change from the USR and SR sides.
		 
		Let us first consider the fast case in the top panel. Around a central trajectory with large $h$ (upper trajectories), 
		the crossing rate $\partial N/\partial\phi_i$  is strongly asymmetric to the sign of $\delta\phi$,
		i.e.~there is a large second derivative $\partial^2 N/\partial \phi_i^2$
		and hence a large $f_{\rm NL}$ according to
		 Eq.~\eqref{eq:deltaN}. This is due to the much larger contribution to the rate of surfaces crossed in
		 the USR side where the asymmetry is larger than the SR side where the asymmetry is small. 

		 On the other hand, the asymmetry around a small $h$ trajectory (lower trajectories) is small and therefore the non-Gaussianity is small. This is due to the smaller  contribution to $\partial N/\partial \phi_i$ on the USR side relative to the SR side.  In other words the power spectrum continues to grow on the
		SR side at small $h$, which suppresses the non-Gaussianity from the USR side.
		
		By comparing the fast and slow cases, we can visually see that the transition duration sets an effective maximum transition amplitude $h_{\rm eff}$. Above a certain value of $h$, the trajectory joins the slow-roll attractor before the transition and therefore will have a highly suppressed non-Gaussianity comparable to the small-slow transition of \S\ref{subsec:inflection}.

		\begin{figure}[t]
		\centering
		\includegraphics[width=.65\linewidth]{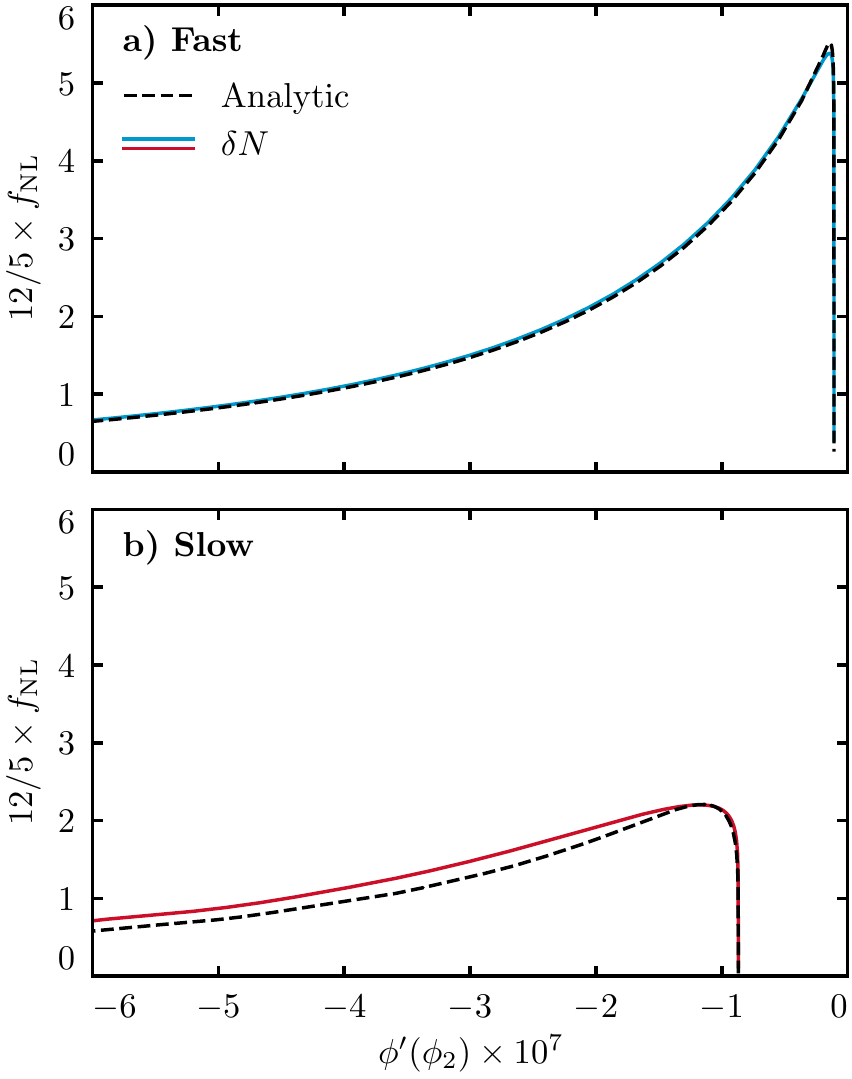}
		\caption[Non-Gaussianity from $\delta N$]{Non-Gaussianity for trajectories in the corresponding phase spaces of Fig.~\ref{fig:phasespaceTransition} computed via the $\delta N$ formula~\eqref{eq:deltaN} compared to our calibrated analytic result Eqs.~\eqref{eq:ansatz} and~\eqref{eq:fNLhEff}. The horizontal axis here corresponds to the vertical axis of Fig.~\ref{fig:phasespaceTransition}. Only for fast and large transitions does the transition model reproduce the USR result, and for fixed $\delta_2$ there is a maximum value of $\fNL$ attainable. 	
		}
		\label{fig:deltaN}
		\end{figure}\clearpage

		In Fig.~\ref{fig:deltaN} we formalize these heuristic  arguments by computing the $\delta N$ formula~\eqref{eq:deltaN} for different trajectories in these phase spaces. We organize the trajectories by their velocity at $\phi_2$, and thus the horizontal axis of Fig.~\ref{fig:deltaN} corresponds to the vertical axis of the Fig.~\ref{fig:phasespaceTransition} at the transition point. We then also compare the $\delta N$ result to our analytic expectation for $\fNL$, Eqs.~\eqref{eq:ansatz} and \eqref{eq:fNLhEff}.

		The $\delta N$ computation cross-validates our analytic formula which was calibrated to the in-in
		calculations, showing excellent agreement for all methods of computation across the fast-slow and 
		large-small transition space.  The $\delta N$ computation thus also confirms that a large $\fNL$ requires a large-fast transition. 
		
		For large $\phi'(\phi_2)$, after the peak non-Gaussianity, there is a sharp cliff in Fig.~\ref{fig:deltaN} beyond which $\fNL$ becomes suppressed. This cliff corresponds to trajectories which reach the attractor before the transition, thus inhabiting the space-filling regions of Fig.~\ref{fig:phasespaceTransition}, and the edge of the
		cliff is given by the  black dashed envelope of Fig.~\ref{fig:phasespaceTransition}. These trajectories are slow transitions even though they have the same narrow field space width $\delta_2$ and appear in the `fast' panel. These
		cases behave in the same way as those in the `slow' panel once the duration of the transition in $e$-folds $d_N$ is accounted for. 

		Of course even for a large and fast transition, for which $5/12 \ \fNL \rightarrow 6$, the response of the small scale power spectrum to the long-wavelength mode is still dependent on the value of the long-wavelength mode, and in particular the argument of \S\ref{sec:usr} still holds, that 
		\begin{equation}
		\frac{\Delta P_\curv}{P_\curv} \lesssim 1,
		\end{equation}
		unless the model already produces PBHs with a Gaussian distribution.

\section{Discussion}
	\label{sec:conclusion}
	Canonical slow-roll inflation cannot produce primordial black holes in a large enough quantity to be the dark matter. While perturbations do exhibit a small level of local non-Gaussianity which couples short-wavelength PBH fluctuations to the long-wavelength modes they live in and can in principle enhance local abundances at peaks of long-wavelength modes, transforming to a freely-falling coordinate system shows that locally measured PBH abundances are completely insensitive to this non-Gaussianity because it is generated by the reverse coordinate transformation to begin with. 

	Any confirmation that the dark matter is in the form of PBHs would rule out canonical slow-roll inflation. The only way to rescue canonical inflation would be to violate slow roll, and a phase of ultra-slow-roll inflation after CMB scales exit the horizon is the natural way to do this. We showed by gauge transformation and by the $\delta N$ formalism, which can be illustrated graphically and contrasted to the SR case, that local non-Gaussianities are large in the USR phase when perturbations freeze out instantly at some fixed field position. 

	The same coordinate transformation machinery as in slow roll confirms that squeezed USR non-Gaussianities can locally enhance PBH abundances. However the effect is very mild, giving at most an order unity enhancement of the local power spectrum. Such enhancements can only make models that are already on the borderline of succeeding to produce PBHs as the dark matter under Gaussian assumptions  actually succeed.
	For such cases, generally a small change in parameters that prolong the USR phase would equally well produce  PBHs under Gaussian assumptions. 

	Even more importantly, the USR phase has to end in some way. Ref.~\cite{Cai:2017bxr} established that the non-Gaussianity is very sensitive to how this period ends using cases where the transition is
	infinitely fast. By exact computation in the in-in formalism and validation with the $\delta N$ formalism, we mapped the entire range of possible endings to USR to show that only a small class of transitions conserves the large USR non-Gaussianity through the transition to slow roll. These are the transitions which are fast, in that the potential exhibits a sharp feature that is traversed by the inflaton in much less than an $e$-fold, and large, in that the inflaton needs to gain significant velocity after transiting the feature. All other types of transitions suppress the non-Gaussianity significantly.

	Note that while we have computed the full local non-Gaussianity, it is only the squeezed non-Gaussianity which we have shown has a negligible effect on whether a model produces PBH dark matter. Understanding the effect of other mode couplings requires a full knowledge of the probability distribution function of the density perturbation taking into account the contributions from all orders of non-Gaussian correlators of every shape. Work in this direction has been pursued in a variety of contexts (see, e.g., \cite{Saito:2008em,Smith:2011,Byrnes:2012yx,Ferraro:2014jba,Young:2015kda,Tada:2015noa,Franciolini:2018vbk}), but it remains a challenging problem and direction for future work.

	Nonetheless the conclusion that producing primordial black holes as dark matter in canonical single-field inflation requires a complicated and fine-tuned potential shape with a transient violation of slow roll is robust to the inclusion of local squeezed non-Gaussian effects.

\chapter{The Higgs Field During Inflation}
\label{chap:higgs}
We have dealt in detail with single-field inflation and its generalizations, and we can now complete this thesis by studying the behavior of extra spectator fields during inflation and the conditions that must be satisfied for them to contribute to primordial black hole dark matter production, based on Ref.~\cite{Passaglia:2019ueo}.

We will focus our study on the Standard Model Higgs field, which exhibits the unusual property that its effective potential turns over at large field values. This instability can lead to complicated behavior during and after inflation which has been claimed to lead to PBH DM production.

We will first provide a general overview how the Higgs instability could possibly lead to PBH production. We will then focus first on the Higgs' evolution during inflation and compute its spectrum of perturbations on CMB scales, which should be small relative to its value on PBH scales. We then track the Higgs and its fluctuations nonlinearly through reheating to compute whether Higgs fluctuations are converted to large curvature fluctuations. We will find that the Higgs fails on both counts and therefore does not produce PBH DM.

\section{The Higgs Instability Mechanism for Primordial Black Holes}
\label{sec:mechanism}

In this section, we review the general features of the Higgs instability mechanism for producing PBH dark matter and the principles governing the spectrum of perturbations it generates.

The Higgs field acts as a spectator during inflation, which is driven by an inflaton field, but converts its quantum fluctuations into curvature fluctuations once it decays after inflation. If these quantum fluctuations are amplified into large enough curvature fluctuations  by the Higgs instability, they will form PBHs when they re-enter the horizon in the radiation dominated epoch.

In particular, after inflation the spatial metric on a uniform total energy density hypersurface,
\begin{equation}
\label{eq:sign_convention}
g_{ij} = a^2 e^{2 \zeta} \delta_{i j},
\end{equation}
 possesses a  curvature perturbation $\zeta$ that can be decomposed as
\begin{equation} 
\label{eq:zeta_tot_intro}
\zeta =\left(1-   \frac{\rho_h'}{\rho_{\rm tot}'} \right)  \zeta_{\rm r} + \frac{\rho_h'}{\rho_{\rm tot}'} \zeta_h,
\end{equation}
in linear theory.  Here
$\zeta_{\rm r}$ is the curvature perturbation on hypersurfaces of uniform energy density $\rho_{\rm r}$ of reheat products from the inflaton, $\zeta_h$ is the curvature perturbation on hypersurfaces of uniform Higgs energy density $\rho_h$, and $\rho_{\rm tot} \equiv \rho_{\rm r} + \rho_h$ is the total energy density. Throughout primes $'$\, denote a derivatives with respect to the $e$-folds $N$, where $N=0$ marks the end of inflation.

For this mechanism to succeed observationally, the variance per logarithmic interval in $k$ of the Fourier modefunction $\zeta^k$, 
\begin{equation}
\Delta^2_{\zeta}(k)\equiv \frac{k^3}{2\pi^2}|\zeta^k|^2,
\end{equation}
must be small on the large scales probed by anisotropies in the CMB, 
\begin{equation}
\label{eq:delta_2_cmb}
\Delta^2_{\zeta}(k_{\rm CMB}) \sim 10^{-9},
\end{equation}
while on small scales associated with primordial black holes the power spectrum must reach
\begin{equation}
\label{eq:delta_2_pbh}
\Delta^2_{\zeta}(k_{\rm PBH})  \sim 10^{-2},
\end{equation}
so that enough regions are over the collapse threshold $\zeta_c \sim 1$ to form PBHs in sufficient abundance to be the dark matter \cite{Motohashi:2017kbs,Passaglia:2018ixg}. 
Moreover, the large-scale modes should be sourced predominantly by a single degree of freedom in order to comply with isocurvature constraints from the CMB. 
The working assumption for a successful model is that inflaton perturbations lead to a
$\zeta$ which dominates on CMB scales while Higgs fluctuations produce one which
dominates on PBH scales. We use a superscript $k$ to denote relations that are exclusively in
Fourier space as opposed to real space quantities or linear relations that apply to both.

Although the Higgs field $h$ is a spectator during inflation, the mechanism works by enhancing its impact on the total $\zeta$ by exploiting the unstable, unbounded nature of the Higgs potential $V(h)$ at large field values $h>\hcr$ in the Standard Model. 
In particular, the effective Higgs potential at field values far larger than its electroweak vacuum expectation value can be approximated as \cite{Gross:2018ivp,Espinosa:2015qea}
\begin{equation}
V(h) = \frac{1}{4} \lambda h^4,
\end{equation}
with $\lambda=\lambda^{\SM}$ and
\begin{equation} \label{eq:lamsm}
\lambda^{\SM}  \simeq -b \ln \left(\frac{h^2}{\hcr^2 \sqrt{e}}\right),
\end{equation}
where $\hcr$ is the location of the maximum of the Higgs potential which separates the familiar metastable electroweak vacuum from the unstable region, and $b$ controls the flatness of the potential around the maximum. $\hcr$ and $b$ are computable given the parameters of  the SM, and in this work, we choose to fix them at representative values $\hcr = 4 \times 10^{12} \GeV$ and $b = 0.09 / (4 \pi)^2$, corresponding to a top quark mass $M_t \simeq 172 \GeV$, following Refs.~\cite{Espinosa:2017sgp,Espinosa:2018euj,Gross:2018ivp} to facilitate comparisons. 
The Higgs instability exists for $M_t \gtrsim 171 \GeV$ \cite{Buttazzo:2013uya}, which includes
the range from the most recent constraints by the Tevatron and the LHC \cite{ATLAS:2014wva,Khachatryan:2015hba,Aaboud:2016igd,TevatronElectroweakWorkingGroup:2016lid}. Here we have neglected an effective mass term for the Higgs generically generated by a nonminimal Higgs coupling to the Ricci scalar, since at the level expected from quantum corrections it does not change the qualitative features of the mechanism \cite{Gross:2018ivp}. 

\begin{figure}[t]
\centering
\includegraphics[width=.65\linewidth]{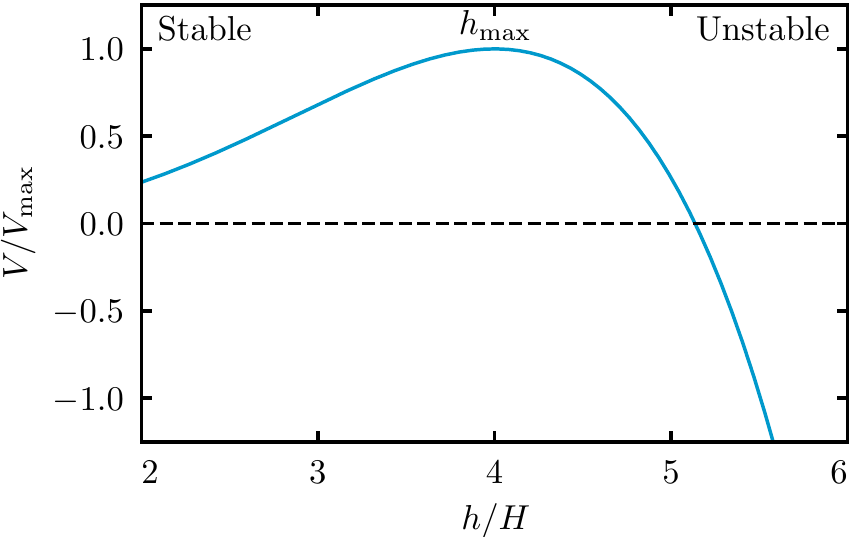}
\caption[Higgs instability]{Beyond the field value $\hcr$, the Standard Model Higgs effective potential turns over and decreases rapidly, with an unbounded true vacuum of negative energy density. In the PBH scenario, the spectator Higgs rolls down the unstable potential, amplifying its stochastic field fluctuations.
}
\label{fig:potential_turn_over}
\end{figure}

\begin{figure}[t]
\centering
\includegraphics[width=.65\linewidth]{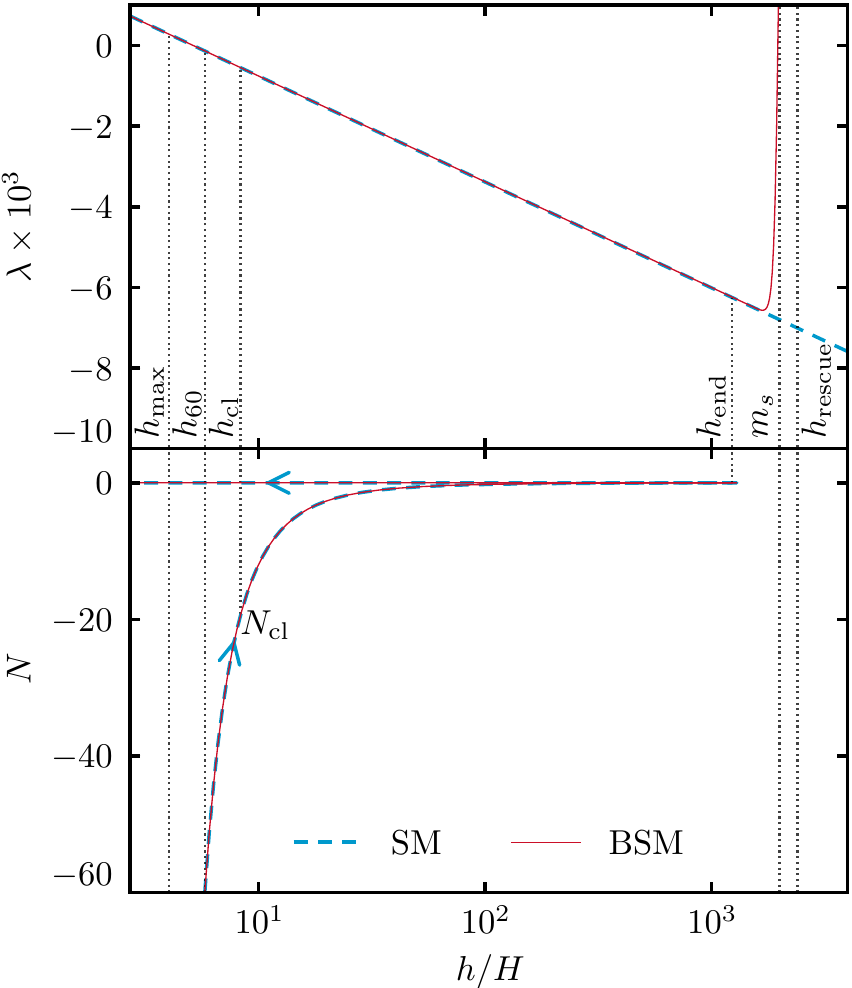}
\caption[Position of the Higgs during inflation]{Quartic coupling $\lambda$ and $e$-folds $N$ corresponding
to the Higgs field position $h$, with marked special values as computed in \S\ref{sec:inflation} and \S\ref{sec:reheating}: $\hcr$, the maximum of the Higgs potential; $h_{60}$, its position 60 $e$-folds before the end of inflation;  $\hcl$, its classical roll position; $\hend$, its position at the end of inflation; $\hrescue$, beyond which the SM Higgs cannot be ``rescued" by reheating so that it rolls back and oscillates around the origin after inflation. The BSM Higgs adds a coupled scalar of suitable mass $m_s$ to eliminate the runaway instability. 
}
\label{fig:lambdaN}
\end{figure}

We show the potential just around its maximum in Fig.~\ref{fig:potential_turn_over}, and across a wider range of scales in Fig.~\ref{fig:lambdaN} on the unstable side in order to illustrate the
field values that will be crucial to the Higgs instability phenomenology detailed in the following sections. 
Specifically, for  a representative choice for the Hubble scale during inflation $H=10^{12} \GeV$ which we employ throughout for illustration,
the potential maximum is at
\begin{equation}
\label{eq:hmax}
\hcr = 4H.
\end{equation}
When $h<\hcr$, the minimum corresponds to our familiar electroweak vacuum, while for $h>\hcr$, the potential decreases and is unbounded from below. 

On the unstable side of the potential, the Higgs at first rolls slowly relative to the Hubble rate before accelerating as it rolls down the instability.  The location where the Higgs' roll in one $e$-fold becomes comparable to $H/2\pi$ defines the classical roll scale $\hcl$, which we shall define precisely in \S\ref{sec:inflation}. For our parameter choices, it lies at
\begin{equation}
\label{eq:hcl}
\hcl \simeq 8.3 H.
\end{equation}
At the scale $k_{\rm cl}$ which crosses the horizon at $N_{\rm cl} \equiv N(\hcl)$, the power spectrum of $\zeta_h$ at horizon crossing becomes order unity,
\begin{equation}
\label{eq:delta_2_h_rough}
\Delta^2_{\zeta_h} (k_{\rm cl}) \sim 1.
\end{equation}
Our working assumption is that the classical-roll scale $k_{\rm cl}$ will be the one to produce primordial black holes, $k_{\rm cl} = k_{\rm PBH}$. We will show in \S\ref{sec:inflation} that this implies that the Higgs is in fact on the unstable side of the potential during all phases of inflation relevant for observation. In particular, the CMB scales left the horizon during inflation a few $e$-folds after
our Hubble patch crossed the horizon, which we will assume is 60 $e$-folds before the end of inflation. 
At this time, the Higgs is on the unstable side of the potential at
\begin{equation}
\label{eq:h60}
h_{60} \simeq 5.8 H,
\end{equation}
if we take the field value at the end of inflation to be
\begin{equation}
\label{eq:hend}
\hend \simeq 1200 H,
\end{equation}
which we will see below is approximately the largest value possible.  This also leads to $N_{\rm cl} \sim -20$.
If PBHs are produced on that scale then they  have a small mass $M_{\rm PBH} \simeq 10^{-15} M_{\odot}$ at formation, and after mergers and accretion could today lie in the region $M_{\rm PBH} \simeq 10^{-12} M_{\odot}$ where 
all the dark matter could be in the form of PBHs \cite{Montero-Camacho:2019jte,Niikura:2017zjd}.

\clearpage

The Higgs continues to roll to larger field values until the end of reheating when interactions with the thermal bath lift the effective Higgs potential. If the Higgs lies within a maximum rescuable distance $\hrescue$,
\begin{equation}
\hrescue \simeq 2400 H,
\end{equation}
which we compute in \S\ref{sec:reheating}, then after inflation it rolls safely back to the metastable electroweak vacuum, oscillating in a roughly quadratic potential with a temperature-dependent mass $\propto T \propto 1/a$ until it decays to radiation on a uniform Hubble surface. 

The behavior of Higgs perturbations through reheating is complicated. However, all relevant physical scales are at this stage far outside the horizon. So long as a gradient expansion holds,
under which such perturbations can be absorbed into an approximate FLRW background for a local observer, then the curvature field after horizon crossing evolves locally, with no explicit scale-dependence \cite{Wands:2000dp}. Therefore the Higgs contribution to the fully nonlinear curvature field $\zeta$ on a uniform total density slice after the Higgs decays is related to the curvature $\zeta_h$ on constant Higgs slices on superhorizon scales during inflation by 
\begin{equation}
\label{eq:zetaconv}
\left. \zeta \right\vert_{{\rm decay}} = R(\zeta_h) \times \left. \zeta_h \right\vert_{\rm inflation},
\end{equation}
with all of the complicated physics of reheating absorbed into a local remapping $R(\zeta_h)$.
More generally, this local remapping would also involve inflaton curvature fluctuations but
these are statistically independent and can be calculated separately in the usual way.
In linear theory, the mapping becomes a simple rescaling factor
\begin{equation}
\label{eq:Rdef}
R \equiv \lim_{\zeta_h \rightarrow 0} R(\zeta_h).
\end{equation} 
The power spectrum after Higgs decay is then related mode-by-mode to the power spectrum during inflation in a scale-independent fashion
\begin{equation}
\label{eq:D2conv}
\left. \Delta^2_\zeta (k) \right\vert_{{\rm decay}}  = R^2 \left. \Delta^2_{\zeta_h}(k)\right\vert_{\rm inflation},
\end{equation}
in which the number $R$ encodes all of the details of reheating.

It is therefore important to emphasize here that if the Higgs instability mechanism is successful, such that $\Delta_\zeta^2 \sim 10^{-2}$ on PBH scales after the Higgs decays, then in linear theory $R$ must reach at least $0.1$ for the order unity Higgs perturbations~\eqref{eq:delta_2_h_rough} to be converted to curvature perturbations with the correct amplitude~\eqref{eq:delta_2_pbh} to form sufficient PBHs and the details of how this is achieved through reheating and Higgs decay are irrelevant for the prediction of the linear power spectrum on other scales. In particular, its value on CMB scales depends solely on the inflationary $\Delta_{\zeta_h}^2(k)$.
This form is controlled by the Higgs potential itself and by the evolution of the Hubble rate during inflation.
Therefore, the viability with respect to CMB anisotropies of the PBH formation scenarios introduced in the literature \cite{Espinosa:2017sgp,Gross:2018ivp,Espinosa:2018euj}, which all assume mode evolution can be calculated linearly through reheating, can be assessed independently of the
details of the reheating model.

On the other hand, we shall see that the nonlinear nature of the Higgs instability plays
an important role in the mapping between $\zeta_h$ and $\zeta$ in Eq.~\eqref{eq:zetaconv}. 
Here, though the mapping remains local in that a given value of $\zeta_h(\vec{x})$ at a given position
$\vec{x}$ is mapped into 
a specific value of $\zeta(\vec{x})$, Fourier modes no longer evolve independently.   
Instead, we will compute the mapping using the nonlinear $\delta N$ formalism.  This mapping does depend on the specifics of how inflation ends,  but is independent of physical 
scale.  CMB scale fluctuations are in principle calculable from the spatial field  $\zeta_h(\vec{x})$ determined by modes that froze out during inflation. 

To illustrate these concepts, we make a few simplifying assumptions about how inflation and reheating proceed.  
We show in \S\ref{sec:inflation} that the most optimistic case for the scenario occurs when $H$ is effectively constant through inflation (see Eq.~\eqref{eq:zeta_k_h_comparison}).  Therefore rather than introducing a specific inflaton potential we assume that inflation occurs at a fixed $H$ and ends after an appropriate number of $e$-foldings.
The constant Hubble scale during inflation $H$
and the position $\hend$ 
of the Higgs at the end of inflation then together control the number of $e$-folds between the classical-roll scale $\hcl$ and the end of inflation, and therefore they control the physical scales on which PBHs are formed. 

We then assume that at the end of inflation, the inflaton decays instantly into radiation
and that the Higgs later also suddenly decays into radiation, as in the  model proposed in Ref.~\cite{Espinosa:2017sgp}. Maximizing $R$ in linear theory requires that the position of the Higgs at the end of inflation, $\hend$, is as close as possible to the maximum rescuable distance $\hrescue$. This {criticality} requirement motivates the various choices of scale in Eqs.~\eqref{eq:hmax}-\eqref{eq:hend}, following Ref.~\cite{Espinosa:2017sgp}.  Once $\hend$ is set in this way,
the value $H = 10^{12} \GeV$ is chosen to give a certain mass scale to PBHs by fixing
$N_{\rm cl} \sim -20$.

However, evolving the Higgs on the unstable side of its potential during inflation is dangerous, and the required proximity of $\hend$ to $\hrescue$ aggravates the situation beyond linear theory.
Due to quantum fluctuations of the Higgs during inflation, there are regions in which the local Higgs value at the end of inflation exceeds the background value, overshoots $\hrescue$, and cannot be restored by reheating to the metastable electroweak vacuum created thermally. Such vacuum decay bubbles, with infinitely growing $|\rho_h|$, expand even after the end of inflation and eventually engulf our current horizon.  These quantum fluctuations occur independently in the $e^{120}$ causally disconnected regions at ${N_{\rm cl}}\sim-20$ which make up our current horizon,  and therefore avoiding the vacuum decay bubbles requires extreme fine-tuning
\cite{Gross:2018ivp}. In \S\ref{ssec:nonlin}, we will cast this fine-tuning in terms of a breakdown in linear theory at the end of inflation, and we will show using the nonlinear $\delta N$ formalism that fine-tuning away the vacuum decay bubbles directly tunes away the PBH abundance.

Vacuum decay bubbles can be avoided by stabilizing the Higgs at some large field value between $\hend$ and $\hrescue$.
By adding a singlet heavy scalar of mass $m_s$ with appropriate couplings to the theory, a threshold effect can be exploited to lift the Higgs effective potential during inflation and induce a new true minimum at $h \sim m_s$, preventing unbounded runaway~\cite{EliasMiro:2012ay,Espinosa:2018euj}. For the purposes of this mechanism this Higgs potential beyond the Standard Model can be modeled as
\begin{equation}
\lambda^{\BSM} \simeq \lambda^{\SM} + \frac{\delta \lambda}{2} \left[ 1 + \tanh\left( \frac{h - m_s}{\delta} \right) \right],
\end{equation}
such that for $h \ll m_s$ the potential is as in the SM, given in \eqref{eq:lamsm}, while for $h\gg m_s$ the potential is increased by $\delta \lambda ( h^4/4)$. The step height $\delta \lambda$ should be such that the Higgs potential is stabilized, the step position $m_s$ should be close to $\hrescue$, and the step width $\delta$ sufficiently narrow to not interfere with $\hend$. In Fig.~\ref{fig:lambdaN}, we plot the Higgs potential with the representative choices $\left\{ \delta \lambda, m_s, \delta \right\} = \left\{0.02,\ 2000 H ,\ 100 H \right\}$.  
While the BSM potential does not suffer from vacuum decay bubbles, it still experiences a breakdown in linearity at the end of inflation. We will therefore also use the nonlinear $\delta N$ formalism to compute the conversion of $\zeta_h$ to $\zeta$ in this case.

Despite this difference at the end of inflation, the SM and BSM potentials are identical until large field values and therefore $\Delta_{\zeta_h}^2$ during inflation is the same in both potentials.
Fluctuations at this stage can be locally
remapped onto $\Delta_{\zeta}^2$.  
For a successful PBH model, this remapping must still achieve
$\Delta_{\zeta}^2 \sim 10^{-2}$ in both cases.
We will therefore focus on the SM potential until we begin discussing nonlinear effects at the end of inflation in \S\ref{sec:reheating}.

\section{Higgs During Inflation}
\label{sec:inflation}

In this section, we compute the power spectrum of the Higgs fluctuations from their production inside the horizon through to a common epoch when all modes relevant for observation are superhorizon in scale.

In \S\ref{ssec:classicalroll}, we present the equation of motion for the Higgs and describe the local competition between stochastic kicks and classical roll which governs its evolution. In \S\ref{ssec:background}, we argue that a well-defined background for the Higgs exists during the inflationary epochs relevant for observations, and that Higgs fluctuations during inflation can be computed by linearizing around this background mode-by-mode. In \S\ref{ssec:attini}, we follow this procedure and compute the Higgs power spectrum during inflation at all scales relevant for observations, 
regarding the Higgs field as a spectator and hence dropping metric perturbations.

Combined, these arguments will show that during inflation the Higgs power spectrum at CMB scales is larger than the Higgs power spectrum at the primordial black hole scales
\begin{equation}
\Delta^2_{\zeta_h}(k_{\rm CMB})  > \Delta^2_{\zeta_h}(k_{\rm PBH}).
\end{equation}

After the conversion of these superhorizon Higgs fluctuations during inflation into curvature fluctuations after inflation through Eq.~\eqref{eq:zetaconv}, this leads to
\begin{equation}
\Delta^2_{\zeta}(k_{\rm CMB})  > \Delta^2_{\zeta}(k_{\rm PBH}),
\end{equation}
in linear theory. 
Accounting for nonlinearity, we shall see that
a similar relation between the scales holds as long as the mapping between the Higgs and curvature fluctuations
is local.
Therefore the first conclusion of the present paper is that in the Higgs vacuum instability scenario, a large amplitude of the power spectrum on small scales generating PBHs is ruled out by the CMB normalization.
Conversely,
if one chooses a different set of parameters in this scenario in order to satisfy the CMB normalization, one ends up with a small-scale power spectrum of at most $\O(10^{-9})$ which fails to form PBHs. 

\subsection{Classical Roll vs Stochastic Kicks}
\label{ssec:classicalroll}

The equation of motion for the position- and time-dependent Higgs field $h(\vec{x}, N)$ is the Klein-Gordon equation
\begin{equation}
\Box h(\vec{x},N) = \frac{\pa V}{\pa h}\bigg\vert_{h(\vec{x},N)} \equiv \left. V_{,h}\right\vert_{h(\vec{x},N)} ,
\label{eq:KG}
\end{equation}
where here and throughout we denote partial derivatives with comma subscripts for compactness.

An important scale in this equation is the classical-roll scale $\hcl$, defined as follows. Every $e$-fold, the potential derivative leads $h(\vec{x},N)$ to roll by
\begin{equation}
\Delta h \simeq -\frac{1}{3 H^2} \left.V_{,h}\right\vert_{h(\vec{x},N)} .
\end{equation}
Meanwhile, if one splits the field into a piece averaged on scales larger than a fixed proper distance $\sim 1/H$ and small scale modes which continually cross the averaging scale, the small scale modes can be viewed as providing a local stochastic noise term to the equation for the coarse-grained superhorizon field~\cite{Starobinsky:1986fx,Starobinsky:1994bd}. The rms of this noise term each $e$-fold is
\begin{equation}
\langle \Delta h ^2 \rangle^{\frac{1}{2}} \simeq \frac{H}{2 \pi}.
\end{equation}
In the language of perturbation theory, this is the per $e$-fold rms of the free field fluctuation $\delta h$ and leads to a stochastic behavior of $h(\vec{x}, N)$. There are no subtleties involved in using $N$ as a time coordinate since the number of $e$-folds is not a stochastic quantity so long as the Higgs remains a spectator. 

The location $\hcl$ in the potential where the roll contribution and the stochastic contribution are equal,
\begin{equation}
\label{eq:classical_roll}
-\frac{1}{3 H^2} \left. V_{,h}\right\vert_{\hcl}   = \frac{H}{2 \pi},
\end{equation}
defines the `classical roll' scale $\hcl$ beyond which the classical term dominates the evolution of $h(\vec{x},N)$. We show this scale in Fig.~\ref{fig:lambdaN}, where it lies at $\hcl \simeq 8.3 H$. 

The classical-roll scale is important because in slow-roll, Higgs modes which cross the horizon when the background satisfies Eq.~\eqref{eq:classical_roll} generically have a large power spectrum. In particular, the Higgs power spectrum at the scale $k_{\rm cl}$ which crosses the horizon at $N_{\rm cl}=N(\hcl)$ is order one at horizon crossing,
\begin{equation}
\Delta^2_{\zeta_h} (k_{\rm cl}) \sim \frac{\langle \Delta h ^2 \rangle}{(\Delta h)^2} \sim 1.
\end{equation} 
If these Higgs fluctuations are converted to large curvature fluctuations, they can satisfy the requirements of \S\ref{sec:mechanism} such that $k_{\rm cl} = k_{\rm PBH}$. 

\subsection{Background and Linearization}
\label{ssec:background}

We split $h(\vec{x},N)$ equation into a background and perturbations
\begin{equation}
h(\vec{x},N) = h(N) + \delta h(\vec{x},N).
\end{equation}
Here we define the background to be the part of the field representing
 the spatial average over our  Hubble patch.  Therefore  at 
$N\sim-60$, the spatial average for the perturbation vanishes, $\langle \delta h(\vec{x}, -60 )\rangle = 0$.  The fluctuations are then generated by kicks from quantum fluctuations at
$N > -60$.

To evolve the Higgs field under Eq.~\eqref{eq:KG} we need to evaluate its position on the potential after $N=-60$, as established by its classical roll or quantum kicks. To do this, we need to establish whether the perturbations $\delta h(\vec{x},N)$ are linear around the background $h(N)$.

For the mechanism to work, there should be a well-defined classical roll to the Higgs field
at $N_{\rm cl}$, and we 
can linearize the Higgs fluctuations at that epoch as usual \cite{Espinosa:2017sgp,Gross:2018ivp}.
Between $-60 \lesssim N \lesssim N_{\rm cl}$, there is a competition between the local stochastic kicks and the bulk classical roll, and we need to check whether the kicks destabilize the average field in our Hubble patch.  

\begin{figure}[t]
\centering
\includegraphics[width=.65\linewidth]{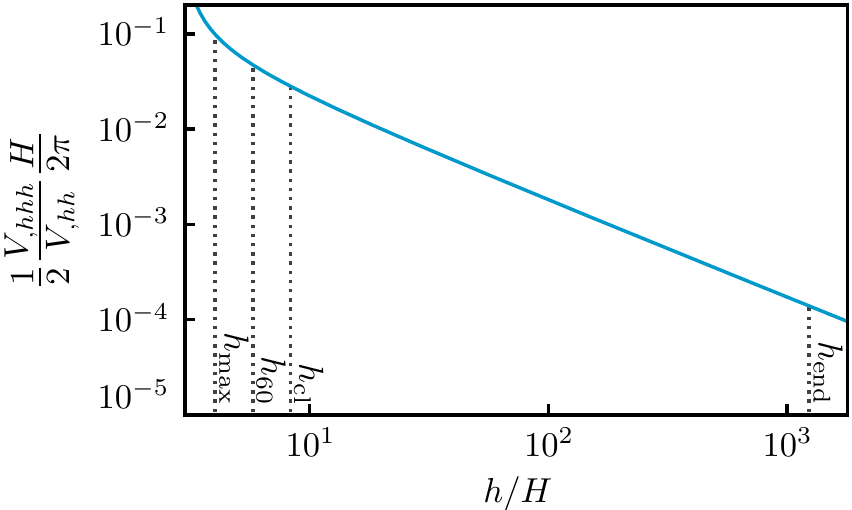}
\caption[Linearity of Higgs perturbations]{Interaction strength for modes of the typical horizon crossing amplitude $H/2\pi$  from \eqref{eq:potential_linearity}. Since it is far less than one at the relevant scales $h_{60}$ and $\hcl$,
modes evolve independently and the cumulative background roll \eqref{eq:background_roll} dominates 
over the stochastic displacement \eqref{eq:stochastic_displacement}.
Linear perturbation theory holds until $\delta h$ grows much larger than $H/2\pi$, near $h_{\rm end}$
(see \S\ref{ssec:nonlin}). }
\label{fig:potential_linearity}
\end{figure}

In linear theory,
stochastic kicks at the same $\vec{x}$ but subsequent times evolve independently from each other. In particular, the potential term in the Klein-Gordon equation controls their interactions. When the Higgs is a spectator field, and we can expand this term around the homogeneous piece as
\begin{equation}
\left.V_{,h}\right\vert_{h(\vec{x},N)} = \left.V_{,h}\right\vert_{h(N)} + \delta h(\vec{x},N) \left. V_{,hh} \right\vert_{h(N)} + \ldots
\end{equation}
where `$\ldots$' contains terms higher order in $\delta h(\vec{x},N)$. If we neglect the higher order terms, then each subsequent kick evolves as a free field and, as previously mentioned, has rms $H / 2 \pi$ at horizon crossing.    The higher order terms then are suppressed relative to the linear term by
\begin{equation}
\label{eq:potential_linearity}
\frac{1}{2} \frac{V_{,hhh}}{V_{,hh}} \sqrt{\langle \delta h^2 \rangle}  \simeq \frac{1}{2} \frac{V_{,hhh}}{V_{,hh}} \frac{H}{2 \pi},
\end{equation}
in which we approximate all modes by their value at horizon crossing, which is appropriate while both the inflaton and the Higgs fields 
are slowly rolling. We plot this quantity for the Standard Model Higgs potential in Fig.~\ref{fig:potential_linearity} and show that it is less than one at all scales in the unstable region. 

Therefore, the distance traveled due to stochastic kicks between $N=-60$ and $N_{\rm cl}$ accumulates as a random walk.  For $N_{\rm cl}=-20$,
they therefore lead to a displacement
\begin{equation}
\label{eq:stochastic_displacement}
|\Delta h|_{\rm stochastic} = \frac{H}{2\pi} \times \sqrt{40} \simeq H.
\end{equation}
This is significantly less than the distance between the classical-roll scale $\hcl=8.3 H$ and the maximum of the potential $\hcr=4 H$. This means that stochastic kicks do not, over $40$ $e$-folds, kick our horizon into the other side of the potential. Therefore our whole Hubble volume was on the unstable side of the potential when it crossed the inflationary horizon at $N=-60$.

Moreover, the total displacement from stochastic kicks~\eqref{eq:stochastic_displacement} is less than the amount the background field rolls in these $40$ $e$-folds as we shall now see.
For a homogeneous field $h(N)$, Eq.~\eqref{eq:KG} becomes
\begin{equation}
\label{eq:higgs_bg_eom}
h'' + (3 - \e) h' + \frac{V_{,h}}{H^2} = 0 ,
\end{equation}
where $\e= -H'/H$ is the first Hubble slow-roll parameter, which is zero during the exact de Sitter inflation in our fiducial model. 

The initial conditions $h_{60}$ and $h'_{60}$ for this equation should be such that the Higgs reaches the desired field position $\hend$ at the end of inflation close enough to $\hrescue$ such that
$R$ is maximized. With one constraint and two initial values, a range of  $h'_{60}$ and $h_{60}$ can lead to $h(N=0) = \hend$. Assuming attractor initial conditions for the Higgs, we choose
\begin{equation}
\label{eq:attractorICs}
h'_{60} = -\frac{1}{(3-\e) H^2} \left.V_{,h}\right\vert_{h_{60}},
\end{equation}
making the initial field position given by Eq.~\eqref{eq:h60} $h_{60} \simeq 5.8 H$, when $\hend$ is set by Eq.~\eqref{eq:hend}.

Therefore the classical roll from $-60$ to $N_{\rm cl}$ is 
\begin{equation}
\label{eq:background_roll}
\left\vert\Delta h \right\vert_{\rm roll} = \left\vert h_{60}-\hcl \right\vert = 2.5 H,
\end{equation}
which is significantly larger than the stochastic displacement \eqref{eq:stochastic_displacement} but is nonetheless safely on the unstable side of the potential. This occurs despite the fact that each stochastic kick is larger than the per $e$-fold roll because the roll is coherent across our Hubble volume while the kicks are random.

Therefore we have a consistent picture where if we begin with an average field in our horizon volume around $h_{60} \sim 5.8 H$, then our local background will reach $\hcl$ at $N\sim-20$, unspoiled by stochastic kicks. Between these scales perturbations are linear, thanks to Fig.~\ref{fig:potential_linearity}, and subdominant over the background roll. Our background will then continue to roll to $\hend$, where the Higgs will be uplifted. We plot this background in the lower panel of Fig.~\ref{fig:lambdaN}. App.~A of Ref.~\cite{Passaglia:2019ueo} shows that this picture is consistent with the creation of our background from superhorizon stochastic fluctuations.

We can now use this background to solve for $\delta h$ mode by mode during inflation for all relevant observational scales as in linear theory. This linearization depends on ignoring the interaction of Higgs fluctuations rather than the full machinery of linear perturbation theory for the metric and the matter, and in particular its validity does not assume $|\zeta_h | \ll 1$. Higgs nonlinearities become important in the last $e$-fold of inflation and beyond as the field fluctuations are amplified by the Higgs instability. Such nonlinear effects will affect the superhorizon CMB and PBH modes equally as we shall show in  \S\ref{sec:reheating}.

\subsection{Higgs Power Spectrum}
\label{ssec:attini}

The linearized Klein-Gordon equation for the Fourier mode $\delta h^k(N)$ of $\delta h(\vec{x},N)$ in spatially flat gauge is
\begin{equation}
\label{eq:linKG}
\left(\frac{d^2}{d \eta^2} + 2 \frac{\dot a}{a} \frac{d}{d \eta} + k^2   \right) \delta h^k + a^2  \delta V^k_{,h} \simeq 0 ,
\end{equation}
where $\delta V^k_{,h} = V_{,hh} \delta h^k$ during inflation, and note that, unlike in the rest of this thesis, in this chapter an overdot denotes a derivative with respect to the conformal time $\eta$.  Here we have dropped metric perturbations, which are suppressed when the Higgs is a spectator; these are restored in App.~B of Ref.~\cite{Passaglia:2019ueo} for completeness.

The Klein-Gordon equation can then be conveniently expressed in terms of the auxiliary variable $u^k \equiv a \delta h^k$,  
\begin{align*}
\ddot{u}^k + \left(k^2 - \frac{\ddot{a}}{a} + a^2 V_{,hh} \right) u^k = 0,
\numberthis
\end{align*}
which holds at all orders of background and Higgs slow roll parameters. 
To order $\O(\e^2)$ and $\O(\e \eta_H)$, where $\eta_H$ is the second Hubble roll parameter (see, e.g., Ref.~\cite{Miranda:2012rm}), but fully general in terms of the Higgs roll, we can write
\begin{equation}
\label{eq:MukhanovSasakiHiggs}
\ddot{u}^k + \left(k^2 - \frac{\ddot{z}}{z}\right) u^k = 0,
\end{equation}
where 
\begin{equation}
z \equiv H \dot{h}.
\end{equation}
This equation, of the Mukhanov-Sasaki type, is conveniently solved in the variable $s \equiv
\eta_{\rm end}-\eta$, the positive decreasing conformal time to the end of inflation (see, e.g., Refs.~\cite{Motohashi:2015hpa,Motohashi:2017gqb}). 

First, let us focus on the evolution in the superhorizon regime.
In that limit, the analytic solution for the Mukhanov-Sasaki equation~\eqref{eq:MukhanovSasakiHiggs} is given by
\begin{equation}
\frac{u^k}{z} = c_0 + c_1 \int \frac{d \eta}{z^2},
\end{equation}
where $c_0$ and $c_1$ are constants. So long as the second mode is decaying, we therefore have that on superhorizon scales
\begin{equation}
\label{eq:ms_uoverz}
\frac{\delta h^k}{h'} = c_0 H^2 + \O(\e, \eta_H).
\end{equation}
In linear theory, the curvature perturbation on uniform Higgs density slices is obtained by gauge transformation as
\begin{equation}
\label{eq:infinitesimal}
\zeta^k_h = -\frac{\delta \rho^k_h}{\rho_h'} \simeq -\frac{\delta h^k}{h'},
\end{equation}
where first the approximate equality indicates that when the Higgs is slowly rolling, uniform Higgs density and uniform Higgs field slicing coincide to order $\e$ (see \cite{Passaglia:2019ueo}). More generally $\zeta_h$ is defined as the change in $e$-folds from a spatially flat surface to a constant density Higgs surface.  This linear approximation holds so long as $\rho_h''/\rho_h'^2 \delta \rho_h \ll 1$, as it is here (see Eq.~\eqref{eq:delta_N_linear}).

Using the superhorizon evolution equation \eqref{eq:ms_uoverz}, we therefore have that if $H$ evolves during inflation, the curvature on uniform Higgs density slices is not conserved on superhorizon scales and in particular decays according to
\begin{equation}\label{eq:sup}
\frac{{\zeta_h^k}'}{\zeta_h^k} = -2\e,
\end{equation}
at leading order in $\e$. This estimate of superhorizon evolution assumes only background slow roll.

This superhorizon evolution is due to a pressure perturbation on uniform density slices for the Higgs, in other words a nonadiabatic pressure, induced because the uniform Higgs density slicing is not a uniform Hubble slicing when $\e\neq0$.  This phenomenon is studied in detail in App.~B of Ref.~\cite{Passaglia:2019ueo}.  Conversely, if $H$ is constant, then the fact that the Higgs field evolves onto an attractor solution implies that nonadiabatic stress vanishes thereafter
and $\zeta_h$ is conserved nonlinearly.   In this case, much like single-field slow-roll inflation, 
the Higgs field supplies the only clock and field perturbations are equivalent to changing 
that clock on the background trajectory. 

Next, let us focus on the evolution from subhorizon scales to the superhorizon regime.
This evolution can be tracked by solving the Mukhanov-Sasaki equation~\eqref{eq:MukhanovSasakiHiggs} with Bunch-Davies initial conditions deep inside the horizon 
\begin{equation}
\label{eq:bunch-davies}
u^k ( s ) = \frac{1}{\sqrt{2 k}} \left( 1 + \frac{i}{k s} \right) e^{i k s}.
\end{equation}
For analytic estimates, we can assume slow-roll evolution of $z$, in which case we can take the de Sitter modefunction \eqref{eq:bunch-davies} to the superhorizon limit, and find that each field fluctuation crosses the horizon with amplitude
\begin{equation}
\label{eq:hkds}
\delta h^k \simeq \frac{i H}{\sqrt{2 k^3}},
\end{equation}
and the field fluctuation power spectrum at that time is
\begin{equation}
\Delta^2_{\delta h} (k) = \frac{k^3}{2 \pi^2} \left\vert \delta h^k \right\vert^2 \simeq \left(\frac{H}{2 \pi}\right)^2 . 
\end{equation}
Using the gauge transformation Eq.~\eqref{eq:infinitesimal} with the field fluctuation Eq.~\eqref{eq:hkds} and the field velocity from the slow-roll solution of Eq.~\eqref{eq:higgs_bg_eom}, the curvature perturbation on uniform Higgs density hypersurfaces at horizon crossing is 
\begin{align*}
\label{eq:zeta_k_h_HC}
\zeta^k_h (\eta_k) &\simeq -\frac{i H}{\sqrt{2 k^3}} \frac{1}{h'} \bigg\vert_{\eta_k}\\
&\simeq -\frac{i H}{\sqrt{2 k^3}}\frac{(3-\e) H^2}{-V_{,h}} \bigg\vert_{\eta_k} . \numberthis
\end{align*}
To lowest order in Higgs- and background-slow-roll $\eta_k$
is chosen to be the epoch of horizon-crossing $k \eta_k = 1$, but to next order
can be optimized to $k \eta_k = \exp{\left[7/3 - \ln 2 - \gamma_E\right]}$, with $\gamma_E$ the Euler-Mascheroni constant  \cite{Motohashi:2015hpa,Motohashi:2017gqb}.

We can now estimate the relative amplitude of $\Delta^2_{\zeta_h}$ on CMB and PBH scales. Choosing some comparison time $\eta_*$ once both scales have exited the horizon but far enough from the end of inflation that slow-roll parameters are still small, we have 
\begin{align*}
\label{eq:zeta_k_h_comparison}
\left.\frac{\Delta^2_{\zeta_h} (k_{\rm CMB})}{\Delta^2_{\zeta_h} (k_{\rm PBH})}\right\vert_{\eta_*}  &\simeq \left( \frac{H_{\rm PBH}}{H_{\rm CMB}}\right)^4 
\frac{k_{\rm CMB}^{3} \left\vert\zeta^{k_{\rm CMB}}_h\right\vert^2_{\eta_k}}{k_{\rm PBH}^{3} \left\vert\zeta^{k_{\rm PBH}}_h\right\vert_{\eta_k}^2}\\
&\simeq \left(\frac{H_{\rm CMB}}{H_{\rm PBH}}\right)^2 \left(\frac{V_{,h} |_{\rm PBH}}{V_{,h} |_{\rm CMB}}\right)^2,
\numberthis
\end{align*}
where we have used at horizon crossing the Higgs slow-roll expression \eqref{eq:zeta_k_h_HC} and outside the horizon the Hubble slow-roll expression \eqref{eq:ms_uoverz}. Thus in the generic situation where $H$ is decreasing and the Higgs rolls downhill, we find that $\Delta^2_{\zeta_h} (k_{\rm CMB}) / \Delta^2_{\zeta_h} (k_{\rm PBH}) > 1$. The most optimistic case for the scenario is therefore the one where $H$ is strictly constant between the CMB and PBH scales, and it still results in $\Delta^2_{\zeta_h} (k_{\rm CMB}) / \Delta^2_{\zeta_h} (k_{\rm PBH}) > 1$.

\begin{figure}[t]
\centering
\includegraphics[width=.65\linewidth]{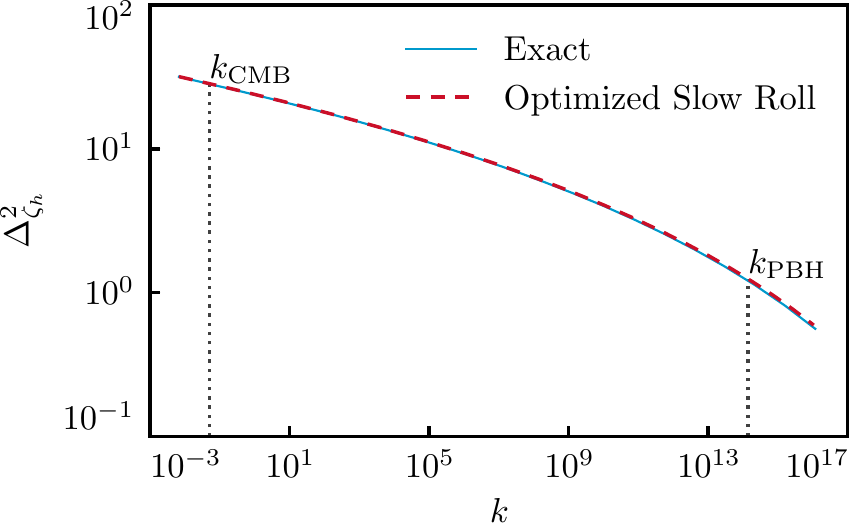}
\caption[Inflationary Higgs power spectrum]{
The Higgs power spectrum during inflation computed as described in \S\ref{ssec:attini} on a uniform Higgs energy density slice by exact solution of the Mukhanov-Sasaki equation \eqref{eq:MukhanovSasakiHiggs} (solid blue) and by the optimized slow-roll approximation \eqref{eq:zeta_k_h_HC} (dashed red), at some time $\eta_*$ after the relevant modes have crossed the horizon in the optimistic scenario where $H$ is constant until $\eta_*$. The Higgs power spectrum is larger on CMB scales than on the classical-roll scales.
}
\label{fig:higgspowerspec}
\end{figure}
We show in Fig.~\ref{fig:higgspowerspec} the  Higgs power spectrum computed by solving the Mukhanov-Sasaki equation exactly in the optimistic case where $H$ is constant between CMB and classical-roll scales, evaluated at the convenient time $\eta_*$ when all relevant modes have crossed the horizon. We compare this exact solution to the slow-roll expression \eqref{eq:zeta_k_h_HC} with the optimized freeze-out epoch. 
For decreasing $H$, the ratio $\Delta^2_{\zeta_h} (k_{\rm CMB})/\Delta^2_{\zeta_h} (k_{\rm PBH})$ would be larger than the one estimated from Fig.~\ref{fig:higgspowerspec}.

These Higgs fluctuations at $\eta_*$ will be converted to a total curvature fluctuations after Higgs decay by the factor $R(\zeta_h)^2$ which we discuss in \S\ref{sec:reheating}.
The key is that this mapping affects all mode contributions to $\zeta_h$ uniformly.  For example in linear theory $R$ is a constant whose value must be $\sim 0.1$ for successful PBH formation at $k_{\rm PBH}$.  Eq.~\eqref{eq:zeta_k_h_comparison} determines the total curvature power relative to this scale.  In particular, the power spectrum at CMB scales is an order of magnitude larger than the power spectrum at the classical-roll scale. It is simply a feature of the Higgs potential that the field slope increases as the Higgs goes farther into the unstable region, and therefore that the Higgs fluctuation shrinks as $k$ increases. Thus, a model with $\Delta^2_{\zeta_h} (k_{\rm CMB}) / \Delta^2_{\zeta_h} (k_{\rm PBH}) > 1$ that forms
PBHs at $k_{\rm PBH}$ will necessarily violate CMB constraints.

The results of this section hold equally for the SM and BSM potential. The difference between the two potentials enters only into $R(\zeta_h)$ which converts these results into the final curvature perturbation after Higgs decay. More generally as long as this mapping depends only on field amplitude and not on $k$ explicitly, PBHs cannot be formed from
{the} Higgs instability without violating CMB constraints.

\begin{center}
  $\ast$~$\ast$~$\ast$
\end{center}

In summary, we have shown that during inflation
\begin{equation}
\Delta^2_{\zeta_h}(k_{\rm CMB}) > \Delta^2_{\zeta_h}(k_{\rm PBH}).
\end{equation}
As we argued in \S\ref{sec:mechanism}, the conversion of the inflationary $\zeta_h$ to the final $\zeta$ depends only on the amplitude of $\zeta_h$ and thus all the information about reheating can be encoded in a scale-independent function $R(\zeta_h)$. 

This means that in linear theory, where $R$ is a constant, if primordial black holes are produced on small scales then on large scales
\begin{equation}
\Delta^2_{\zeta}(k_{\rm CMB})  > \Delta^2_{\zeta}(k_{\rm PBH}) \gtrsim 10^{-2},
\end{equation}
which is incompatible with measurements of the CMB. 

Nonlinearly, when CMB and PBH modes cannot be tracked independently through the final $e$-folding of inflation and reheating, the $\Delta^2_{\zeta_h}(k)$ results in this section provide the superhorizon initial conditions which can be mapped to the final $\zeta$. Given that this  local mapping $R(\zeta_h)$ does not distinguish
between Higgs fluctuations of  different physical scales, we will argue in \S\ref{sec:reheating} that even nonlinearly, Higgs induced curvature
fluctuations
on CMB scales will be larger than those on PBH scales.

We now study in \S\ref{sec:reheating} the specific values taken by the conversion function $R(\zeta_h)$ itself. This will allow us to determine whether or not the Higgs fluctuations computed here can be transferred into large enough curvature perturbations to form PBHs, regardless of the compatibility with the CMB. Moreover, given that we have produced large Higgs fluctuations on CMB scales, this will allow us to determine under which conditions Higgs 
criticality is incompatible with the small curvature fluctuations observed in the CMB.

\section{Higgs and Reheating}
\label{sec:reheating}

We now track the curvature perturbations $\zeta_h$ on uniform Higgs density slices during inflation, computed in \S\ref{sec:inflation}, through the end of inflation, reheating, and Higgs decay to compute the final curvature perturbations $\zeta$ on uniform total density hypersurfaces relevant for PBH formation.

In \S\ref{ssec:instantaneous}, we discuss how the Higgs evolves near the end of inflation and present the basic features of the instantaneous reheating model proposed by Ref.~\cite{Espinosa:2017sgp}.

In \S\ref{ssec:deltaN}, we discuss how to use the nonlinear $\delta N$ formalism to convert $\zeta_h$ to $\zeta$ for any local reheating scenario, and we discuss jump conditions which much be satisfied during instantaneous reheating. We also present linearized $\delta N$ formulae which yield results corresponding to those of linear perturbation theory, allowing an important crosscheck of the computation.

In \S\ref{ssec:linear}, we follow the assumption of Refs.~\cite{Espinosa:2017sgp,Gross:2018ivp,Espinosa:2018euj,Espinosa:2018eve} that linear theory holds through reheating and we compute explicitly the conversion of the inflationary $\zeta_h$ to the final $\zeta$. We show that energy conservation at reheating, neglected in previous works, prevents the model from achieving the required $R=0.1$ for both the SM and BSM potentials and thus PBH 
are not produced in sufficient quantities to be the dark matter in linear theory.

In \S\ref{ssec:nonlin} we show that linear theory is in fact violated at the end of inflation and we explicitly compute the full nonlinear conversion $R(\zeta_h)$ for the SM and BSM Higgs effective potentials. We show that PBHs in sufficient abundance to be the dark matter are never formed, second-order gravitational waves are suppressed, and only for a special class of criticality scenarios can observable perturbations be produced on CMB scales.

\subsection{Instantaneous Reheating}
\label{ssec:instantaneous}

\begin{figure}[t]
\centering
\includegraphics[width=.65\linewidth]{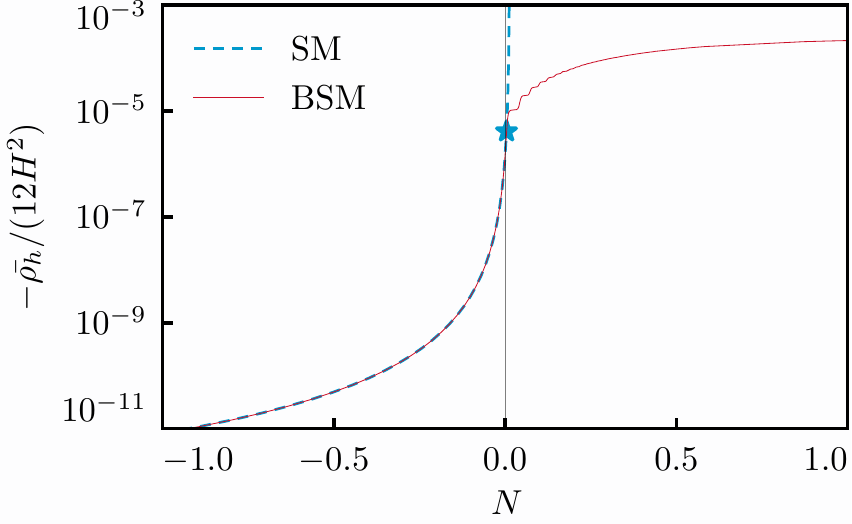}
\caption[Higgs energy density around reheating]{The (negative) Higgs energy density evolution during inflation, with $H =\,$const. $N=0$ corresponds to the end of inflation for the background, and $N>0$ shows how the Higgs energy would evolve in
local regions in advance of the background.  
The SM Higgs (dashed blue) at $N=0$ is close to saturating the rescue condition \eqref{eq:rescue} (blue star).
The BSM Higgs (solid red) hits a wall during inflation and therefore is always rescued.
Due to the Higgs attractor behavior, the exact value of $\zeta(0^+)$ for any given local shift $\delta N =\zeta_h$
can be read off using Eq.~\eqref{eq:nonlin_jump}.
}
\label{fig:rho}
\end{figure} 

As the Higgs travels farther and farther on the unstable side of the SM potential, it rolls faster and faster and if inflation never ended its energy density would diverge in finite time. In Fig.~\ref{fig:rho}, we show this $\rho_h$ during this last phase of the instability as a function of the number of $e$-folds.
Of course inflation does end and for our chosen example this occurs at $N=0$  and a field
value  $\hend$ in the background.  In this case, $N>0$ then shows how the Higgs energy would continue to evolve if inflation did not end at $N=0$. This range around $N=0$ will also be useful when we consider perturbations that can
be ahead of or behind the background value.

As we can see, the fiducial position of the background SM Higgs we have chosen is near-critical. Its energy is increasing rapidly and if inflation lasts much longer it will gain sufficient energy such that it is no longer a spectator.    On this edge, the {\it background} Higgs field experiences the same evolution in the SM and BSM
cases by construction 
(see Fig.~\ref{fig:rho}).

Once inflation ends, the process of reheating transfers
the inflaton's energy to radiation. The Higgs then no longer evolves in vacuum and the Klein-Gordon equation becomes
\begin{equation}
\Box h = \VT_{,h},
\label{eq:KG_eff}
\end{equation}
where $\VT$ is the thermal effective potential in a bath of temperature $T$ \cite{Espinosa:2015qea},
\begin{equation} \label{eq:VT}
\VT  = V + \frac{1}{2} M_{\rm T}^2 h^2 e^{-h/2 \pi T}, \quad M_{\rm T}^2 \simeq 0.12 T^2.
\end{equation}

If reheating is instantaneous, as proposed by Ref.~\cite{Espinosa:2017sgp}, then the total energy is conserved across it,
\begin{equation}
\left(\rho_{\phi} + \rho_h\right) (0^-) = \left(\rho_{\rm r} + \rho_h\right)(0^+).
\label{eq:energyconback}
\end{equation}

The division of the total energy density after inflation into a radiation piece $\rho_{\rm r}$ and a thermal component to the Higgs  $\rho_h$ is somewhat arbitrary.  Since to leading order in $\rho_h/\rho_{\rm tot}$, the curvature perturbation after the Higgs has decayed depends only on the conservation of the total energy, for convenience we choose to define $\rho_{\rm r}$ as a separately conserved thermal piece obeying the equation of motion
\begin{equation}
\label{eq:rho_r_eom}
\rho_{\rm r}' = - 4 \rho_{\rm r},
\end{equation}
and the thermal bath temperature $T$ is then
\begin{equation}
T = \left(\frac{30 \rho_{\rm r} }{ \pi^2 g_*}\right)^{1/4},
\end{equation}
where $g_*=106.75$ is the number of degrees of freedom in the Standard Model. Conservation of the total stress-energy along with the Higgs equation of motion~\eqref{eq:KG_eff} then imply that the separately conserved Higgs energy is
\begin{equation}
\label{eq:rho_h}
\rho_h \equiv \frac{1}{2} \frac{\dot h^2}{a^2} + \VT.
\end{equation}
Changes in $\VT$ as the Higgs rolls down the effective potential then provide kinetic energy for the field
as if it were a true potential energy.

Since $3 H^2(0^-) = (\rho_\phi + \rho_h){(0^-)}$, neglecting the Higgs' energy density contribution to $H$ during inflation but including it after entails a dynamically negligible $\O(\rho_h/\rho_\phi) \sim \O(\rho_h/3H^2)$  discontinuity
in the Hubble rate at the end of inflation $H_{\rm end}$ (see Fig.~\ref{fig:rho}). On the other hand, strict energy conservation~\eqref{eq:energyconback} at reheating is important because we will evaluate perturbations on constant density surfaces.

After reheating, the thermal effective potential in Eq.~\eqref{eq:KG_eff} will rescue the Higgs from the unbounded SM minimum so long as
\begin{equation}
\label{eq:rescue}
{\rm Max}\left[h\right] < \hrescue,
\end{equation}
where ${\rm Max}\left[h\right]$ is the maximum displacement of the Higgs field and $\hrescue$ is the peak of the thermal potential $\VT$.  Due to the nonzero kinetic energy of the Higgs at the end of inflation, ${\rm Max}\left[h\right]$ is larger than the field displacement at the end of inflation $\hend$. We mark the maximal point which saturates this bound with a star in Fig.~\ref{fig:rho}.

Neglecting the exponential term, we find that the peak of the uplifted potential at reheating is
\begin{equation}
\hrescue^{(0)} = \frac{M_T}{\sqrt{|\lambda^{\SM}|}},
\end{equation}
a solution which can be iterated to account for the exponential term, yielding
\begin{equation}
\label{eq:hrescue}
\hrescue^{(1)} = \hrescue^{(0)} e^{-\hrescue^{(0)} / {4 \pi T}} \sim 1.6 \sqrt{H_{\rm end} \Mpl},
\end{equation}
where we have used $|\lambda| \sim 0.007$ and which with $H_{\rm end}=10^{12} \GeV$ evaluates to $ \sim 2500 H$.
The value of the uplifted Higgs potential at this approximate maximum is
\begin{equation}
\label{eq:potential_max}
\VT(\hrescue^{(1)}) \sim 0.02 H_{\rm end}^2 \Mpl^2.
\end{equation}
These scalings are in good agreement with the exact calculation for $\hrescue^{(0)}/4\pi T \ll 1$ and they serve to highlight the dependence of the results with parameter choices. For our fiducial parameter set, the exact calculation yields  $\hrescue \simeq 2400 H_{\rm end}$, $\VT(\hrescue)\simeq 0.02 H_{\rm end}^2 \Mpl^2$.
 
If the Higgs is rescued, then it oscillates in its uplifted temperature-dependent potential, redshifting as radiation on the cycle-average up to corrections from the non-quadratic components of its potential, until it decays on the $e$-fold timescale on constant Hubble surfaces.  The rescue point is therefore relevant even for the BSM potential.   In our example shown in Fig.~\ref{fig:rho}, we set the $m_s$ barrier
close to $\hrescue$ to maximize the instability while ensuring that the field returns to the electroweak
vacuum after reheating.

We now describe in \S\ref{ssec:deltaN} how to track perturbations through the end of inflation and this instantaneous reheating epoch.

\subsection{Nonlinear Curvature Evolution}
\label{ssec:deltaN}

The PBH abundance depends on the probability that the local horizon averaged density field exceeds some collapse threshold $\delta_c$. We approximate this by the Gaussian probability that the curvature on uniform total density slices $\zeta$ lies above some threshold $\zeta_c$. We therefore need to compute $\zeta$ after the Higgs decays. 

Nonlinearly in the Higgs field perturbations, the transformation of the curvature on uniform Higgs density slicing during inflation $\zeta_h$ to the curvature on uniform total density slicing after inflation $\zeta$ can be performed in the $\delta N$ formalism \cite{Starobinsky:1985aa,Salopek:1990jq,Sasaki:1995aw,Lyth:2004gb}, which allows us to evolve superhorizon perturbations by counting the number of $e$-folds of expansion from an initial flat slice at some convenient initial time $N_i$ to a uniform total density slice at a final time $N$,
\begin{equation}
\label{eq:delta_N}
\zeta(N, \vec{x}) = \N(\rho_\alpha(N_i,\vec{x}) ;\  \rho_{\rm tot} (N)) - \medbar\N( \bar{\rho}_\alpha(N_i);\ \rho_{\rm tot} (N)),
\end{equation}
where $\N$ is the local number of $e$-folds of expansion from the initial flat hypersurface at $N_i$ on which any fields $\alpha$
have energy densities $\rho_\alpha(N_i,\vec{x}) = \bar{\rho}_\alpha(N_i) + \delta \rho_\alpha(N_i, \vec{x})$ to a final surface of uniform total density $\rho_{\rm tot} (N)$, and $\medbar\N$ is the corresponding expansion of the unperturbed universe. 
Using the separate universe assumption, $\N$ can be computed in terms of background FLRW equations for a universe with the labeled energy contents. 

If we chose $N_i$ to be some time during inflation when we know the superhorizon density fluctuation $\delta \rho_h$ (or $\zeta_h$) from our computation in \S\ref{sec:inflation}, then since we are considering only perturbations sourced by the Higgs we can make the inflaton density at $N_i$ implicit and keep only the dependence on the initial $\delta \rho_h$. We will perform this fully general calculation in \S\ref{ssec:nonlin}.

To understand these results, it is also useful to have a simple analytic approximation for the impact of reheating
given conditions just before reheating.  In this case we can set the initial time just after reheating at $N_i=0^+$.
The Higgs energy density is a small component of the total energy budget, and therefore to leading order in $\zeta$ we can evaluate the $\delta N$ formula assuming that $\rho_{\rm tot} \propto a^{-4}$  to find 
\begin{align*}
\zeta(0^+, \vec{x}) &\simeq \frac{1}{4} \ln{\left(\frac{\rho_{\rm tot}(0^+,\vec{x})}{\bar{\rho}_{\rm tot}(0^+)}\right)}\\
&\simeq \frac{\rho_{\rm tot} (0^+,\vec{x})-\bar{\rho}_{\rm tot}(0^+)}{12 H_{\rm end}^2}.
\numberthis
\end{align*}
Conservation of energy at reheating \eqref{eq:energyconback} then implies the $N=0$ jump condition
\begin{equation}
\label{eq:nonlin_jump}
\zeta(0^+, \vec{x}) = \frac{\rho_{h} (0^-,\vec{x})-\bar{\rho}_{h}(0^-)}{12 H_{\rm end}^2}.
\end{equation}
Note that this condition applies to nonlinear Higgs density fluctuations $|(\rho_h -\bar\rho_h)/\bar \rho_h| \gg 1$
so long as $|(\rho_h -\bar\rho_h)/\bar \rho_{\rm tot}| \ll 1$.

We can see from \eqref{eq:nonlin_jump} that the post-inflationary energy partitioning chosen in \eqref{eq:energyconback} does not enter into the total curvature just after reheating. Instead, $\zeta(0^+, \vec{x})$ is determined solely by energy conservation.

We can further simplify this condition by noting that to the extent that $H$ is constant during inflation, which we have shown by \eqref{eq:zeta_k_h_comparison} is the most optimistic scenario for PBH production,
the shift in $e$-folds to a constant Higgs energy density $\delta N_h =\zeta_h$ is conserved nonlinearly.\footnote{If $H$ evolves, the leading order effect will be simply to shrink the Higgs $\delta N_h$.}
Therefore $\rho_{h} (0^-,\vec{x}) = \bar{\rho}_{h} (-\delta N_h)$, 
with this Higgs density computed as though inflation did not end at $N=0$ as in Fig.~\ref{fig:rho}.
We can therefore read off $\zeta(0^+, \vec{x})$ for a given $\zeta_h$ from $\bar\rho_h(N)$ as
\begin{equation}
\label{eq:nonlin_jump2}
\zeta(0^+, \vec{x}) = \frac{\bar\rho_{h} ({-}\zeta_h)-\bar{\rho}_{h}(0^-)}{12 H^2} \Big|_{\rm inf},
\end{equation}
where $|_{\rm inf}$ denotes this convention of evaluating the background as if inflation never ends.

Before using these nonlinear formulae \eqref{eq:delta_N} and \eqref{eq:nonlin_jump2} in \S\ref{ssec:nonlin}, we will in \S\ref{ssec:linear} perform the calculation using linear perturbation theory. To validate the linear theory calculations, below we derive linear approximations to the $\delta N$ formulae. 

The full $\delta N$ formula \eqref{eq:delta_N} can be linearized in $\delta \rho_{h_i} \equiv \delta \rho_h(N_i,\vec{x})$ to obtain
\begin{equation}
\label{eq:delta_N_linear}
\zeta \simeq \frac{\pa \N(\rho_{h_i};\  \rho_{\rm tot} )}{\pa \rho_{h_i}}\delta \rho_{h_i} \simeq  -\frac{\pa \N(\rho_{h_i};\  \rho_{\rm tot})}{\pa \rho_{h_i}}\rho'_{h_i} \zeta_{h_i},
\numberthis
\end{equation}
where $\rho_{h_i}$ and the $\rho'_{h_i}$ are Higgs density and its derivative at $N_i$ and $\zeta_{h_i}$ is the Higgs curvature at that time.
Likewise, a linear Taylor expansion of the jump condition \eqref{eq:nonlin_jump2} is given by
\begin{align*}
\label{eq:linear_jump}
\zeta(0^+, \vec{x}) &\simeq -\zeta_h (0^-) \times \frac{\bar\rho_h'(0^-)}{12 H^2}.
\numberthis
\end{align*}
The ratio $-\rho_h' (0^-) /12 H^2$ is the rescaling factor $R(0^+)$ if $H$ is constant through to the end of inflation.

As we shall see below, these linear $\delta N$ formulae provide an important point of contact between the nonlinear $\delta N$ and linear perturbation theory approaches.

\subsection{Linear Conversion}
\label{ssec:linear}

We now follow the assumption of Refs.~\cite{Espinosa:2017sgp,Gross:2018ivp,Espinosa:2018euj,Espinosa:2018eve} that linear theory holds through reheating, and we show that under this assumption primordial black holes cannot be the dark matter.

While Higgs field values $h$ and $\delta h$ and their derivatives $h'$ and ${\delta h}'$ are all continuous through reheating, the Higgs potential and its slope change instantaneously when the Higgs potential is uplifted. Therefore the Higgs energy density \eqref{eq:rho_h}, its derivative
\begin{equation}
{\rho}_h' = -3 \frac{\dot{h}^2}{a^2}  -  \VT_{,T} T,
\end{equation}
and its perturbation
\begin{equation}
\label{eq:drhohafter}
\delta \rho_h = \frac{1}{a^2} \dot{h} \delta \dot{h}  + \VT_{,h} \delta h +  \VT_{,T} \delta T,
\end{equation}
are not continuous with their values 
at $N=0^-$.
$\delta T$ here is any perturbation in the bath temperature correlated with the Higgs, which we shall see is generically induced at reheating. For simplicity, we have omitted here a contribution to the perturbed energy density coming from the metric lapse perturbation. Again, these are restored in App.~B of \cite{Passaglia:2019ueo}. 
Its relative contribution is negligible.

The jump in $\rho_h'$ 
\begin{equation}
\Delta \left[\rho_h'\right] =  -\VT_{,T} T,
\end{equation}
and in the energy density perturbation
\begin{equation}
\label{eq:higgs_energy_jump}
\Delta \left[\delta \rho_h\right] = \delta h \left( \VT_{,h}  - V_{,h} \right) +\VT_{,T} \delta T,
\end{equation}
imply that the curvature perturbation~\eqref{eq:infinitesimal} on constant Higgs energy density slices is discontinuous at reheating. This instantaneous change in $\zeta_h$ is due to an instantaneous source in the conservation equation from the interaction of the Higgs with the thermal bath. 

However, the instantaneous increase in the Higgs energy density perturbation $\delta \rho_h$ does not come for free. Conservation of energy, which we imposed at the level of the background in Eq.~\eqref{eq:energyconback}, also holds locally. It implies that the increase in the Higgs energy density is counterbalanced by an induced perturbation in the radiation field
\begin{equation}
\label{eq:delta_rho_r}
\delta \rho_{\rm r} (0^+) = 
- \Delta \left[\delta \rho_h\right],
\end{equation}
and therefore that the Higgs and radiation energy densities after uplift are nearly canceling.
In other words, the uplift creates a Higgs-radiation isocurvature fluctuation rather than a 
net curvature fluctuation. To the extent that the Higgs fluctuation then redshifts like radiation, the isocurvature mode does not subsequently contribute to the curvature fluctuation.

The conserved curvature on uniform radiation density slices which corresponds to this induced radiation perturbation is
\begin{equation}
\label{eq:zeta_r}
\zeta_{\rm r} = - \frac{\delta \rho_{\rm r}}{\rho_{\rm r}'} = \frac{\delta T}{T},
\end{equation}
and solving for the radiation perturbation \eqref{eq:delta_rho_r} using the jump in Higgs energy \eqref{eq:higgs_energy_jump} and 
Eq.~\eqref{eq:zeta_r},
we find
\begin{equation}
\label{eq:radcomp}
\delta \rho_{\rm r} (0^+)  = -\delta h \left( \VT_{,h}  - V_{,h} \right) \left(1- \VT_{,T} \frac{T}{\rho_{\rm r}'}\right)^{-1}.
\end{equation}
This induced radiation perturbation comes from the direct interaction of the Higgs with the radiation during the thermal uplift. It is distinct from radiation perturbations corresponding to intrinsic inflaton fluctuations, which are uncorrelated and can be computed separately, or to inflaton perturbations produced by the gravitational influence of the Higgs perturbations during inflation, which are suppressed. 

This radiation perturbation was omitted in Refs.~\cite{Espinosa:2017sgp,Espinosa:2018eve,Espinosa:2018euj,Gross:2018ivp}, though it in fact its role in conserving total energy has a large impact on the final curvature perturbation $\zeta$. In particular, on a constant total density surface, the curvature perturbation is given by Eq.~\eqref{eq:zeta_tot_intro}, reproduced here for convenience,
\begin{equation} 
\label{eq:zeta_tot}
\zeta = \left(1-   \frac{\rho_h'}{\rho_{\rm tot}'} \right)  \zeta_{\rm r} + \frac{\rho_h'}{\rho_{\rm tot}'} \zeta_h.
\end{equation}
Immediately after the uplift of the Higgs potential, $\zeta$ therefore satisfies
\begin{align*}
\label{eq:upliftR}
\zeta (0^+) &= \zeta_h (0^-) \times \frac{\rho_h' (0^-)}{\rho'_{\rm tot} (0^+)} \\ 
&\simeq -\zeta_h (0^-) \times \frac{\rho_h' (0^-)}{12 {H_{\rm end}^2}},\numberthis
\end{align*}
by virtue of Eq.~\eqref{eq:delta_rho_r}. This equation is nothing but the linear jump condition \eqref{eq:linear_jump}, this time derived from linear perturbation theory rather than the $\delta N$ formalism.

To evolve $\zeta(N)$ from its value at $\zeta(0^+)$, we must solve for the Higgs perturbations after uplift. 
Again the most important aspect of this calculation is energy conservation.  
Energy conservation guarantees that the cancellation responsible for the suppression in the starting value
$\zeta(0^+)$ is maintained on a timescale short compared to an $e$-fold.

Although during inflation we assumed the Higgs is a spectator, after potential uplift we jointly solve the Higgs background equation  \eqref{eq:higgs_bg_eom}, now with $V\rightarrow \VT$, and the radiation background equation \eqref{eq:rho_r_eom}, 
with the Hubble rate determined by Friedmann equation. 
For the perturbations we solve the linearized Klein-Gordon equation \eqref{eq:linKG} for the Higgs, now with a perturbed potential
\begin{equation}
\label{eq:perturbed_potential}
\delta V^{Tk}_{,h} = \VT_{,hh} \delta h^k + \VT_{,hT} \delta T^k,
\end{equation}
which accounts for the effect of temperature perturbations.
Again the metric terms are negligible and we exploit here that $\zeta_{\rm r}$ is constant to avoid solving perturbation equations for the radiation component; both are shown to be good approximations in App.~B of \cite{Passaglia:2019ueo}.

To quantify the importance of the induced radiation perturbation \eqref{eq:delta_rho_r}, we chose $\hend$ such that the post-inflationary Higgs contribution to the total $\Delta_\zeta^2 (k_{\rm PBH})$ is $\sim (0.1)^2$. This is the calculation performed in the literature which suggests that primordial black holes can be formed at the classical-roll scale. We then add the induced radiation contribution and see how $\Delta_\zeta$ is affected.

\begin{figure}[t]
\centering
\includegraphics[width=.65\linewidth]{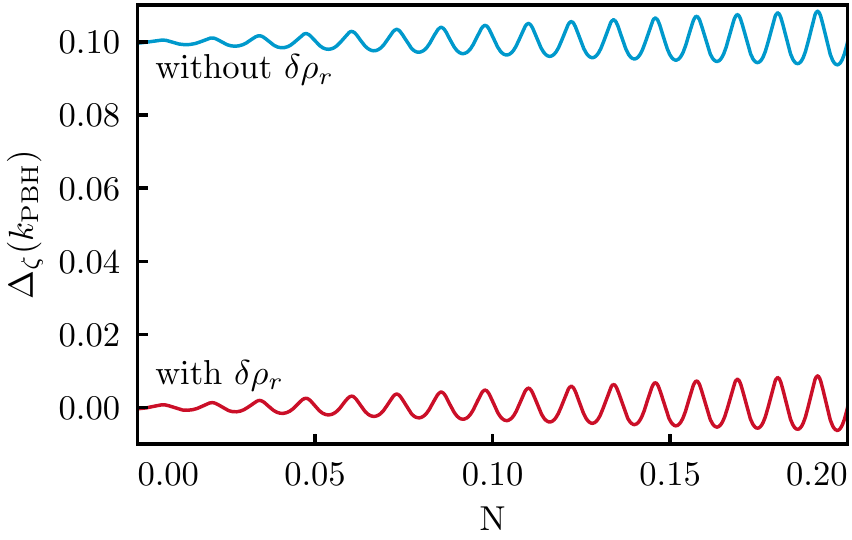}
\caption[Induced radiation perturbations]{The curvature perturbation on uniform total energy 
slices computed in linear perturbation theory 
with and without including radiation perturbations required by energy conservation at
reheating.  These induced perturbations suppress the cycle-averaged curvature by several orders of magnitude (see \S\ref{ssec:linear}).}
\label{fig:zeta_comparison}
\end{figure}

We show these numerical results in Fig.~\ref{fig:zeta_comparison}.  
Here we plot $\Delta_\zeta(k_{\rm PBH})$ with a phase convention $\varphi$ such that the analogous superhorizon Higgs fluctuation 
\begin{equation}
\Delta_{\zeta_h} \equiv e^{i\varphi} \sqrt{ \Delta_{\zeta_h}^2},
\end{equation}
is negative real  during inflation. Note that $\Delta_{\zeta_h}$ changes sign at the potential uplift and becomes positive real.
After inflation, $\Delta_\zeta(k_{\rm PBH})$ oscillates between negative and positive values but stays real.

It is immediately clear from Fig.~\ref{fig:zeta_comparison} that the induced radiation perturbation $\delta \rho_{\rm r}$ suppresses the amplitude of the total curvature $\zeta$ by orders of magnitude, making it much more difficult to achieve the required $R = 0.1$ in this model. For the fiducial background, which was claimed to produce $R=0.1$, by taking into account $\delta \rho_{\rm r}$ we instead have $R(0^+)\simeq 3\times10^{-4}$.

\begin{figure}[t]
\centering
\includegraphics[width=.65\linewidth]{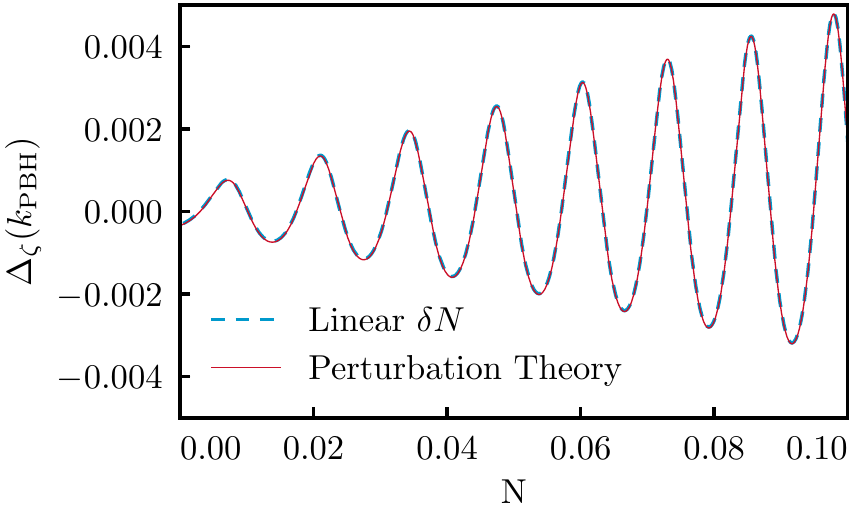}
\caption[Linear theory curvature perturbation after inflation]{The linear perturbation theory calculation including induced radiation fluctuations agrees well with the linearized $\delta N$ result based on \eqref{eq:delta_N_linear}, validating our result that energy conservation suppresses curvature fluctuations (see \S\ref{ssec:linear}). 
}
\label{fig:zeta_validation}
\end{figure}

In Fig.~\ref{fig:zeta_validation}, we validate our calculation of $\Delta_\zeta(k_{\rm PBH})$ by also computing $\zeta$ from the linearized $\delta N$ equation \eqref{eq:delta_N_linear}. The $\delta N$ result relies solely on the behavior of the background equations and thus is an independent check on the rather involved perturbation theory calculations. The $\delta N$ result agrees closely with our perturbation theory calculation and confirms that the induced radiation perturbation is crucial in this mechanism. This test would fail if the radiation compensation in
Eq.~(\ref{eq:radcomp}) were omitted as in Refs.~\cite{Espinosa:2017sgp,Espinosa:2018eve,Espinosa:2018euj,Gross:2018ivp}.

$\zeta$ is not conserved after reheating, and in particular it oscillates due to the changing nonadiabatic pressure induced as the Higgs oscillates.  Even though oscillations in $\delta\rho_h$ are relatively small, the
initial near cancellation between $\delta\rho_h$ and $\delta\rho_{\rm r}$ make them prominent in $\zeta$. Moreover, deviations of the Higgs potential from a simple quadratic with a 
temperature-dependent mass make these oscillations grow in time.
These effects are discussed in detail in App.~B of Ref.~\cite{Passaglia:2019ueo}.
So long as the Higgs decay time, generally of order an $e$-fold, is larger than the oscillation timescale, it is the cycle-averaged $\zeta$ that matters. Because of the initial outgoing trajectory of the Higgs, the instantaneous value of $|\zeta|$ at $N=0^+$ is always larger than the cycle average of the first oscillations. 

To the extent that the cycle-averaged Higgs energy redshifts as radiation, the cycle-averaged value of
$\zeta$ will be conserved. However, the nonquadratic terms in the Higgs potential also cause deviation from this behavior which leads the near cancellation between the Higgs and radiation energy densities gradually to break down.

In particular, Higgs perturbations redshift slightly slower than radiation on the cycle average,\footnote{The rate  at which Higgs perturbations redshift \eqref{eq:redshifting_perts} is different from the rate at which the background Higgs redshifts,
\begin{equation}
\langle \rho_h \rangle \propto a^{-4 -3 \Delta \widebar{w}},
\end{equation}
with $\Delta \widebar{w} \sim -0.002$. This means that there is also internal nonadiabatic stress in the
Higgs field itself and so the cycle averaged $\zeta_h^k$ would also evolve.}
\begin{equation}
\label{eq:redshifting_perts}
\langle \delta \rho_h^k\rangle \propto a^{-4 - 3 \Delta w},
\end{equation}
with $\Delta w \equiv \langle w \rangle - 1/3 \sim -0.004$.
Therefore the cancellation between the radiation piece and the Higgs piece gradually becomes undone, 
\begin{equation}
\zeta^k = - \frac{\delta \rho^k_{\rm r} + \delta \rho^k_h}{\rho_{\rm tot}'} \sim - \frac{\delta \rho^k_{\rm r} + \delta \rho^k_h}{\rho_{\rm r}'} = \zeta^k_{\rm r} + \frac{\delta \rho^k_h}{4 \rho_{\rm r}},
\end{equation}
and $\langle \zeta^k \rangle$ grows gradually. 

Once the Higgs piece dominates, the cycle-average becomes
\begin{equation}
\label{zetagrow}
\langle \zeta^k \rangle (N) \sim - \frac{3}{4} \frac{\delta \rho_h^k(0^+)}{\rho_{\rm r}(0^+)} \Delta w \times N \sim 10^{-3} N.
\end{equation}
Thus the curvature grows to $\O{(10^{-1})}$ only on a timescale
\begin{equation}
\label{eq:many_efolds}
\Delta N \sim 100 \text{ $e$-folds} .
\end{equation}
The Higgs must decay to radiation well before this, and therefore the curvature perturbations cannot become large enough in this scenario to form PBHs.

Note that the details of the post-reheating evolution and in particular the de-cancellation rate derived here do depend on the radiation-Higgs split in \eqref{eq:energyconback} through the temperature dependence of the thermal potential \eqref{eq:VT}. However, conservation of total energy on the timescale of Higgs oscillations imposes that the total curvature can grow only on the $e$-fold timescale, and only due to deviations in the redshifting rate of the different components. Therefore the qualitative result that this growth will take many $e$-folds is robust to our specific implementation here.

In summary, under the assumption that linear theory is valid
through reheating, the Higgs instability mechanism falls far short of being able to form PBHs as the
dark matter.  Models that were previously thought to achieve the required $R=0.1$ in fact produce
$R \lesssim 10^{-3}$ once the radiation density perturbations required by energy conservation at 
reheating are properly accounted for. 

\subsection{Nonlinear Conversion}
\label{ssec:nonlin}

In \S\ref{ssec:linear}, we computed curvature fluctuations assuming that the Higgs perturbations remain linear 
through reheating.  In fact, the Higgs instability induces a breakdown of linear theory when the background position of the Standard Model Higgs at the end of inflation $\hend$ is close to the maximum rescue scale $\hrescue$.  

This breakdown can be seen immediately from Fig.~\ref{fig:rho}. With $\delta N = \zeta_h \sim \pm1$, a typical outwardly perturbed region of the Standard Model Higgs field crosses $\hrescue$ during inflation, gains exceedingly large negative energy and will inevitably backreact on the background trajectory. Reheating will be disrupted, the perturbed Higgs will not be rescued from the unbounded vacuum, and our universe will be destroyed.  Even for smaller $\delta N$ which do not cross $\hrescue$,  the perturbed Higgs energy density is not well represented by the linear Taylor expansion (\ref{eq:linear_jump})  around the background value
due to the extremely rapid evolution of $\rho_h$.

In terms of field interactions, linear theory itself also reveals its own breakdown. At the end of inflation, an order unity $\zeta_h$ leads to an RMS Higgs fluctuation of roughly
\begin{equation}
\delta h_{\rm end} \simeq h'_{\rm end} \simeq -\frac{1}{3 H^2} \lambda \hend^3,
\end{equation}
where we have used the Higgs slow-roll approximation throughout. With $\hend\sim10^3 H$ and $|\lambda|\sim 10^{-2}$ we have 
\begin{equation}
\delta h_{\rm end} \simeq 10^7 H,
\end{equation}
which as we have seen is orders of magnitude larger than the distance between $\hend\simeq 1200H$ and $\hrescue \simeq 2400H$. Moreover, the potential interaction ratio \eqref{eq:potential_linearity}  is   
\begin{equation}
\frac{1}{2} \delta h \frac{V_{,hhh}}{V_{,hh}} \simeq  \frac{\delta h}{4 h} \simeq  10^4 \gg 1,
\end{equation} 
and thus field fluctuations interact. 
Nevertheless, this breakdown has no effect on our previous computation of $\Delta_{\zeta_h}^2$ during inflation since $\zeta_h$ is conserved nonlinearly so long as $H \sim$ const.

Though it was not phrased in terms of a breakdown of linear theory, Ref.~\cite{Gross:2018ivp} noted that the Standard Model Higgs is generally not rescued in this scenario. It was argued in Ref.~\cite{Espinosa:2018euj} that the background $\hend$ can be placed near $\hrescue$ while multiverse and anthropic considerations justify tuning the local Higgs field at the end of inflation such that ${\rm Max}\left[h(\vec{x}) \right] < \hrescue$ everywhere. However, tuning $\delta h (\vec{x}) $ at the end of inflation is  equivalent to tuning $\delta N = \zeta_h$ to be small at the end of inflation and so it directly tunes away the ability to form PBHs as we shall now show.

\begin{figure}[t]
\centering
\includegraphics[width=.65\linewidth]{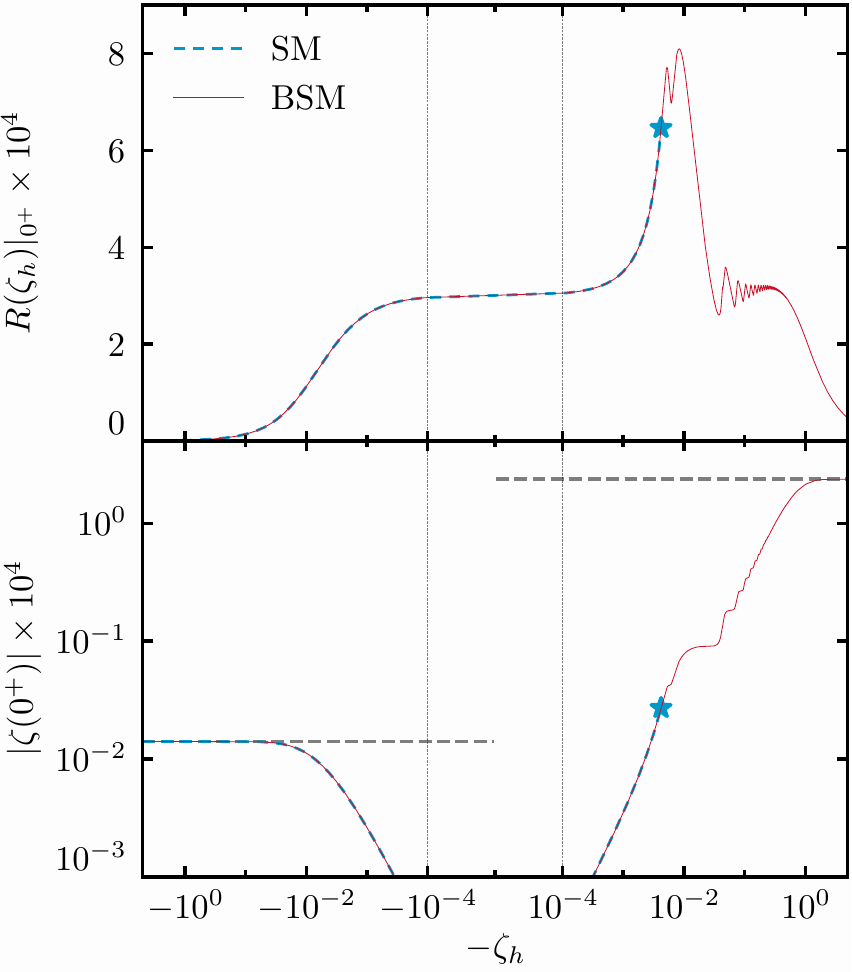}
\caption[Non-linear curvature perturbation mapping]{The  nonlinear mapping of the inflationary $\zeta_h$ to the post-reheating $\zeta (0^+) = \left.R(\zeta_h)\right\vert_{0^+} \zeta_h$. Fluctuations outwards,
toward the instability, correspond to $-\zeta_h>0$. The horizontal axis scale is linear between $\pm 10^{-4}$ and logarithmic elsewhere. 
$R$ deviates from the linear theory 
value $\simeq 3\times10^{-4}$ for fluctuations larger
than about $\pm 10^{-3}$. The horizontal dashed lines indicate analytic saturation values
\eqref{eq:BSMsatin} and \eqref{eq:BSMsatout}.
For the SM Higgs, the maximal $\zeta$ satisfying the rescue condition \eqref{eq:rescue} is marked with a star.
For the BSM Higgs,  $\zeta$ saturates to a maximum.   Neither value is large enough to form PBHs
}
\label{fig:zeta_zetah}
\end{figure}\clearpage

In Fig.~\ref{fig:zeta_zetah} we show $\zeta(0^+)$ as a function of $\zeta_h$ as computed using the nonlinear $\delta N$ formalism 
using Eq.~(\ref{eq:nonlin_jump2}) and Fig.~\ref{fig:rho}.
Linear theory holds for small enough inflationary $\left\vert\zeta_h\right\vert \lesssim 10^{-3}$, but breaks down for perturbations of the typical amplitude produced during inflation. 

Large inwards perturbations away from the instability, shown on the left-hand side of Fig.~\ref{fig:zeta_zetah}, 
saturate to a constant $\zeta(0^+)$ that is independent of $\zeta_h$. These uphill kicks produce
a local Higgs energy density at $N=0^-$ that has a much smaller magnitude than its background value as can be
seen in Fig.~\ref{fig:rho}.
Using Eq.~\eqref{eq:nonlin_jump}, the left-hand side saturation can therefore be written as
\begin{equation}
\zeta^{\rm in}_{\SM} (0^+) =  \frac{- \bar{\rho}_h(0^-)}{12 H^2} \simeq +1.4 \times 10^{-6},
\end{equation}
which we show as a horizontal dashed line on the left-hand side in Fig.~\ref{fig:zeta_zetah}.

Outward perturbations of the SM Higgs toward the instability, shown on the right-hand side of Fig.~\ref{fig:zeta_zetah}, are enhanced relative to linear theory. This is because the amplitude of the energy density of the Higgs shown in Fig.~\ref{fig:rho} grows much faster than expected from a linear approximation. The largest outward perturbations that satisfy the rescue condition \eqref{eq:rescue} produce a curvature
\begin{equation}
\zeta_{\SM}^{\rm out} (0^+) = \frac{\rho_h (\hrescue) - \bar{\rho}_h(0^-)}{12 H^2} \simeq -2.7 \times 10^{-6}.
\end{equation}

Despite the enhancement of the Higgs perturbation relative to linear theory, $\zeta_{\SM}^{\rm out}$ evolves after inflation much like the linear theory $\zeta$ computed in \S\ref{ssec:linear}. The cycle average of $\zeta_{\SM}^{\rm out}(N)$ is smaller than $\zeta_{\SM}^{\rm out} (0^+)$.
So long as the Higgs redshifts like radiation after inflation, this value is then conserved.   Nonlinear evolution
does not change the conclusion of linear theory on PBHs with the SM potential. 

So far in this nonlinear calculation we have kept the background trajectory of the Higgs fixed. We might wonder whether a different background Higgs trajectory, at fixed $H$, can achieve a larger value for the saturating $\zeta$. Moving the background away from the instability suppresses $\bar{\rho}_h(0^-)$, and so the largest curvature perturbation which can be produced in the SM, maximized over the choice of background trajectory, is 
\begin{equation}
\zeta_{\SM}^{\rm max} (0^+) = \zeta_{\SM}^{\rm out}(0^+) - \zeta_{\SM}^{\rm in}(0^+) \simeq -4.1 \times 10^{-6},
\end{equation}
which is still insufficient to form PBHs.  
Therefore since
$\zeta$ is uniquely determined by $\zeta_h$ in this way,
anthropically tuning away $\zeta_h({\vec x})$ or setting it to the edge of rescueability forbids
the Standard Model Higgs from forming  PBHs in sufficient abundance to be the dark matter.
  
Ref.~\cite{Espinosa:2018euj} proposed that the mechanism could function with a BSM potential derived from the addition of a massive scalar as detailed in \S\ref{sec:mechanism}. The massive scalar adds a wall in the potential between the field value where the background Higgs ends inflation $\hend$ and the maximum rescuable point $\hrescue$, preventing the local Higgs from reaching parts of the potential from which it cannot be rescued by thermal uplift at reheating. 

When CMB and PBH modes cross the horizon, the BSM potential behaves like the SM potential and as we have seen in \S\ref{sec:inflation} this means that it generates larger inflationary $\zeta_h$ perturbations on CMB scales than on PBH scales. In addition, so long as the background trajectory never encounters the wall, this model behaves like the SM in linear theory and yields $R\ll 0.1$ as shown in \S\ref{ssec:linear}.

However, whereas typical regions with outward field fluctuations were not rescued in the SM, in the BSM case such regions oscillate in a new minimum of the potential during inflation and then can be safely rescued at reheating. 

To evolve fluctuations through this highly nonlinear process, we again use the nonlinear $\delta N$ formalism described in \S\ref{ssec:deltaN}, just as in the Standard Model case, and the BSM results for $\zeta(0^+)$ are also shown in Fig.~\ref{fig:zeta_zetah}. We compute results using the representative parameter set for the BSM scalar described in \S\ref{sec:mechanism}, and we will later show how our results scale with different choices of model parameters. 

Inward perturbations of the BSM Higgs act just like inward perturbations in the SM and thus again lead to the same saturation
\begin{equation}
\label{eq:BSMsatin}
\zeta^{\rm in}_{\BSM}(0^+) = \zeta^{\rm in}_{\SM}(0^+) \simeq +1.4 \times 10^{-6}.
\end{equation}

Large outward Higgs perturbations hit the BSM potential wall, become trapped in the new minimum at $h\sim m_s$,
lose their kinetic energy, and end inflation with a potential dominated Higgs with energy $V_{\rm min}$.
This leads to a saturating curvature
\begin{equation}
\label{eq:BSMsatout}
\zeta^{\rm out}_{\BSM}(0^+) = \frac{V_{\rm min} - \bar{\rho}_h(0^-)}{12 H_{\rm end}^2} \simeq -2.4 \times 10^{-4},
\end{equation}
which we show with a horizontal dashed line on the right-hand side in Fig.~\ref{fig:zeta_zetah}. 
This value is still too small to form enough PBHs to be the DM. 

For perturbations which do not fully saturate this limit, $\zeta(0^+)$ has a stepped behavior and $R(0^+)$ an oscillatory one as depicted in Fig.~\ref{fig:zeta_zetah}. These features correspond to the energy density oscillations for the BSM Higgs seen in Fig.~\ref{fig:rho}, induced because the Higgs has large oscillations around the potential minimum before settling on the $e$-fold timescale. Note that the approximate equality of the linear theory $R(0^+)$ and the nonlinear $R(0^+)$ corresponding to $\vert{\zeta_h}\vert \simeq 0.1$ is a coincidence: changes to the background position change the linear theory $R(0^+)$ while leaving $R(0^+)$ on this brief plateau fixed.

Once more we might wonder if a different choice of background trajectory could enhance the saturation value, but in fact the maximum over background trajectories can again be computed as
\begin{equation}
\zeta_{\BSM}^{\rm max}(0^+) =  \zeta_{\BSM}^{\rm out}(0^+) - \zeta_{\BSM}^{\rm in}(0^+) \simeq \zeta^{\rm out}_{\BSM}(0^+),
\end{equation}
and therefore PBHs cannot be formed for our fiducial BSM potential no matter the position of the background or the size of initial fluctuation.

Note that so far we have only computed $\zeta(0^+)$ for BSM. We should check whether $\zeta$ evolves significantly after $N=0^+$. We do so again with the $\delta N$ formalism, using the full Eq.~\eqref{eq:delta_N}. Since inward fluctuations lead to a negligible $\zeta(0^+)$, we can select a typical outward field fluctuation with $\zeta_h(\vec{x}) = \Delta_{\zeta_h}(k_{\rm PBH}) \simeq -1$ as an example.

\begin{figure}[t]
\centering
\includegraphics[width=.65\linewidth]{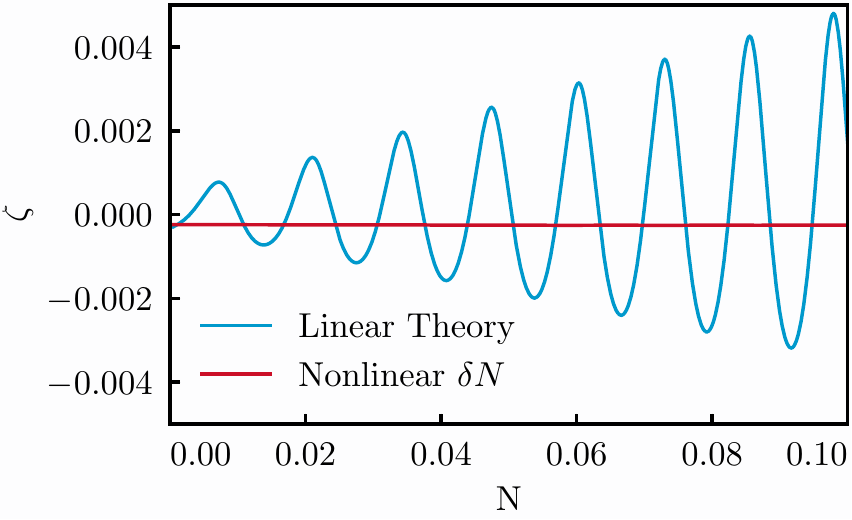}
\caption[Curvature evolution after reheating]{Curvature evolution after reheating. The  BSM nonlinear $\delta N$ result for the local $\zeta(\vec{x})$, from an example inflationary 
$\zeta_h(\vec{x}) = \Delta_{\zeta_h}(k_{\rm PBH})$, is compared to the linear theory approximation $\Delta_{\zeta}(k_{\rm PBH})$ of Fig.~\ref{fig:zeta_validation}. The nonlinear evolution of $\zeta$ after reheating is too small to form PBHs (see \S\ref{ssec:nonlin}). 
}
\label{fig:zeta_nonlin}
\end{figure}

We show this case in Fig.~\ref{fig:zeta_nonlin}. As expected from Fig.~\ref{fig:zeta_zetah}, the nonlinearly evolved $\zeta$ is small, comparable in amplitude  to the linear $\zeta(0^+)$ but not in its evolution.   In fact $\zeta$ evolves negligibly after $N=0^+$ 
and we can robustly conclude that  PBHs are generically not formed nonlinearly in this case. 

This lack of nonlinear evolution can be explained by the difference in the impact of
uplift on the perturbations.
In linear theory the small amplitude of $\zeta$ resulted from large cancellations between the Higgs and radiation perturbations due to energy conservation and
the large impact of uplift. Nonlinearly the impact of uplift is much smaller, bounded by the BSM modification, and so the Higgs energy density fluctuations after reheating are no longer
as dominated by the uplift contribution.   In particular,
\begin{equation}
\frac{\rho_h(0^+,\vec{x}) - \bar{\rho}_h(0^+)}{\rho_h(0^-,\vec{x}) - \bar{\rho}_h(0^-)} \ll \left.\frac{\delta \rho_h (0^+)}{\delta \rho_h(0^-)}\right\vert_{\rm linear},
\end{equation}
where the right-hand side is in linear theory. Therefore the cancellation with radiation is less dramatic than in linear theory. As discussed in \S\ref{ssec:linear}, the cancellation and subsequent decancellation are responsible for the linear theory oscillations and slow secular drift~\eqref{zetagrow}, and therefore all these effects are suppressed in the nonlinear case. For the same reason, the details of the split of the total energy density into Higgs and radiation pieces are also less important for the nonlinear curvature. Our
nonlinear results therefore essentially depend only on energy conservation during reheating.

Finally, the fiducial BSM potential used to compute the results of Fig.~\ref{fig:zeta_zetah} was constructed according to the specifications of Ref.~\cite{Espinosa:2018euj}: it uplifts the Standard Model potential somewhere between $\hend$ and $\hrescue$. We might wonder whether PBHs could be formed by optimizing the position of the uplift so that it as close as possible to $\hrescue$, maximizing the criticality of the scenario. 

The maximum position of $m_s$ will be just before $\hrescue$. This leads to a maximum curvature for this entire scenario of 
\begin{equation}
\label{eq:deltaN_max_BSM}
\textrm{Max}[\zeta_{\BSM}^{\rm max}] =  \frac{V(\hrescue)}{12 H_{\rm end}^2}.
\end{equation}
Using the approximate maximum rescueable field value \eqref{eq:hrescue} and $\lambda = \lambda^{\SM} \sim - 0.007$, we have
\begin{equation}
V(\hrescue^{(1)}) = \frac{1}{4} \lambda^{\SM} \left(\hrescue^{(1)}\right)^4 = -0.012 H_{\rm end}^2,
\end{equation}
which yields
\begin{equation}
\label{eq:zeta_max_max}
\zeta^{\rm max}_{\BSM} \simeq -1.0 \times 10^{-3},
\end{equation}
which depends on $H$ only through the logarithmic evolution of $\lambda^{\SM}$ evaluated at $\hrescue$. This estimate is in good agreement with the computation using the exact value of $\hrescue$ which yields $\zeta^{\rm max}_{\BSM} = -8.2 \times 10^{-4}$.

Therefore no matter the size of inflationary Higgs perturbations, the position of the background Higgs, the SM or BSM nature of the Higgs potential, the position of the BSM wall, or the Hubble rate, this mechanism does not produce perturbations large enough to form PBHs in sufficient abundance to be the dark matter.

Moreover, the largest possible curvature perturbations produced by this model, Eq.~\eqref{eq:zeta_max_max}, are so small that the second-order gravitational waves predicted by Ref.~\cite{Espinosa:2018eve} will be undetectable with LISA.

\begin{figure}[t]
\centering
\includegraphics[width=.65\linewidth]{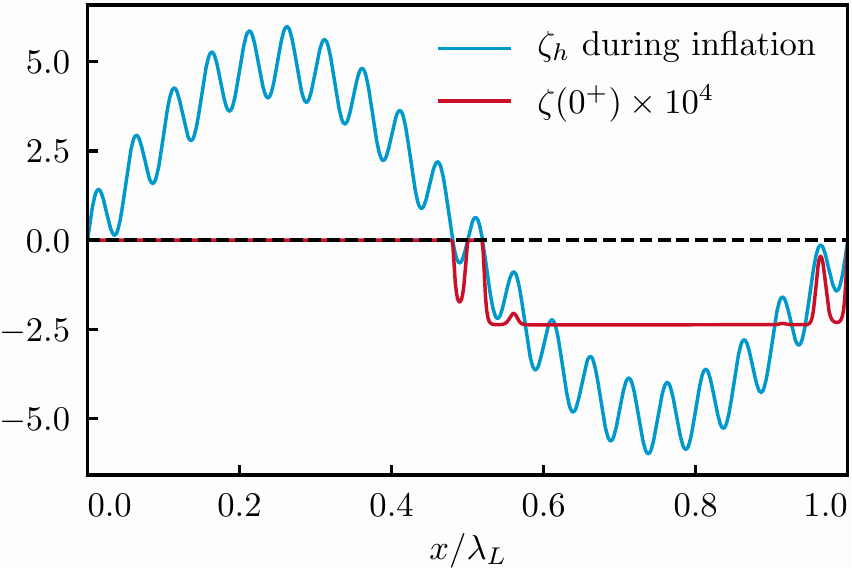}
\caption[Non-linear transformation of real-space curvature field]{The real-space post-inflationary curvature field $\zeta$ (red) produced by the nonlinear transformation of the inflationary $\zeta_h$ (blue). Perturbations on small scales are suppressed when they occur where the long-wavelength $\lambda_L$ mode has saturated the curvature field, and therefore CMB scale perturbations are larger than PBH scale perturbations fully nonlinearly in this mechanism (see \S\ref{ssec:nonlin}). 
}
\label{fig:zeta_real_space}
\end{figure}

We can also ask what happens to CMB scale fluctuations nonlinearly in this model. We showed in \S\ref{sec:inflation} that during inflation $\Delta^2_{\zeta_h}(k_{\rm CMB})  > \Delta^2_{\zeta_h}(k_{\rm PBH})$. In Fig.~\ref{fig:zeta_real_space}, we show the effect of the highly nonlinear local transformation of $\zeta_h$ to $\zeta$ shown for the BSM potential in Fig.~\ref{fig:zeta_zetah} on a cartoon realization of the curvature field on a constant Higgs density surface during inflation.

For visualization purposes, we have generated two modes apart by a factor of only $20$ in scale rather than the $\sim35$ $e$-folds which separate the CMB and PBH modes. We see that fluctuations on the long-wavelength scales cause a saturation of the short-wavelength fluctuations, and therefore fully nonlinearly we have that perturbations are larger on long wavelengths than on small wavelengths in this model as long as this is true of the inflationary $\zeta_h$ itself. Therefore if the reheating scenario is changed somehow to achieve large PBH scale fluctuations, the CMB scale fluctuations will still be larger than PBH modes unless the functional form of the transformation shown in Fig.~\ref{fig:zeta_zetah} is radically altered.

Beyond the specific motivation of PBH DM formation, we can now return to the question of whether  Higgs instability is compatible with the small curvature fluctuations observed in the CMB.

For the SM Higgs, suppressing $\zeta^{\rm out}_{\SM}$ requires placing the background $\hend$ far from $\hrescue$. Specifically, since the typical CMB scale perturbation has $|\zeta_h| \sim 5$, $\hend$ should be moved at least $\sim5$ $e$-folds backwards along its trajectory.

For the BSM potential, with the scalar mass set near $\hrescue$ to ensure that no regions ever fall into the unrescuable region, there are two situations which are compatible with the CMB and one which is not.

If $\hend$ does not approach $\hrescue$, then we return to the linear theory, Standard Model result, where Higgs fluctuations on CMB scales do not lead to large curvature fluctuations since
$\rho_h/H^2$ decreases sharply in Fig.~\ref{fig:rho} and predicts $\zeta$ through Eq.~(\ref{eq:nonlin_jump2}).

Conversely, if the Higgs travels far down the unstable region early in inflation, 
the Higgs becomes uniform in the potential well during inflation and thus leads to no curvature perturbations after inflation. In this region as well the Higgs instability is compatible with the CMB.

It is only in the region of parameter space near our fiducial model, where the background Higgs $\hend$ approaches but does not reach the minimum induced by the BSM massive scalar near $\hrescue$, that Higgs fluctuations on CMB scales can be converted to curvature fluctuations which are large enough to disturb the CMB. 

To avoid this possibility completely, one should set the mass $m_s$ of the scalar field to be slightly smaller than $\hrescue$. In particular to achieve $|\zeta|\lesssim10^{-5}$, using Eq.~\eqref{eq:deltaN_max_BSM} and Eq.~\eqref{eq:zeta_max_max}, one requires
\begin{equation}
\frac{m_s}{\hrescue} \lesssim \left(\frac{1}{100}\right)^{\frac{1}{4}} \lesssim \frac{1}{3}.
\end{equation}

In summary, it is only a special class of Higgs criticality scenarios where the parameters are arranged so
that the regions of the universe fluctuate near the edge of rescuable instability which would
be testable in the CMB and even that class cannot form PBHs as the majority of the dark matter,
nor generate second-order gravitational waves at an amplitude detectable with LISA.

\section{Discussion}
\label{sec:conclusions}

We have definitively shown that the dark matter is not composed exclusively of primordial black holes produced by the collapse of density perturbations generated by a spectator Higgs field during inflation.

While a spectator Higgs evolving on the unstable side of its potential can generate large Higgs fluctuations on PBH scales, even larger Higgs fluctuations are produced on CMB scales. 
This result is obtained using linear perturbation theory during inflation, which we show holds even though the CMB modes cross the horizon at an epoch when the Higgs' per $e$-fold classical roll is smaller than its per $e$-fold stochastic motion, because the stochastic motion is incoherent and does not backreact on the Higgs background. 

Inflation ends well after all relevant modes have crossed the horizon, and when reheating occurs the Higgs potential is uplifted by the interaction between the Higgs and the thermal bath. If the Higgs is rescued from the unstable region by this thermal uplift then the Higgs redshifts as, and eventually decays to, radiation. The CMB and PBH modes are superhorizon at these epochs and evolve in the same way through these processes. Therefore if the Higgs fluctuations are converted into sufficiently large curvature perturbations such that enough PBHs were produced to explain the dark matter, CMB constraints would necessarily be violated.

In fact though, a sufficient abundance of PBHs is never produced and CMB constraints are only violated in cases of near criticality. We first showed that this is true under the assumption that linear theory holds through reheating, where we correct an error in local energy conservation made in the literature.

We then showed that linear theory is violated because the model requires the Higgs to be as close as possible to the maximum value beyond which it cannot be rescued at reheating. This criticality condition leads typical perturbations to evolve nonlinearly, and using the nonlinear $\delta N$ formalism we also show that the Standard Model Higgs, regardless of fine-tuning or anthropic arguments, can never produce enough PBHs to be the DM. Modifying the Higgs potential at large field values can eliminate fine-tuning or anthropic issues, but cannot enhance curvature perturbations significantly enough to explain the dark matter.

\chapter{Conclusion}
\label{chap:conclusion}

Primordial black holes are extreme objects that push inflationary theory to its limits. Over the course of this thesis, we have endeavored to use the breaking points to understand the more general principles that govern the physics of inflation.

To do so, we had to take a wide variety of theoretical approaches. We used the geometric quantities of ADM to build an effective field theory for perturbations, and we solved equations of motion using generalized slow-roll; We computed non-Gaussianities with in-in, and we evolved regions nonlinearly using $\delta N$; And we explored the arcane secrets of perturbation theory in various gauges and reference frames.

Along the way we broached some fascinating topics, like the history of the Higgs in our universe, while other ideas were touched on only briefly, like reheating, stochasticity, tensor modes, and true multi-field effects. These, and further frontiers, are certainly worthy of more detailed study.

Yet in the end, only observation can bring definitive answers in our quest to understand inflation. Guided by theory, existing and upcoming experiments around the world are searching for primordial non-Gaussianity, hunting for primordial black holes, and hoping to catch inflationary gravitational waves. With these clues, or as yet unexpected ones, we will uncover the birth of our universe.

\makebibliography


\begin{thebibliography}{100}

\bibitem{Passaglia:2018afq}
Samuel Passaglia and Wayne Hu.
\newblock {Scalar Bispectrum Beyond Slow-Roll in the Unified EFT of Inflation}.
\newblock {\em Phys. Rev.}, D98(2):023526, 2018.

\bibitem{Ramirez:2018dxe}
H\'ector Ram\'irez, Samuel Passaglia, Hayato Motohashi, Wayne Hu, and Olga
  Mena.
\newblock {Reconciling tensor and scalar observables in G-inflation}.
\newblock {\em JCAP}, 1804(04):039, 2018.

\bibitem{Passaglia:2018ixg}
Samuel Passaglia, Wayne Hu, and Hayato Motohashi.
\newblock {Primordial black holes and local non-Gaussianity in canonical
  inflation}.
\newblock {\em Phys. Rev.}, D99(4):043536, 2019.

\bibitem{Passaglia:2019ueo}
Samuel Passaglia, Wayne Hu, and Hayato Motohashi.
\newblock {Primordial Black Holes as Dark Matter through Higgs Criticality}.
\newblock {\em Phys. Rev. D}, 101:123523, 2020.

\bibitem{Penzias:1965wn}
Arno~A. Penzias and Robert~Woodrow Wilson.
\newblock {A Measurement of excess antenna temperature at 4080-Mc/s}.
\newblock {\em Astrophys. J.}, 142:419--421, 1965.

\bibitem{Smoot:1992td}
George~F. Smoot et~al.
\newblock {Structure in the COBE differential microwave radiometer first year
  maps}.
\newblock {\em Astrophys. J. Lett.}, 396:L1--L5, 1992.

\bibitem{Aghanim:2018eyx}
N.~Aghanim et~al.
\newblock {Planck 2018 results. VI. Cosmological parameters}.

\bibitem{Motohashi:2017kbs}
Hayato Motohashi and Wayne Hu.
\newblock {Primordial Black Holes and Slow-Roll Violation}.
\newblock {\em Phys. Rev.}, D96(6):063503, 2017.

\bibitem{Montero-Camacho:2019jte}
Paulo Montero-Camacho, Xiao Fang, Gabriel Vasquez, Makana Silva, and
  Christopher~M. Hirata.
\newblock {Revisiting constraints on asteroid-mass primordial black holes as
  dark matter candidates}.
\newblock 2019.

\bibitem{Motohashi:2017gqb}
Hayato Motohashi and Wayne Hu.
\newblock {Generalized Slow Roll in the Unified Effective Field Theory of
  Inflation}.
\newblock {\em Phys. Rev.}, D96(2):023502, 2017.

\bibitem{Gao:2014soa}
Xian Gao.
\newblock {Unifying framework for scalar-tensor theories of gravity}.
\newblock {\em Phys. Rev.}, D90:081501, 2014.

\bibitem{Gao:2014fra}
Xian Gao.
\newblock {Hamiltonian analysis of spatially covariant gravity}.
\newblock {\em Phys. Rev.}, D90:104033, 2014.

\bibitem{Horava:2009uw}
Petr Horava.
\newblock {Quantum Gravity at a Lifshitz Point}.
\newblock {\em Phys. Rev.}, D79:084008, 2009.

\bibitem{Blas:2009yd}
D.~Blas, O.~Pujolas, and S.~Sibiryakov.
\newblock {On the Extra Mode and Inconsistency of Horava Gravity}.
\newblock {\em JHEP}, 10:029, 2009.

\bibitem{Blas:2010hb}
Diego Blas, Oriol Pujolas, and Sergey Sibiryakov.
\newblock {Models of non-relativistic quantum gravity: The Good, the bad and
  the healthy}.
\newblock {\em JHEP}, 04:018, 2011.

\bibitem{BenAchour:2016fzp}
Jibril Ben~Achour, Marco Crisostomi, Kazuya Koyama, David Langlois, Karim Noui,
  and Gianmassimo Tasinato.
\newblock {Degenerate higher order scalar-tensor theories beyond Horndeski up
  to cubic order}.
\newblock {\em JHEP}, 12:100, 2016.

\bibitem{Langlois:2017mxy}
David Langlois, Michele Mancarella, Karim Noui, and Filippo Vernizzi.
\newblock {Effective Description of Higher-Order Scalar-Tensor Theories}.
\newblock {\em JCAP}, 1705(05):033, 2017.

\bibitem{Motohashi:2020wxj}
Hayato Motohashi and Wayne Hu.
\newblock {Effective field theory of degenerate higher-order inflation}.
\newblock {\em Phys. Rev. D}, 101:083531, 2020.

\bibitem{Motohashi:2016prk}
Hayato Motohashi, Teruaki Suyama, and Kazufumi Takahashi.
\newblock {Fundamental theorem on gauge fixing at the action level}.
\newblock {\em Phys. Rev.}, D94(12):124021, 2016.

\bibitem{Maldacena:2002vr}
Juan~Martin Maldacena.
\newblock {Non-Gaussian features of primordial fluctuations in single field
  inflationary models}.
\newblock {\em JHEP}, 05:013, 2003.

\bibitem{Chen:2006nt}
Xingang Chen, Min-xin Huang, Shamit Kachru, and Gary Shiu.
\newblock {Observational signatures and non-Gaussianities of general single
  field inflation}.
\newblock {\em JCAP}, 0701:002, 2007.

\bibitem{Pajer:2016ieg}
Enrico Pajer, Guilherme~L. Pimentel, and Jaap V.~S. Van~Wijck.
\newblock {The Conformal Limit of Inflation in the Era of CMB Polarimetry}.
\newblock {\em JCAP}, 1706(06):009, 2017.

\bibitem{Arroja:2011yj}
Frederico Arroja and Takahiro Tanaka.
\newblock {A note on the role of the boundary terms for the non-Gaussianity in
  general k-inflation}.
\newblock {\em JCAP}, 1105:005, 2011.

\bibitem{Rigopoulos:2011eq}
Gerasimos Rigopoulos.
\newblock {Gauge invariance and non-Gaussianity in Inflation}.
\newblock {\em Phys. Rev.}, D84:021301, 2011.

\bibitem{Burrage:2011hd}
Clare Burrage, Raquel~H. Ribeiro, and David Seery.
\newblock {Large slow-roll corrections to the bispectrum of noncanonical
  inflation}.
\newblock {\em JCAP}, 1107:032, 2011.

\bibitem{RenauxPetel:2011sb}
Sebastien Renaux-Petel.
\newblock {On the redundancy of operators and the bispectrum in the most
  general second-order scalar-tensor theory}.
\newblock {\em JCAP}, 1202:020, 2012.

\bibitem{Seery:2005gb}
David Seery and James~E. Lidsey.
\newblock {Primordial non-Gaussianities from multiple-field inflation}.
\newblock {\em JCAP}, 0509:011, 2005.

\bibitem{Adshead:2011bw}
Peter Adshead, Wayne Hu, Cora Dvorkin, and Hiranya~V. Peiris.
\newblock {Fast Computation of Bispectrum Features with Generalized Slow Roll}.
\newblock {\em Phys. Rev.}, D84:043519, 2011.

\bibitem{Cheung:2007sv}
Clifford Cheung, A.~Liam Fitzpatrick, Jared Kaplan, and Leonardo Senatore.
\newblock {On the consistency relation of the 3-point function in single field
  inflation}.
\newblock {\em JCAP}, 0802:021, 2008.

\bibitem{Creminelli:2004yq}
Paolo Creminelli and Matias Zaldarriaga.
\newblock {Single field consistency relation for the 3-point function}.
\newblock {\em JCAP}, 0410:006, 2004.

\bibitem{Creminelli:2011rh}
Paolo Creminelli, Guido D'Amico, Marcello Musso, and Jorge Norena.
\newblock {The (not so) squeezed limit of the primordial 3-point function}.
\newblock {\em JCAP}, 1111:038, 2011.

\bibitem{Adshead:2013zfa}
Peter Adshead, Wayne Hu, and Vinícius Miranda.
\newblock {Bispectrum in Single-Field Inflation Beyond Slow-Roll}.
\newblock {\em Phys. Rev.}, D88(2):023507, 2013.

\bibitem{Cheung:2007st}
Clifford Cheung, Paolo Creminelli, A.~Liam Fitzpatrick, Jared Kaplan, and
  Leonardo Senatore.
\newblock {The Effective Field Theory of Inflation}.
\newblock {\em JHEP}, 03:014, 2008.

\bibitem{Woodard:2015zca}
Richard~P. Woodard.
\newblock {Ostrogradsky's theorem on Hamiltonian instability}.
\newblock {\em Scholarpedia}, 10(8):32243, 2015.

\bibitem{Solomon:2017nlh}
Adam~R. Solomon and Mark Trodden.
\newblock {Higher-derivative operators and effective field theory for general
  scalar-tensor theories}.
\newblock {\em JCAP}, 1802(02):031, 2018.

\bibitem{Horndeski:1974wa}
Gregory~Walter Horndeski.
\newblock {Second-order scalar-tensor field equations in a four-dimensional
  space}.
\newblock {\em Int. J. Theor. Phys.}, 10:363--384, 1974.

\bibitem{Gleyzes:2014qga}
Jérôme Gleyzes, David Langlois, Federico Piazza, and Filippo Vernizzi.
\newblock {Exploring gravitational theories beyond Horndeski}.
\newblock {\em JCAP}, 1502:018, 2015.

\bibitem{Kase:2014cwa}
Ryotaro Kase and Shinji Tsujikawa.
\newblock {Effective field theory approach to modified gravity including
  Horndeski theory and Hořava–Lifshitz gravity}.
\newblock {\em Int. J. Mod. Phys.}, D23(13):1443008, 2014.

\bibitem{DeFelice:2013ar}
Antonio De~Felice and Shinji Tsujikawa.
\newblock {Shapes of primordial non-Gaussianities in the Horndeski's most
  general scalar-tensor theories}.
\newblock {\em JCAP}, 1303:030, 2013.

\bibitem{DeFelice:2011uc}
Antonio De~Felice and Shinji Tsujikawa.
\newblock {Inflationary non-Gaussianities in the most general second-order
  scalar-tensor theories}.
\newblock {\em Phys. Rev.}, D84:083504, 2011.

\bibitem{Gao:2011qe}
Xian Gao and Daniele~A. Steer.
\newblock {Inflation and primordial non-Gaussianities of 'generalized
  Galileons'}.
\newblock {\em JCAP}, 1112:019, 2011.

\bibitem{Fasiello:2014aqa}
Matteo Fasiello and Sébastien Renaux-Petel.
\newblock {Non-Gaussian inflationary shapes in $G^3$ theories beyond
  Horndeski}.
\newblock {\em JCAP}, 1410(10):037, 2014.

\bibitem{Weinberg:2005vy}
Steven Weinberg.
\newblock {Quantum contributions to cosmological correlations}.
\newblock {\em Phys. Rev.}, D72:043514, 2005.

\bibitem{Adshead:2009cb}
Peter Adshead, Richard Easther, and Eugene~A. Lim.
\newblock {The 'in-in' Formalism and Cosmological Perturbations}.
\newblock {\em Phys. Rev.}, D80:083521, 2009.

\bibitem{Adshead:2008gk}
Peter Adshead, Richard Easther, and Eugene~A. Lim.
\newblock {Cosmology With Many Light Scalar Fields: Stochastic Inflation and
  Loop Corrections}.
\newblock {\em Phys. Rev.}, D79:063504, 2009.

\bibitem{Stewart:2001cd}
Ewan~D. Stewart.
\newblock {The Spectrum of density perturbations produced during inflation to
  leading order in a general slow roll approximation}.
\newblock {\em Phys. Rev.}, D65:103508, 2002.

\bibitem{Choe:2004zg}
Jeongyeol Choe, Jinn-Ouk Gong, and Ewan~D. Stewart.
\newblock {Second order general slow-roll power spectrum}.
\newblock {\em JCAP}, 0407:012, 2004.

\bibitem{Dvorkin:2009ne}
Cora Dvorkin and Wayne Hu.
\newblock {Generalized Slow Roll for Large Power Spectrum Features}.
\newblock {\em Phys. Rev.}, D81:023518, 2010.

\bibitem{Hu:2011vr}
Wayne Hu.
\newblock {Generalized Slow Roll for Non-Canonical Kinetic Terms}.
\newblock {\em Phys. Rev.}, D84:027303, 2011.

\bibitem{Kadota:2005hv}
Kenji Kadota, Scott Dodelson, Wayne Hu, and Ewan~D. Stewart.
\newblock {Precision of inflaton potential reconstruction from CMB using the
  general slow-roll approximation}.
\newblock {\em Phys. Rev.}, D72:023510, 2005.

\bibitem{Miranda:2012rm}
Vinicius Miranda, Wayne Hu, and Peter Adshead.
\newblock {Warp Features in DBI Inflation}.
\newblock {\em Phys. Rev.}, D86:063529, 2012.

\bibitem{Adshead:2012xz}
Peter Adshead and Wayne Hu.
\newblock {Fast Computation of First-Order Feature-Bispectrum Corrections}.
\newblock {\em Phys. Rev.}, D85:103531, 2012.

\bibitem{Ohashi:2012wf}
Junko Ohashi and Shinji Tsujikawa.
\newblock {Potential-driven Galileon inflation}.
\newblock {\em JCAP}, 1210:035, 2012.

\bibitem{Miranda:2015cea}
Vinicius Miranda, Wayne Hu, Chen He, and Hayato Motohashi.
\newblock {Nonlinear Excitations in Inflationary Power Spectra}.
\newblock {\em Phys. Rev.}, D93(2):023504, 2016.

\bibitem{Baumann:2011su}
Daniel Baumann and Daniel Green.
\newblock {Equilateral Non-Gaussianity and New Physics on the Horizon}.
\newblock {\em JCAP}, 1109:014, 2011.

\bibitem{Senatore:2009gt}
Leonardo Senatore, Kendrick~M. Smith, and Matias Zaldarriaga.
\newblock {Non-Gaussianities in Single Field Inflation and their Optimal Limits
  from the WMAP 5-year Data}.
\newblock {\em JCAP}, 1001:028, 2010.

\bibitem{Bartolo:2010di}
Nicola Bartolo, Matteo Fasiello, Sabino Matarrese, and Antonio Riotto.
\newblock {Large non-Gaussianities in the Effective Field Theory Approach to
  Single-Field Inflation: the Trispectrum}.
\newblock {\em JCAP}, 1009:035, 2010.

\bibitem{Bartolo:2013exa}
Nicola Bartolo, Dario Cannone, and Sabino Matarrese.
\newblock {The Effective Field Theory of Inflation Models with Sharp Features}.
\newblock {\em JCAP}, 1310:038, 2013.

\bibitem{Kinney:2005vj}
William~H. Kinney.
\newblock {Horizon crossing and inflation with large eta}.
\newblock {\em Phys. Rev.}, D72:023515, 2005.

\bibitem{Manasse:1963}
F.~K. Manasse and C.~W. Misner.
\newblock {Fermi normal coordinates and some basic concepts in differential
  geometry}.
\newblock {\em J. Math. Phys.}, 4(6):735, 1963.

\bibitem{Senatore:2012ya}
Leonardo Senatore and Matias Zaldarriaga.
\newblock {The constancy of $\zeta$ in single-clock Inflation at all loops}.
\newblock {\em JHEP}, 09:148, 2013.

\bibitem{Senatore:2012wy}
Leonardo Senatore and Matias Zaldarriaga.
\newblock {A Note on the Consistency Condition of Primordial Fluctuations}.
\newblock {\em JCAP}, 1208:001, 2012.

\bibitem{Pajer:2013ana}
Enrico Pajer, Fabian Schmidt, and Matias Zaldarriaga.
\newblock {The Observed Squeezed Limit of Cosmological Three-Point Functions}.
\newblock {\em Phys. Rev.}, D88(8):083502, 2013.

\bibitem{Namjoo:2012aa}
Mohammad~Hossein Namjoo, Hassan Firouzjahi, and Misao Sasaki.
\newblock {Violation of non-Gaussianity consistency relation in a single field
  inflationary model}.
\newblock {\em EPL}, 101(3):39001, 2013.

\bibitem{Martin:2012pe}
Jerome Martin, Hayato Motohashi, and Teruaki Suyama.
\newblock {Ultra Slow-Roll Inflation and the non-Gaussianity Consistency
  Relation}.
\newblock {\em Phys. Rev.}, D87(2):023514, 2013.

\bibitem{Starobinsky:1985aa}
Alexei~A. Starobinsky.
\newblock {Multicomponent de Sitter (inflationary) stages and the generation of
  perturbations}.
\newblock {\em JETP Lett.}, 42:152--155, 1985.
\newblock \href{http://www.jetpletters.ac.ru/ps/98/article_1729.shtml}{Pis'ma v
  ZhETF. 42, 124 (1985)}.

\bibitem{Salopek:1990jq}
D.~S. Salopek and J.~R. Bond.
\newblock {Nonlinear evolution of long wavelength metric fluctuations in
  inflationary models}.
\newblock {\em Phys. Rev.}, D42:3936--3962, 1990.

\bibitem{Sasaki:1995aw}
Misao Sasaki and Ewan~D. Stewart.
\newblock {A General analytic formula for the spectral index of the density
  perturbations produced during inflation}.
\newblock {\em Prog. Theor. Phys.}, 95:71--78, 1996.

\bibitem{Sugiyama:2012tj}
Naonori~S. Sugiyama, Eiichiro Komatsu, and Toshifumi Futamase.
\newblock {$\delta$N formalism}.
\newblock {\em Phys. Rev.}, D87(2):023530, 2013.

\bibitem{Chen:2013eea}
Xingang Chen, Hassan Firouzjahi, Eiichiro Komatsu, Mohammad~Hossein Namjoo, and
  Misao Sasaki.
\newblock {In-in and $\delta N$ calculations of the bispectrum from
  non-attractor single-field inflation}.
\newblock {\em JCAP}, 1312:039, 2013.

\bibitem{Cai:2017bxr}
Yi-Fu Cai, Xingang Chen, Mohammad~Hossein Namjoo, Misao Sasaki, Dong-Gang Wang,
  and Ziwei Wang.
\newblock {Revisiting non-Gaussianity from non-attractor inflation models}.
\newblock {\em JCAP}, 1805(05):012, 2018.

\bibitem{Pattison:2017mbe}
Chris Pattison, Vincent Vennin, Hooshyar Assadullahi, and David Wands.
\newblock {Quantum diffusion during inflation and primordial black holes}.
\newblock {\em JCAP}, 1710(10):046, 2017.

\bibitem{Cabass:2016cgp}
Giovanni Cabass, Enrico Pajer, and Fabian Schmidt.
\newblock {How Gaussian can our Universe be?}
\newblock {\em JCAP}, 1701(01):003, 2017.

\bibitem{Ade:2015ava}
P.~A.~R. Ade et~al.
\newblock {Planck 2015 results. XVII. Constraints on primordial
  non-Gaussianity}.
\newblock {\em Astron. Astrophys.}, 594:A17, 2016.

\bibitem{Garcia-Bellido:2017mdw}
Juan Garc\'ia-Bellido and Ester Ruiz~Morales.
\newblock {Primordial black holes from single field models of inflation}.
\newblock {\em Phys. Dark Univ.}, 18:47--54, 2017.

\bibitem{Pattison:2018bct}
Chris Pattison, Vincent Vennin, Hooshyar Assadullahi, and David Wands.
\newblock {The attractive behaviour of ultra-slow-roll inflation}.
\newblock 2018.

\bibitem{Hu:2016wfa}
Wayne Hu and Austin Joyce.
\newblock {Separate Universes beyond General Relativity}.
\newblock {\em Phys. Rev.}, D95(4):043529, 2017.

\bibitem{Cicoli:2018asa}
Michele Cicoli, Victor~A. Diaz, and Francisco~G. Pedro.
\newblock {Primordial Black Holes from String Inflation}.
\newblock {\em JCAP}, 1806(06):034, 2018.

\bibitem{Ozsoy:2018flq}
Ogan {\"O}zsoy, Susha Parameswaran, Gianmassimo Tasinato, and Ivonne Zavala.
\newblock {Mechanisms for Primordial Black Hole Production in String Theory}.
\newblock {\em JCAP}, 1807(07):005, 2018.

\bibitem{Byrnes:2018txb}
Christian~T. Byrnes, Philippa~S. Cole, and Subodh~P. Patil.
\newblock {Steepest growth of the power spectrum and primordial black holes}.

\bibitem{Atal:2018neu}
Vicente Atal and Cristiano Germani.
\newblock {The role of non-gaussianities in Primordial Black Hole formation}.

\bibitem{Saito:2008em}
Ryo Saito, Jun'ichi Yokoyama, and Ryo Nagata.
\newblock {Single-field inflation, anomalous enhancement of superhorizon
  fluctuations, and non-Gaussianity in primordial black hole formation}.
\newblock {\em JCAP}, 0806:024, 2008.

\bibitem{Smith:2011}
Kendrick~M. {Smith} and Marilena {LoVerde}.
\newblock {Local stochastic non-Gaussianity and N-body simulations}.
\newblock {\em Journal of Cosmology and Astro-Particle Physics}, 2011:009,
  November 2011.

\bibitem{Byrnes:2012yx}
Christian~T. Byrnes, Edmund~J. Copeland, and Anne~M. Green.
\newblock {Primordial black holes as a tool for constraining non-Gaussianity}.
\newblock {\em Phys. Rev.}, D86:043512, 2012.

\bibitem{Ferraro:2014jba}
Simone Ferraro and Kendrick~M. Smith.
\newblock {Using large scale structure to measure $f_{NL}, g_{NL}$ and
  $\tau_{NL}$}.
\newblock {\em Phys. Rev.}, D91(4):043506, 2015.

\bibitem{Young:2015kda}
Sam Young and Christian~T. Byrnes.
\newblock {Signatures of non-gaussianity in the isocurvature modes of
  primordial black hole dark matter}.
\newblock {\em JCAP}, 1504(04):034, 2015.

\bibitem{Tada:2015noa}
Yuichiro Tada and Shuichiro Yokoyama.
\newblock {Primordial black holes as biased tracers}.
\newblock {\em Phys. Rev.}, D91(12):123534, 2015.

\bibitem{Franciolini:2018vbk}
G.~Franciolini, A.~Kehagias, S.~Matarrese, and A.~Riotto.
\newblock {Primordial Black Holes from Inflation and non-Gaussianity}.
\newblock {\em JCAP}, 1803(03):016, 2018.

\bibitem{Gross:2018ivp}
Christian Gross, Antonello Polosa, Alessandro Strumia, Alfredo Urbano, and Wei
  Xue.
\newblock {Dark Matter in the Standard Model?}
\newblock {\em Phys. Rev.}, D98(6):063005, 2018.

\bibitem{Espinosa:2015qea}
Jose~R. Espinosa, Gian~F. Giudice, Enrico Morgante, Antonio Riotto, Leonardo
  Senatore, Alessandro Strumia, and Nikolaos Tetradis.
\newblock {The cosmological Higgstory of the vacuum instability}.
\newblock {\em JHEP}, 09:174, 2015.

\bibitem{Espinosa:2017sgp}
J.~R. Espinosa, D.~Racco, and A.~Riotto.
\newblock {Cosmological Signature of the Standard Model Higgs Vacuum
  Instability: Primordial Black Holes as Dark Matter}.
\newblock {\em Phys. Rev. Lett.}, 120(12):121301, 2018.

\bibitem{Espinosa:2018euj}
J.~R. Espinosa, D.~Racco, and A.~Riotto.
\newblock {Primordial Black Holes from Higgs Vacuum Instability: Avoiding
  Fine-tuning through an Ultraviolet Safe Mechanism}.
\newblock {\em Eur. Phys. J.}, C78(10):806, 2018.

\bibitem{Buttazzo:2013uya}
Dario Buttazzo, Giuseppe Degrassi, Pier~Paolo Giardino, Gian~F. Giudice,
  Filippo Sala, Alberto Salvio, and Alessandro Strumia.
\newblock {Investigating the near-criticality of the Higgs boson}.
\newblock {\em JHEP}, 12:089, 2013.

\bibitem{ATLAS:2014wva}
ATLAS, CDF, CMS, and {D0~Collaborations}.
\newblock {First combination of Tevatron and LHC measurements of the top-quark
  mass}.
\newblock 2014.

\bibitem{Khachatryan:2015hba}
Vardan Khachatryan et~al.
\newblock {Measurement of the top quark mass using proton-proton data at
  ${\sqrt{(s)}}$ = 7 and 8 TeV}.
\newblock {\em Phys. Rev.}, D93(7):072004, 2016.

\bibitem{Aaboud:2016igd}
Morad Aaboud et~al.
\newblock {Measurement of the top quark mass in the $t\bar{t}\to$ dilepton
  channel from $\sqrt{s}=8$ TeV ATLAS data}.
\newblock {\em Phys. Lett.}, B761:350--371, 2016.

\bibitem{TevatronElectroweakWorkingGroup:2016lid}
Tevatron Electroweak~Working Group and T.~Aaltonen.
\newblock {Combination of CDF and D0 results on the mass of the top quark using
  up $9.7\:{\rm fb}^{-1}$ at the Tevatron}.
\newblock 2016.

\bibitem{Niikura:2017zjd}
Hiroko Niikura et~al.
\newblock {Microlensing constraints on primordial black holes with the
  Subaru/HSC Andromeda observation}.

\bibitem{Wands:2000dp}
David Wands, Karim~A. Malik, David~H. Lyth, and Andrew~R. Liddle.
\newblock {A New approach to the evolution of cosmological perturbations on
  large scales}.
\newblock {\em Phys. Rev.}, D62:043527, 2000.

\bibitem{EliasMiro:2012ay}
Joan Elias-Miro, Jose~R. Espinosa, Gian~F. Giudice, Hyun~Min Lee, and
  Alessandro Strumia.
\newblock {Stabilization of the Electroweak Vacuum by a Scalar Threshold
  Effect}.
\newblock {\em JHEP}, 06:031, 2012.

\bibitem{Starobinsky:1986fx}
Alexei~A. Starobinsky.
\newblock {Stochastic de Sitter (inflationary) stage in the early universe}.
\newblock {\em Lect. Notes Phys.}, 246:107--126, 1986.

\bibitem{Starobinsky:1994bd}
Alexei~A. Starobinsky and Junichi Yokoyama.
\newblock {Equilibrium state of a selfinteracting scalar field in the De Sitter
  background}.
\newblock {\em Phys. Rev.}, D50:6357--6368, 1994.

\bibitem{Motohashi:2015hpa}
Hayato Motohashi and Wayne Hu.
\newblock {Running from Features: Optimized Evaluation of Inflationary Power
  Spectra}.
\newblock {\em Phys. Rev.}, D92(4):043501, 2015.

\bibitem{Espinosa:2018eve}
José~Ramón Espinosa, Davide Racco, and Antonio Riotto.
\newblock {A Cosmological Signature of the SM Higgs Instability: Gravitational
  Waves}.
\newblock {\em JCAP}, 1809(09):012, 2018.

\bibitem{Lyth:2004gb}
David~H. Lyth, Karim~A. Malik, and Misao Sasaki.
\newblock {A General proof of the conservation of the curvature perturbation}.
\newblock {\em JCAP}, 0505:004, 2005.

\end{thebibliography}
\end{document}